\documentclass[prx,twocolumn,aps, secnumarabic,balancelastpage,amsmath,amssymb,superscriptaddress,floatfix,longbibliography]{revtex4-1}
\usepackage[utf8]{inputenc}
\DeclareUnicodeCharacter{2009}{\,}

\usepackage{graphicx}
\usepackage{bm}
\usepackage{float}
\usepackage{todonotes}
\usepackage{dsfont}
\usepackage{hyperref}
\usepackage{soul}
\usepackage{bbold}
\usepackage{amsfonts}
\usepackage{makecell} 
\usepackage{pifont}
\usepackage{enumerate}

\usepackage{multirow}
\usepackage{amsthm}

\newcommand{\rfig}[1]{Fig.~\ref{#1}}
\newcommand{\rfigP}[2]{Fig.~\ref{#1}({#2})}
\newcommand{\refsec}[1]{Sec.~\ref{#1}}

\newcommand{\Tr}{\text{Tr}}

\newcommand{\wUf}{\widetilde{U}_f}

\newtheorem{lemma}{Lemma}
\newtheorem{theorem}{Theorem}

\newcommand{\Uf}{U_f}

\newcommand{\eqnref}[1]{Eq.~(\ref{#1})}
\newcommand{\refthm}[1]{Theorem~(\ref{#1})}

\newcommand{\cH}{\mathcal{H}}

\newcommand{\bra}[1]{\langle #1 |}
\newcommand{\ket}[1]{| #1 \rangle}

\newcommand{\figOne}{
  
  \begin{figure}
    \centering
    \includegraphics[width = 3.2in]{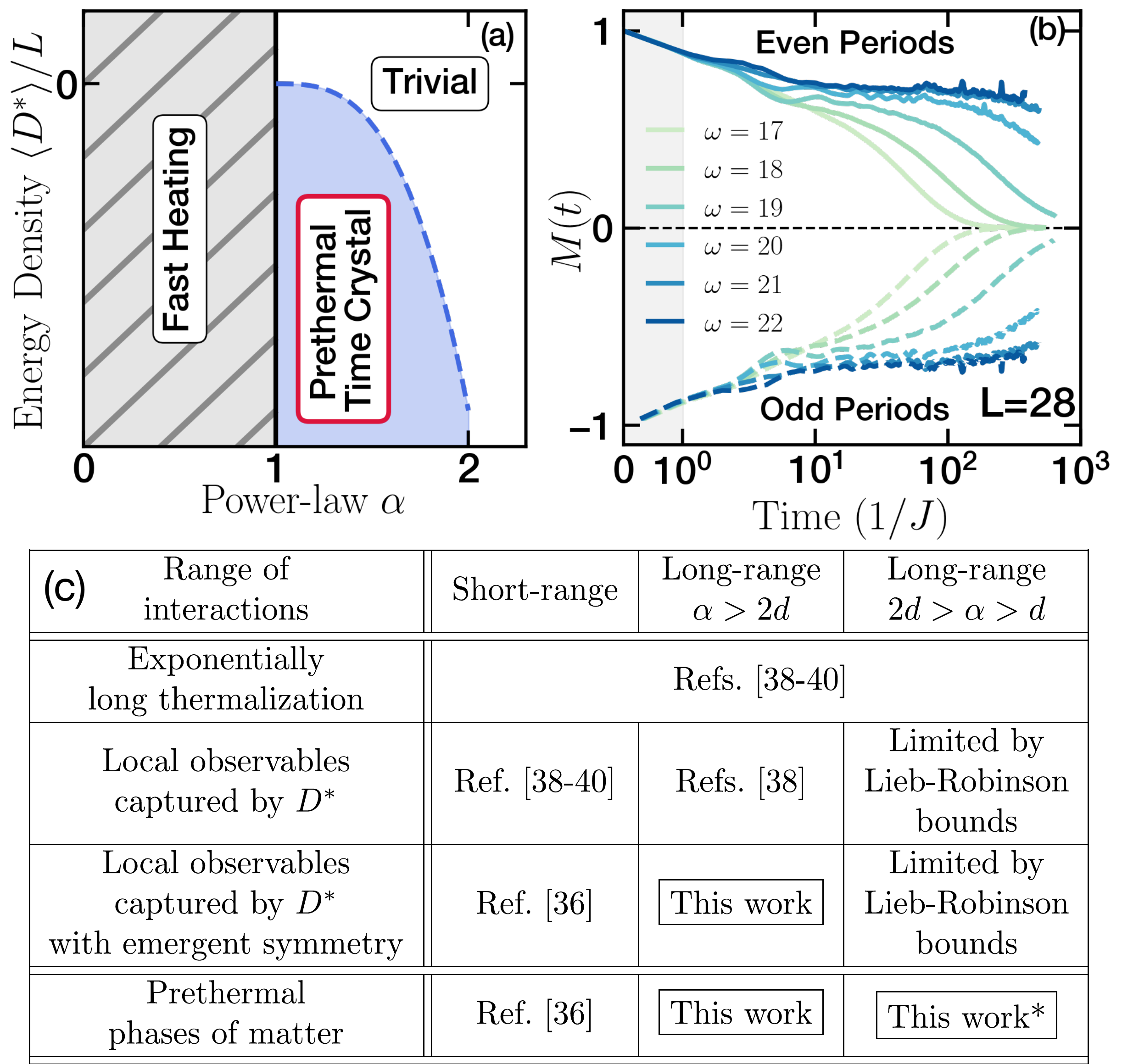}
    \caption{
      {\bf (a)}
      Schematic phase diagram for a one-dimensional prethermal time crystal as a function of interaction power-law and energy density. The 1D PDTC can only exist for long-range interactions (i.e.,~$J_{ij} \propto |i-j|^{-\alpha}$) with power-law $1<\alpha<2$ and an energy density that lies in the symmetry broken phase of the prethermal Hamiltonian $D^*$.
      {\bf (b)} PDTC Floquet dynamics depicting the magnetization $M(t)$ for a system size $L=28$.
      The robust period doubling behavior, which survives for \emph{exponentially} long times in the frequency of the drive $\omega$, signals  prethermal time crystalline order.
      {\bf (c)} Table summarizing our analytical results. The star indicates that for this case, prethermal phases exist provided that we assume that local observables to relax to the Gibbs state of $D^*$, which we expect since this is the state that maximizes the entropy subject to the constraint of conservation of energy.
    }
    \label{fig1}
  \end{figure}
}

\newcommand{\figComparison}{
  \begin{figure*}[t]
    \centering
    \includegraphics[width = 7in]{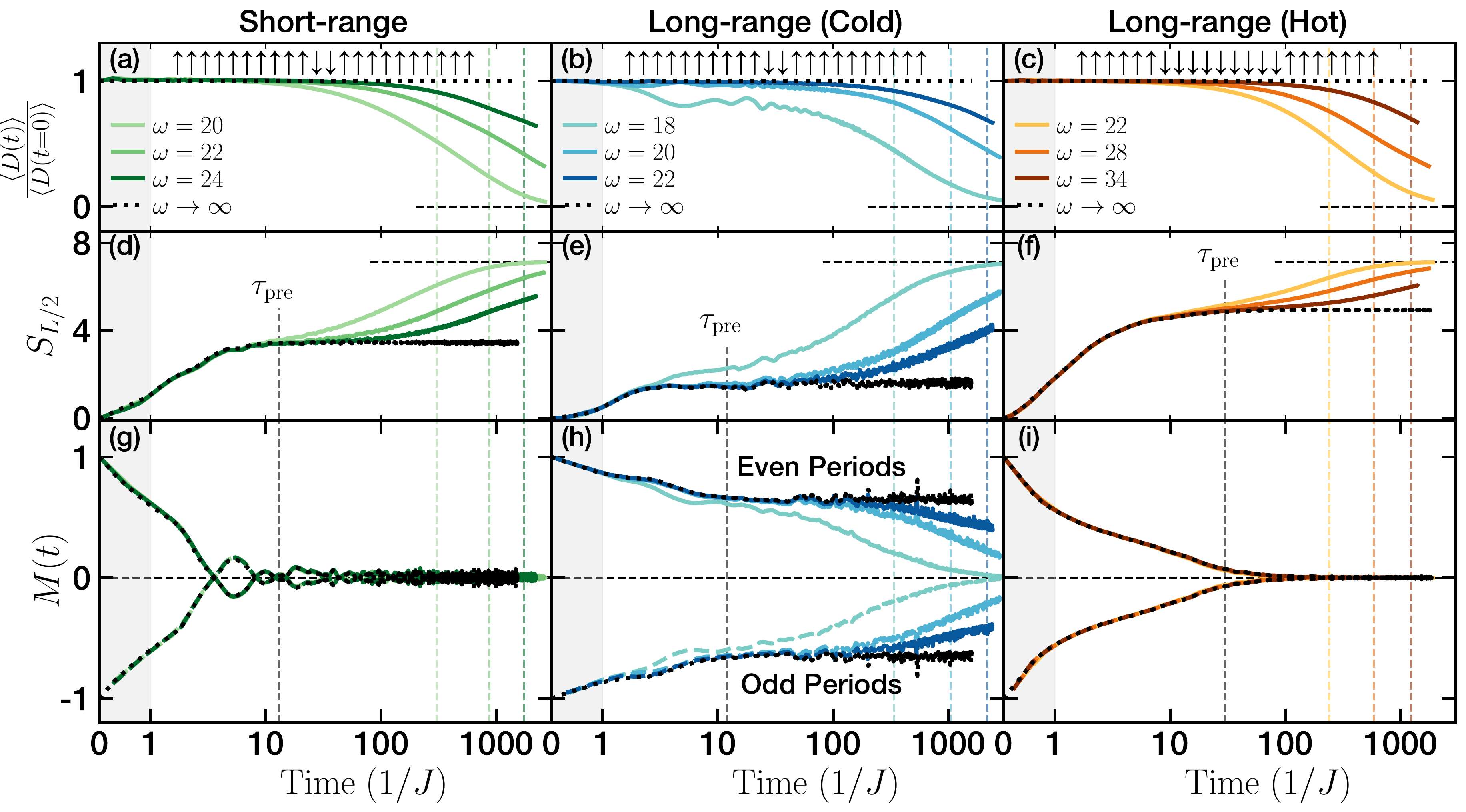}
    \caption{
      Evolution of an L=22 spin chain under the short-range model (left column) and the long-range model. For the latter, we consider a ``cold'' initial state near the top of the spectrum of $D^*$ (center column) and another ``hot'' state near the center of the spectrum (right column).
      {\bf (a-c)} Evolution of the energy density $\langle D\rangle/L$.
      Regardless of the model or initial state, the heating time scale $\tau^*$ (which measures the approach to infinite temperature) scales \emph{exponentially} in the frequency of the drive. 
      {\bf (d-f)} Evolution of the entanglement entropy $S_{L/2}$.
      At intermediate times and large frequencies we observe a plateau corresponding to the entanglement entropy of the prethermal state, as independently corroborated by the evolution under the $\omega \to \infty$ limit of our Floquet evolution (captured on even periods by the evolution under $D$). 
      Analogous to the energy density, at late times ($t>\tau^*$), the entanglement entropy approaches its infinite temperature value of $ (L\log(2) - 1) / 2$ \cite{Page_1993}. 
      {\bf (g-i)} Evolution of $M(t)$ for even (full line) and odd periods (dashed line).
      In both the short-range model (g) and the ``hot'' long-range initial state (i), any period doubling behavior of the magnetization quickly decays as the system approaches, independently of the frequency of, the prethermal state at $\tau_{\text{pre}}$.
      By contrast, in the ``cold'' long-range initial state (h), the magnetization exhibits a robust period doubling behavior for as long as the energy density remains conserved; the decay of both quantities occurs at $\tau^* =\mathcal{O}(e^{\omega/J_{\text{local}}})$ and the prethermal time crystal is robust.
      This distinction is even clearer when considering the $\omega\to\infty$ limit of our Floquet evolution, where the magnetization shows no signs of decay.
    }
  \label{fig:LongvsShort_evo}
  \end{figure*}
}

\newcommand{\figOperatorExpansion}{ 
  \begin{figure}
    \centering
    \includegraphics[width = 3.2in]{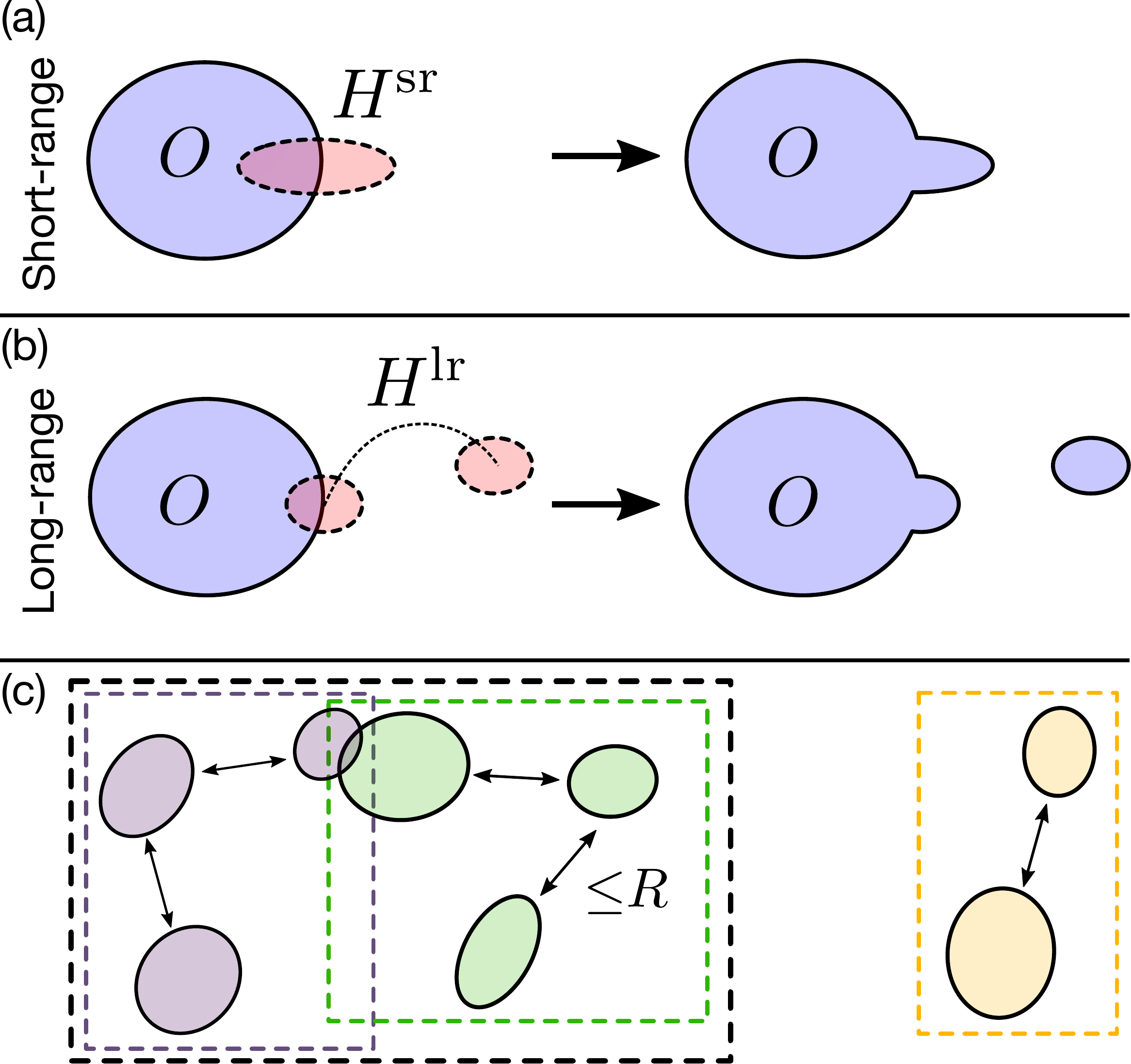}
    \caption{
      {\bf (a) [(b)]}
      Illustration of operator spread via the action of a short- [long]-range Hamiltonian, Eqs.~(\ref{eq:ex_sr_Hamil})~[(\ref{eq:ex_lr_Hamil})]. 
      In the short-range case (a), the operator remains close to its original location.
      For the operator to spread to a far away location, it requires many actions of $H^{\text{sr}}$ which leads to a correspondingly large increase in its support;
      the range and support are closely related notions of size.
      In the long-range case (b), this need not be the case.
      The operator can very quickly spread across the system without a significant increase to its support; the range and the support of the operator capture very different notions of size.
      {\bf c)}
      An $R$-ranged set is a set where any two elements can be connected via a sequence of ``jumps'' (within the set) of size no greater than $R$. 
      We illustrate this concept with the gray, green and orange sets, each representing a different $R$-ranged set.
      Crucially, this definition is closed: when two $R$-ranged sets have a non-empty intersection, their union is also an $R$-ranged set (e.g. the gray and green sets).
      If they do not intersect, the union of two $R$-ranged sets need not form an $R$-ranged set (e.g. the green and orange sets).
}
    \label{fig:OperatorExpansion}
  \end{figure}
}

\newcommand{\figDecayMain}{
  \begin{figure}
  \centering
  \includegraphics[width =3.4in]{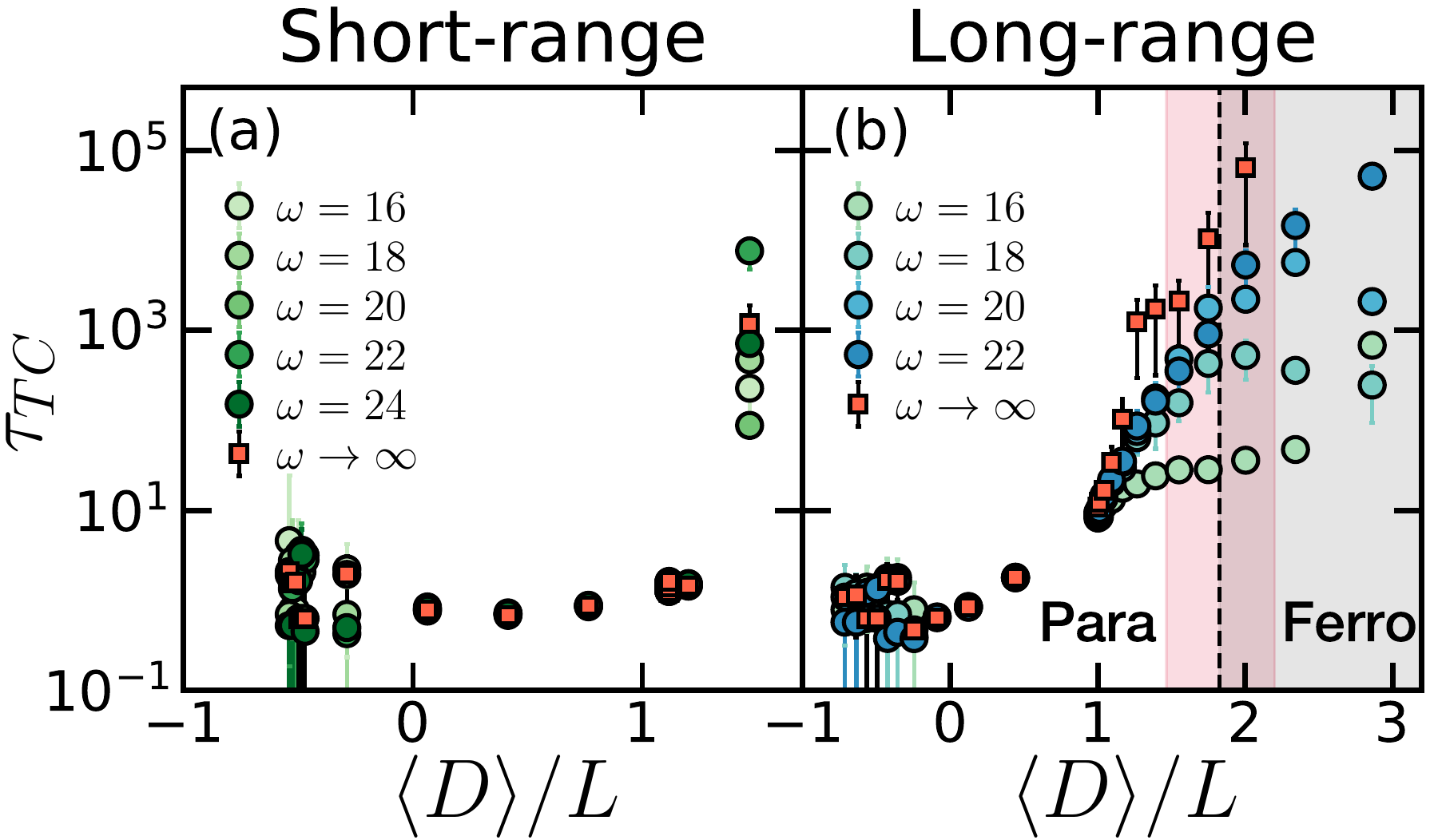} 
  \caption{
   {\bf (a[b])} Decay time scale of the time crystalline order parameter $\tau_{TC}$ as a function of the energy density of the initial state for the short[long]-range model.   In the short-range model (a), $\tau_{\text{TC}}$ is fast, independent of frequency, and in agreement
with the decay time scale of the magnetization $M(t)$ if the system were evolved according to $D$ alone (red squares).
   In the long-range model (b), an analogous  behavior occurs near the center of the spectrum.
   However, as one moves to higher energies across the para- to ferromagnetic phase transition (red shaded region), $\tau_{\text{TC}}$ becomes exponentially dependent on the frequency of the drive and $\tau_{TC}$ approaches $\tau^*$.
   In this regime, $\tau_{\text{TC}}$ is set by the exponentially slow heating rather than the prethermal dynamics for all frequencies---the prethermal time crystal is stable.
  }
  \label{fig:spectrum}
\end{figure}
}

\newcommand{\figSchematic}{
  \begin{figure}[t]
    \centering
    \includegraphics[width = 3.2in]{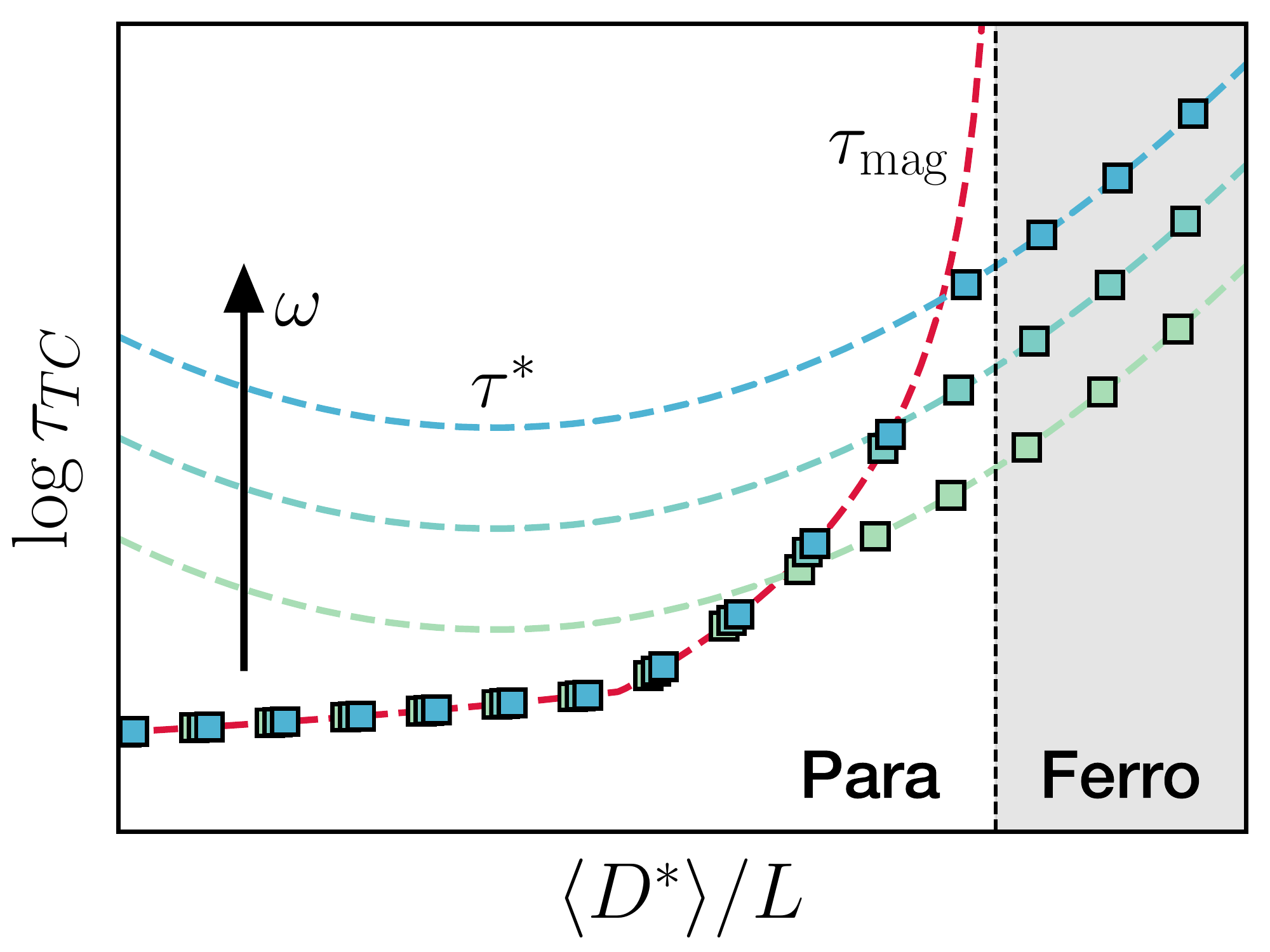} 
    \caption{
      Schematic explanation of the behavior near the transition of the long-range model (Fig.~\ref{fig:spectrum}).
      There are two competing time scales:
      the heating time $\tau^*$,
      and the magnetization decay time $\tau_{\text{mag}}$ of the prethermal Hamiltonian $D^*$ [captured by the red squares in Fig.~\ref{fig:spectrum}(a,b)].
      As the system approaches the phase transition into the ferromagnetic phase (shaded region) from the paramagnetic side, $\tau_{\text{mag}}$ diverges (red dashed line).
      The relaxation time $\tau_{\text{TC}}$ is given by the \emph{smaller} of these two time-scales.
      In (most of) the paramagnetic phase,  $\tau_{\text{mag}}$ is smaller and approximately frequency independent;
      while in the ferromagnetic phase, $\tau^*$ is smaller; $\tau_{TC}$ shares its strong frequency dependence.
    }
    \label{fig:schematic}
  \end{figure}
}

\begin{document}

\title{Long-range Prethermal Phases of Non-equilibrium   Matter}

\author{Francisco Machado}
\affiliation{Department of Physics, University of California, Berkeley, CA 94720, USA}

\author{Dominic V. Else}
\affiliation{Department of Physics, Massachusetts Institute of Technology, Cambridge, MA 02139, USA}
\affiliation{Department of Physics, University of California, Santa Barbara, CA 93106 USA}

\author{Gregory D. Kahanamoku-Meyer}
\affiliation{Department of Physics, University of California, Berkeley, CA 94720, USA}

\author{Chetan Nayak}
\affiliation{Department of Physics, University of California, Santa Barbara, CA 93106 USA}
\affiliation{Station Q, Microsoft Research, Santa Barbara, CA 93106-6105, USA}

\author{Norman Y. Yao}
\affiliation{Department of Physics, University of California, Berkeley, CA 94720, USA}
\affiliation{Materials Sciences Division, Lawrence Berkeley National Laboratory, Berkeley CA 94720, USA}

\begin{abstract}
  We prove the existence of non-equilibrium phases of matter in the prethermal regime of periodically-driven, long-range interacting systems, with power-law exponent  $\alpha > d$, where $d$ is the dimensionality of the system.
  In this context, we predict the existence of a disorder-free, prethermal discrete time crystal in one dimension---a phase strictly forbidden in the absence of long-range interactions. 
  Finally, using a combination of analytic and numerical methods, we highlight key experimentally observable differences between such a prethermal time crystal and its many-body localized counterpart.

\end{abstract}

\maketitle

\section{Introduction}


%
%

%
Periodic driving represents one of the most versatile tools for manipulating quantum systems. 
Classic examples of this abound in magnetic resonance spectroscopy, where it has been used for more than half a century to help narrow spectral line-shapes \cite{Waugh_1968,Mansfield_1971,Rhim_1973}. 
More recently, in the context of cold atomic gases, periodic driving has also helped to enable the realization of novel types of many-body interactions \cite{Jaksch_2003, Aidelsburger_2011, Lindner_2011,Dai_1602}.

Despite this ubiquity, one place where periodically driven (Floquet) systems have traditionally remained absent is in the study of phases of matter \cite{Moessner_1701,Else_1905,Harper_1905}.
Indeed, the usual, statistical mechanical framework for characterizing phases has largely been restricted to the exploration of systems at or near \emph{equilibrium}.
Floquet systems do not fit this category. Rather, they can continuously absorb energy from the driving field, ultimately approaching an infinite-temperature thermal state at late times \cite{Prosen_9707, Prosen_9808,DAlessio_1210,Lazarides_1403,DAlessio_1402,Bukov_1407,Ponte_1403, Bukov_1507, Bordia_1607, Weidinger_1609,Luitz_1706,Haldar_1803,zhu2019dicke}.
As a result, in the thermodynamic limit, the naive conventional wisdom is that all many-body, Floquet systems must behave trivially from the perspective of phases of matter.
However, seminal recent works have called this assumption into question.

For example, the presence of strong disorder in one dimension (and possibly higher dimensions) can prevent thermalization by inducing a many-body localized (MBL) phase \cite{Nandkishore_1404,Abanin_1804}.
When an MBL phase occurs in a Floquet system \cite{DAlessio_1210,Ponte_1403,Ponte_1410,Abanin_1412} it can prevent energy absorption from the drive and lead to novel, intrinsically out-of-equilibrium phases of matter \cite{Khemani_1508, vonKeyserlingk_1602_a, vonKeyserlingk_1602_b,vonKeyserlingk_1605,
Else_1602, Potter_1602, 
Else_1603,Yao_1608}.
%
  %
However, the dual constraints of strong disorder and low dimensionality significantly limit the scope of both the experiments and models that one can consider, naturally raising the question: can interesting Floquet phase structure survive in periodically driven systems \emph{without} disorder?

\figOne


%

An affirmative answer has recently emerged \cite{Else_1607} in the context of Floquet prethermalization \cite{Abanin_1507,Kuwahara_1508, Mori_1509,Abanin_1509,Abanin_1510,Ho_1706}.
For sufficiently large driving frequencies, a many-body Floquet system can enter a so-called ``prethermal regime'', where its dynamics are well captured by an \emph{effective} static Hamiltonian.
 This static Hamiltonian description necessitates the existence of a conserved energy, which prevents the driven system from heating to an infinite temperature state.
Crucially, the \emph{lifetime} of this prethermal regime has been proven to be \emph{exponentially} long in the frequency of the drive,
providing a parametrically robust mechanism to delay the onset of Floquet heating.

Although such results further cement the power of periodic driving as a technique for Hamiltonian engineering \cite{Jotzu_1406, Meinert_1602, Oka_1804,lerose2018quantum},
we hasten to emphasize that these results \emph{are necessary but not sufficient} for proving the existence of intrinsically non-equilibrium, prethermal Floquet phases of matter. 
Let us unpack this last statement. 
Our focus in this manuscript is on phases of matter that \emph{cannot} exist in equilibrium. 
This means that the Floquet nature of the system is not simply being used as an engineering trick to stitch two disparate Hamiltonians together, but rather, as a \emph{prerequisite ingredient} for the existence of a phase with no direct analog in thermal equilibrium. 
This latter point is most easily summarized as follows:  the phase must, at its core, be protected by the discrete time translation symmetry of the drive \cite{Else_1602,Potter_1602,Else_1607}.
%


Thus, in order to prove the existence of prethermal Floquet phases, one must first demonstrate that the prethermal regime can actually preserve the symmetry structure of the driven system.
With this in mind, recent progress has precisely demonstrated the existence of \emph{emergent symmetries} during the prethermal window  \cite{Else_1607}. The existence of these symmetries can be viewed as a direct manifestation of the discrete time-translation symmetry of the drive.
This theoretical framework provides the perfect landscape for realizing prethermal non-equilibrium phases of matter, including prethermal versions of discrete time crystals \cite{Khemani_1508,Else_1603}, Floquet symmetry protected topological phases \cite{Else_1602,vonKeyserlingk_1602_a,Potter_1602,Potirniche_1610}, and possibly many others \cite{Titum_1506,Po_1701,Schuster_1903,Nathan_1712}.
However, this framework leaves open one fundamental challenge, in that it cannot be applied to long-range interacting systems. 
%

More specifically, one cannot ensure that the resulting effective prethermal Hamiltonian possesses any meaningful sense of locality.
Without this notion of locality, the evolution of local operators may not be well-approximated by the prethermal Hamiltonian.
As a result, the usual assumption that the system will evolve to the prethermal Gibbs state and exhibit the phase structure of local and power-law interacting Hamiltonians may not hold.
The overarching goal of our work is to tackle this concern, \emph{proving} the existence of prethermal Floquet phases in many-body systems that exhibit long-range, power-law interactions (i.e.~Coulomb, dipolar, van der Waals, etc) \cite{blatt2012quantum,Laumann_2016,Kucsko_1609,de2019observation,Moses_1610}.

This goal is motivated from two complementary fronts. On the experimental front, many of the platforms most naturally suited for implementing Floquet dynamics exhibit long-range interactions, including dipolar spins in the solid-state, trapped ions, ultracold polar molecules, and Rydberg atom arrays \cite{Zhang_1609, Yan_1305,Bernien_1707,Moses_1610,Choi_1610,Rovny_1802}. 
Understanding the prethermal properties of this broad class of systems could unlock a myriad of new experimental techniques for Floquet quantum simulation. 
On the theoretical front, even in equilibrium, it is well known that long-range interactions can lead to symmetry-breaking in qualitatively different regimes than that allowed by short-range interactions. 
This suggests the possibility of finding prethermal Floquet phases that can \emph{only} be realized in long-range interacting systems.

Our main results are threefold. First, we prove the existence of prethermal Floquet phases of matter (Figure 1) in long-range interacting systems, so long as the interactions decay as a power-law with exponent $\alpha > d$, where $d$ is the dimension of the system.
Second, we predict the existence of a novel, disorder-free, prethermal discrete time crystal (PDTC) in one dimension. 
This phase is strictly forbidden in any of the three contexts that we discussed earlier: equilibrium, Floquet MBL, and \emph{short-range} interacting prethermal Floquet.
Indeed, the 1D PDTC can only be realized in a \emph{long-range} interacting, prethermal Floquet system!
Finally, leveraging large-scale Krylov subspace methods, we perform extensive numerics characterizing the emergence of a 1D PDTC in a long-range interacting spin chain. 
In this context, we highlight one of the key (experimentally observable) differences between the prethermal time crystal and the MBL time crystal, namely, the presence of a phase transition as a function of energy density (Fig.~\ref{fig1} and Table~\ref{tab:MBLvsPretherm}).

Our paper is organized as follows.
In \refsec{sec:Prethermalization}, we lay the framework for understanding Floquet prethermalization both with and without an emergent symmetry (although only the former admits non-equilibrium phases of matter). 
Moreover, we review and contextualize a number of prior results with a particular emphasis on their implications for understanding the dynamics within the prethermal regime. 
This allows us to formalize the two essential properties for proving the existence of long-range interacting, prethermal phases. 
Building upon these discussions, in \refsec{sec:Proof}, we begin by introducing new machinery to carefully keep track of the spatial structure of the long-range interactions. Leveraging these new tools, we ultimately prove three theorems, which in combination, demonstrate the existence of  long-lived, non-equilibrium prethermal phases of matter in long-range interacting systems with power-laws $\alpha > d$.
Within this context, we also introduce a novel phase of matter: the 1D prethermal discrete time crystal. 
In \refsec{sec:Numerics}, we perform an exhaustive numerical  investigation of a one dimensional Floquet spin chain and demonstrate that it exhibits a PDTC phase, only when the system harbors sufficiently long-range interactions.
Using a combination of Krylov subspace methods and quantum Monte Carlo calculations, we identify one of the unique signatures of a PDTC (as compared to an MBL discrete time crystal), namely, that it displays a phase transition as a function of the energy density of the initial state.
Finally, we provide a short summary of some of the implications and interpretations in \refsec{sec:Conclusion}.

\section{Prethermalization}
\label{sec:Prethermalization}

In an interacting, many-body quantum system, one generally expects dynamics to push the local state of the system toward equilibrium via a process known as thermalization \cite{Deutsch_1991, Srednicki_1994, Rigol_0708, fta}. 
However, in certain cases, the time scale, $\tau^*$, at which thermalization occurs can be significantly larger than the timescale associated with the  intrinsic local interactions of the Hamiltonian, $1/J_\mathrm{local}$ \footnote{Throughout this work we work in natural units $\hbar=1$ and thus frequency carries units of energy.}. 
In such cases, before full thermalization actually occurs (i.e.~for times $t<\tau^*$), the system can first approach a \emph{different} equilibrium state determined by an effective Hamiltonian---this process is called \emph{prethermalization};  the time interval associated with it is known as the  \emph{prethermal regime}, while the effective Hamiltonian is referred to as the \emph{prethermal Hamiltonian}.
%

Systems exhibiting prethermalization generally have two distinct energy scales.
In static systems, this typically requires the underlying Hamiltonian to exhibit two very different couplings which lead to both ``fast'' and ``slow'' degrees of freedom.
Prethermalization can then be understood as the equilibration of the ``fast'' degrees of freedom with respect to a slowly varying background arising from the dynamics of the ``slow'' degrees of freedom.
In this case, $\tau^*$ is expected to depend \emph{algebraically} on the ratio of the energy scales \footnote{
  In special cases, however, it can depend exponentially on the ratio of the energy scales.
  An example occurs in models with near integer spectrum~\cite{Abanin_1509, Else_1607, Kemp_1701, Else_1704, Parker_1808}}.




\emph{Exponentially long Floquet heating time}---Unlike static systems, Floquet systems always exhibit two distinct energy scales: the local energy scale, $J_{\mathrm{local}}$, and the frequency of the drive, $\omega$.
To this end, a Floquet system can almost naturally be expected to exhibit a long-lived intermediate prethermal regime when these two energy scales are sufficiently different; our focus is of course, on the case in which $\omega \gg J_{\mathrm{local}}$. 
In that case (typically referred to as Floquet prethermalization), $\tau^*$ scales  \emph{exponentially} with the ratio of these two energy scales, $\omega / J_{\mathrm{local}}$, rather than algebraically \cite{Abanin_1507,Kuwahara_1508, Mori_1509,Abanin_1509,Abanin_1510}.

The physical intuition for this exponential scaling is simple.
Given a local energy scale $J_{\mathrm{local}}$, the many-body system requires $\omega/J_{\mathrm{local}}$ rearrangements in order to absorb a single quantum of energy from the drive. When interactions are local, the system cannot efficiently make
a large number of correlated local rearrangements.
Thus, the associated rate of energy absorption (i.e.~Floquet heating) is exponentially small in $\omega/J_{\mathrm{local}}$, leading to a heating time scale, $\tau^*\sim e^{\omega/J_{\mathrm{local}}}$.
This physical picture also helps to explain why long-range interacting Floquet systems with power-laws $\alpha < d$ cannot exhibit a prethermal regime. 
In such systems, the energy scale associated with a single local rearrangement diverges as a function of the system size (i.e.~the system exhibits a super-extensive many-body spectrum), implying that 
a single rearrangement can, in principle, absorb an energy quantum from the drive regardless of the magnitude of the driving frequency. 


\emph{Approximation of local Floquet dynamics}---While we have focused above on the existence of an exponentially long Floquet prethermal regime, as we alluded to earlier (while emphasizing the importance of locality), this is not the only constraint that one needs to worry about. 
Rather, just as important is whether one can prove that there actually exists a local prethermal Hamiltonian, $D^*$, that approximately generates the dynamics of the Floquet system during the prethermal regime.
A bit more precisely--- to approximate the unitary time evolution operator, $U_f$, that generates the \emph{exact} Floquet dynamics during a single driving period $T$, should be approximated by
\begin{align} \label{eq:Approximation_Simple}
  U_f \approx  U_f^{\text{app}} = e^{-i D^* T}~.
\end{align}
And, more importantly, one hopes that this approximation correctly captures the dynamics of local observables until the Floquet heating time scale.
A priori, this need not be the case and, in fact, the exact Floquet dynamics might not have any effective Hamiltonian description.

Indeed, the difference between proving the existence of a conserved energy (i.e.~measured with respect to the prethermal Hamiltonian) versus proving that the prethermal Hamiltonian correctly generates the local dynamics is stark. 
For example, although the Floquet heating time, $\tau^*$, has been proven to be exponentially long in generic systems with extensive energy scales (including long-range interacting systems \cite{Kuwahara_1508, Mori_1509,Abanin_1509,Abanin_1510,Else_1607} and even classical systems \cite{Mori_1804}), proving that the associated prethermal Hamiltonian describes the dynamics of local observables has only been achieved for a significantly smaller class of systems \cite{Kuwahara_1508,Abanin_1509, Else_1607, ftb}.
%
In fact, in certain systems it has been shown that the prethermal Hamiltonian \emph{does not} generate the actual Floquet dynamics \cite{Mori_1804}.

\emph{Generalizing to the case of an emergent symmetry}---Up to now, we have focused on how an effective \emph{static} description of the Floquet system (governed by the prethermal effective Hamiltonian) can emerge during the prethermal regime, both in the context of a conserved energy as well as in the context of generating local dynamics. 
While powerful in and of itself, this description limits Floquet systems to mimicry of equilibrium-like physics within the prethermal regime.
This is because, at the moment, our effective static description has forgotten about the structure of the original time periodic drive.
Luckily, this need not be the case!

Before formalizing this last statement, let us illustrate it with a simple example. Consider an $S=1/2$ spin undergoing a $\pi/2$ rotation every period $T$.
In the absence of any perturbing field, the spin will return to its original orientation every four periods.
Crucially, it turns out that even in the presence of small interactions (with respect to the driving frequency $\sim \omega = 2 \pi /T$), this picture remains true for an extremely long time scale. 
One can gain some intuition for this by noting that all of the interactions which fail to commute with the $\pi/2$-rotation get ``echoed out'' (i.e.\ they average to zero in the toggling frame that rotates by $\pi/2$ each Floquet period), which means that at leading order in the inverse frequency, they do not contribute to the dynamics. We emphasize, however, that the general results we eventually consider will hold not just at leading order, but also at higher orders.




Armed with this simple example, let us now formalize how extra symmetry-structure can emerge in the prethermal regime of Floquet systems.
In particular, if $U_f$ contains a large rotation, $X$, that returns to itself after $N$ periods, $X^N = \mathds{1}$ (in our example with the $\pi/2$-rotation, $N=4$) and \emph{generic} interactions (whose strength is much smaller than the driving frequency), then $U_f$ can be \emph{exponentially well} approximated by a much simpler evolution \cite{Else_1607}:
\begin{align}\label{eq:approxStatement} 
  U_f \approx U_f^{\text{app}} &= \mathcal{U} \widetilde{U}_f^{\text{app}} \mathcal{U}^{\dagger}, \notag \\ 
  \widetilde{U}^{\text{app}}_f &= X e^{-iTD^*} \quad \text{with} \quad [D^*,X]=0,
\end{align}
where $D^*$ is the effective prethermal Hamiltonian that commutes with the rotation $X$, and  $\mathcal{U}$ is a  time-independent unitary change of frame, which is close to the identity. 
Note that we will often choose to work directly in the rotated frame given by $\mathcal{U}$, so that the evolution is (approximately) given by $\widetilde{U}_f^{\text{app}}$ rather than $U_f^{\text{app}}$.

The above discussion encodes a few important consequences. First, since $D^*$ commutes with $X$, it remains an \emph{exactly} conserved quantity under this approximate evolution. Taking into account the exponentially small error terms (which track the differences between this approximate evolution and the exact Floquet evolution) leads to $D^*$ being exponentially well conserved.
Second, while $X$ was not a symmetry of the original evolution, it has become a $\mathbb{Z}_N$ symmetry of the approximate time evolution, $\wUf^{\mathrm{app}}$;
this emergent symmetry is protected by the underlying discrete time translation symmetry of the Floquet evolution operator.
As we discuss later, one can leverage this emergent symmetry to realize novel Floquet phases within the prethermal regime, including phases like the  time crystal, which break the discrete time translation symmetry of the underlying drive.
Third, let us emphasize that the presence of $X$ within $\wUf^{\mathrm{app}}$ ensures that for every period, the system undergoes a non-trivial rotation that remains finite even in the high-frequency limit, $\omega \to \infty$; this corresponds to the remnant  ``Floquet structure'' that remains within the prethermal regime.
However, when one considers the evolution every $N$ periods, one finds that the dynamics are simply generated by the static prethermal Hamiltonian $D^*$:
\begin{align} \label{eq:everyN} 
  (\wUf^{\text{app}})^N = e^{-iNT D^*}~.
\end{align}
Finally, we emphasize that the emergent $\mathbb{Z}_N$ symmetry is relevant only within the prethermal regime, where the total energy is also exponentially well conserved.
%


\subsection{Prethermal emergent symmetry as a framework for non-equilibrium phases of matter} 
\label{sec:prethermalTC}

In this section, we further elucidate the role of the emergent symmetry and how it provides a natural framework for realizing non-equilibrium phases of matter.
Since the time evolution every $N$ periods is captured by the  prethermal Hamiltonian $D^*$ (\eqnref{eq:everyN}), there exists a time scale, $\tau_{\text{pre}}$, after which the system has ``prethermalized'' into a  Gibbs state of $D^*$ and thus,  is locally described by  $\rho \propto e^{-\beta D^*}$, with a temperature $\beta^{-1}$ determined by the system's initial energy density.

Let us now examine the evolution of this equilibrium state under a single period of $\wUf^{\text{app}}$.
In general, $\rho$ will evolve trivially because the equilibrium state respects the emergent symmetry $X$:
\begin{align}
  \rho ~\to~X e^{-iD^* T}  \rho e^{iD^*T} X^\dag = X \rho X^\dag = \rho~.
\end{align}
However, if $D^*$ exhibits a spontaneously symmetry broken (SSB) phase with respect to $X$, $\rho$ can instead approach the equilibrium state within a particular symmetry-breaking sector;
let us refer to such a spontaneously symmetry broken state as $\rho_{\mathrm{SSB}}$.
In this case, although $\rho_{\mathrm{SSB}}$ evolves trivially under $D^*$, the action of $X$ is to rotate $\rho_{\mathrm{SSB}}$ into a distinct symmetry-breaking sector, $\rho_{\mathrm{SSB}}'$:
\begin{align}
  X e^{-iD^* T} \rho_{\mathrm{SSB}} &e^{i D^* T} X^{\dagger} = \notag \\ 
  &= X \rho_{\mathrm{SSB}} X^\dag = \rho_{\mathrm{SSB}}'\neq \rho_{\mathrm{SSB}}~.
\end{align}
During each period, the state rotates between the different symmetry-breaking sectors, only coming back to its  original  sector after $N$ periods ($X^N = \mathbb{1}$).
The sub-harmonic nature of this behavior becomes transparent by measuring the order parameter, which is a local observable whose expectation value is different in each of the symmetry sectors.

In the language of time crystals, the fact that the underlying Floquet evolution has a period of $T$, while observables exhibit an enlarged periodicity $NT$, precisely corresponds to the discrete breaking of time translation symmetry \cite{Khemani_1508, Else_1603, Yao_1608,vonKeyserlingk_1605,Zhang_1609,Else_1607}. 
For the remainder of this section, we continue to use the example of time crystalline order to highlight some of the unique features of prethermal non-equilibrium phases (Table~\ref{tab:MBLvsPretherm}).

First, in order to meaningfully label the prethermal time crystal as a phase of matter, one needs to show that it remains stable under small perturbations.
This is  guaranteed so long as the discrete time translation symmetry of the drive is not broken; in particular, this symmetry protects the emergent $\mathbb{Z}_N$ symmetry, and we know that a phase that spontaneously breaks a $\mathbb{Z}_N$ symmetry should be stable with respect to perturbations that do not explicitly break the symmetry.

Second, because our construction requires the system to prethermalize to an SSB state of $D^*$, the observation of a prethermal time crystal depends on the choice of initial state (Table~\ref{tab:MBLvsPretherm}).
In particular, the initial energy density must correspond to a temperature below the critical temperature of the SSB phase transition.
We emphasize that, because the underlying transition of $D^*$ is \emph{sharp} in the thermodynamic limit, there is an equally \emph{sharp} transition between the prethermal time crystal and the trivial prethermal regime as
a function of  energy density (as long as $\tau^* \gg \tau_{\mathrm{pre}}$
\footnote{As long as $\tau_{\mathrm{pre}}$ diverges at most algebraically near the transition, we are guaranteed (owing to the exponential scaling of $\tau^*$) that the transition will be exponentially sharp in the frequency of the drive.}).

Third, as the system begins absorbing energy from the drive at $\tau^*$, the temperature of the system will eventually cross the critical temperature of the SSB transition, leading to the loss of  time crystalline order---the prethermal time crystal phase will always have a finite (but large) lifetime.
To this end, depending on the energy density of the initial state, the lifetime of the time crystalline behavior can exhibit two distinct behaviors.
If the energy density is below the critical SSB temperature, the system prethermalizes to the SSB phase and the time scale $\tau_{TC}$ at which the time crystalline order parameter decays is similar to the heating time scale:  $\tau_{TC} \sim \tau^* \sim e^{\omega/J_{\mathrm{local}}}$.
If, on the other hand, the energy density is above the critical SSB temperature, the system will simply prethermalize to the symmetry preserving (trivial) phase and any transient time crystalline order can only occur before prethermalization, $\tau_{\mathrm{TC}} \lesssim  \tau_{\text{pre}} \sim \mathcal{O}(J_{\mathrm{local}}^{-1})$.

\emph{Differences between the many-body localized and prethermal discrete time crystal}---We end this section by juxtaposing the above discussions about the prethermal discrete time crystal with its many-body localized counterpart. Our focus is on highlighting the key differences between the two phases, as summarized in Table~\ref{tab:MBLvsPretherm}.
These differences can be divided into two categories: 1) the stability of the time crystal and 2) the restrictions on systems that can host a time crystal.
Concerning the former, in contrast to the exponentially long lifetime of the PDTC, the ergodicity-breaking properties of Floquet many-body localization enable the MBL time crystal to persist to infinite times. 
Moreover, while the stability of the MBL time crystal can be independent of the initial state, the PDTC can only occur for a finite range of initial energy densities. 

Let us now turn to the restrictions on systems that can realize an MBL versus a prethermal time crystal.
In the MBL case, such systems are required to have strong disorder \footnote{Or a disorder-like potential such as a quasi-periodic potential.} and are unstable to the presence of an external bath \cite{Nandkishore_1402}, long-range interactions \cite{Yao_1311,DeRoeck_1608}, and higher dimensions \cite{DeRoeck_1608}.
By contrast, the prethermal time crystal suffers from none of these restrictions and requires only two ingredients: a Floquet frequency that is larger than the local bandwidth and the existence of a static Hamiltonian $D^*$ with a spontaneously symmetry broken phase.
Crucially, in one dimension, this latter ingredient \emph{requires} us to consider long-range interacting systems with power-law $1 < \alpha < 2$ \cite{Dyson_1969}; for such power-laws, it is known that even a 1D system can exhibit finite temperature SSB phase, skirting the conventional Landau-Peierls argument that discrete symmetry breaking is forbidden for short-range interacting systems in 1D.


\begin{table}[]
  \renewcommand{\arraystretch}{1.5}
  \begin{tabular}{c||c|c}
    & MBL TC & \makecell{Prethermal TC} \\ \hline \hline 
    Lifetime & $\tau \to \infty$ & $\tau \sim e^{\omega/J_{\mathrm{local}}}$ \\  \hline
    \makecell{Initial State} & Any & Below $T_c$ \\ \hline
    \makecell{Requires Disorder} & Yes & No  \\ \hline 
    \makecell{Interaction Range} & Short-range* & \makecell{Long-range \\ $1 < \alpha \le 2$}  \\   \hline 
  \end{tabular}
    \caption{
    Differences between MBL and prethermal discrete time crystalline order in one dimensional systems. The star next to short-range indicates that the range of the interaction must only be sufficiently short so that MBL is preserved. 
  }
  \label{tab:MBLvsPretherm}
\end{table} 

\subsection{Prethermalization in long-range interacting systems}
\label{sec:prethermLR}

Before proving the existence of long-range interacting, prethermal phases of matter, we briefly contextualize a number of prior results with a particular emphasis on their implications for understanding the dynamics within the prethermal regime. 

In particular, we now formalize the two different properties (for which we previously gave intuition) that $U_f^{\text{app}}$ should satisfy in order to be of the broadest interest and most useful.
We simplify the following discussion by focusing on the case without an emergent symmetry, \eqnref{eq:Approximation_Simple}, but our analysis carries over to the case with an emergent symmetry [\eqnref{eq:approxStatement}] by rotating into the frame $\mathcal{U}$: $U_f \to \widetilde{U}_f = \mathcal{U} U_f \mathcal{U}^{\dagger}$ and $U_f^{\text{app}} \to \widetilde{U}_f^{\text{app}}$.

\begin{enumerate}[(a)]
  
\item {\bf Exponentially long heating time}---For $U_f^{\text{app}}$ to be a good approximation to $U_f$, a naive first requirement is that the difference between the two unitaries be small.
  This can be encoded in a bound of the form:
  \begin{equation} \label{eq:errorUnitaries}
    \| U_f - U_f^{\mathrm{app}} \| \leq \mathcal{O}(\Lambda e^{-\omega/J_{\mathrm{local}}})~,
  \end{equation}
  where $\Lambda$ is the volume of the system.
  Such a result would ensure that the error associated with the approximation in \eqnref{eq:Approximation_Simple} is exponentially small in the frequency of the drive.
  
  However, owing to its volume dependence, this bound, at first, suggests that  $U_f^{\text{app}}$ is not meaningful in the thermodynamic limit, $\Lambda \to \infty$.
  In particular, if one simply computes the overlap between wavefunctions evolved under the approximated and the true evolution, it would go to zero:
  \begin{equation}\label{eq:globalstateError}
    \lim_{\Lambda \to \infty} \langle \psi | U_f^\dag  U_f^{\text{app}} | \psi \rangle = 0~. 
  \end{equation}
  But, of course, one is typically not interested in capturing the dynamics of the \emph{full quantum wave-function} (which cannot be measured), but rather in the dynamics of \emph{local observables}.
  %
  Unfortunately, by itself, \eqnref{eq:errorUnitaries} is insufficient to analyze the error in the evolution of generic local observables.

  Nevertheless, it can still be used to prove important results on the dynamics of \emph{extensive quasi-conserved quantities}.
  Of particular interest is the dynamics of the energy density, $D^*/\Lambda$.
  Since it remains constant under $U_f^{\text{app}}$, bounding the error growth of this observable provides an immediate upper bound on the heating rate under the true evolution!

  To this end, by combining knowledge of the structure of the approximate unitary [\eqnref{eq:Approximation_Simple}] with the error in the unitaries [\eqnref{eq:errorUnitaries}], one can immediately conclude that $D^*/\Lambda$ remains exponentially well conserved under the evolution:
\begin{equation} \label{eq:exponentiallylongthermalization}
  \frac{1}{\Lambda}\left| \left\langle U_f^{-m} D^* U_f^{m} \right\rangle - \left \langle D^* \right \rangle \right| = \mathcal{O}(mTe^{-\omega/J_{\mathrm{local}}})~.
\end{equation}
As promised, this  formalizes the statement that the energy of the system is conserved up to an exponentially long time-scale $\tau^*$ and thus, that the infinite temperature state cannot be reached before $\tau^*$.
Note that for other extensive quantities conserved by $D^*$, similar bounds can also be derived.

\item {\bf Approximation of local dynamics}---
  At this point, we have not yet formalized the statement that $U_f^{\text{app}}$ is the correct ``effective'' generator of the true Floquet dynamics, only that the energy density remains conserved
  \footnote{This is a crucial point when attempting to describe phases of matter within the prethermal regime, because without it, one cannot precisely determine the equilibrium properties of the system during prethermalization.}.
%
By filling in this gap, we would be able to rigorously connect the prethermal regime with the equilibrium properties of $D^*$. 
  This can be achieved by bounding the error in the dynamics of a generic local observable $O$ as:
  \begin{align} \label{eq:applocaldynamics}
    \| U_f^{-m} O U_f^{m} - (U_f^{\text{app}})^{-m} & O (U_f^{\text{app}})^{m} \| \le \notag\\
    &\le \mathcal{O}(~(mT)^\delta e^{-\omega/J_{\mathrm{local}}}~)~,
\end{align}
  for some finite $\delta$.
  Crucially, this result is \emph{independent} of the volume of the system, meaning that it remains applicable even in the thermodynamic limit. This formalizes  the intuition that, even if the global wave-function is not perfectly captured by $U_f^{\mathrm{app}}$ [\eqnref{eq:globalstateError}], the local properties remain correct. 
  %
  Supplementing this result with an understanding of the equilibrium properties of $D^*$ as well as the structure of the unitary evolution (i.e.~the emergent symmetry) will ultimately enable us to prove the existence of long-range, prethermal phases of matter.
%

\end{enumerate}


Having formalized these two properties, we are now in a position to contextualize prior results on prethermalization in long-range interacting systems, \emph{without} an emergent symmetry. 
In the case of an exponentially long thermalization time [property (a) above], the approximate unitary $U_f^{\mathrm{app}}$  has been proven to satisfy \eqnref{eq:exponentiallylongthermalization} for power-laws $\alpha > d$ \cite{Kuwahara_1508, Mori_1509}. 
For approximating local dynamics [property (b) above],   the approximate unitary $U_f^{\mathrm{app}}$,  has been proven to satisfy \eqnref{eq:applocaldynamics} for power-laws $\alpha > 2d$ \cite{Kuwahara_1508,Mori_1509}.
The discrepancy between these two regimes arises from the fact that Lieb-Robinson bounds with power-law light-cones have been proven only for $\alpha > 2d$ \cite{FossFeig_1410,Matsuta_1604,Tran_1808,Else_1809}.
When attempting to extrapolate to the case with an \emph{emergent symmetry} in the prethermal regime, the above prior techniques  do not appear readily generalizable~\cite{Kuwahara_1508,Mori_1509}.

Indeed, even for \emph{short-range interactions} \cite{Else_1607}, generalizing to the case of an \emph{emergent symmetry} requires the use of an alternate construction~\cite{Abanin_1509}.
Curiously, although not explicitly discussed, many of the arguments found in this construction~\cite{Abanin_1509}, generalize directly to the long-ranged case with little modification.
In particular, the construction depends on the \emph{number of lattices sites} each interaction term couples, which remains small even for long-range interactions  (e.g.~the long-range Ising interaction found in trapped ion experiments only couples pairs of sites \cite{Zhang_1609}).
As a result, one can directly use this construction for any power-law $\alpha > d$, to create the approximate Floquet unitary $U_f^{\mathrm{app}}$, and to prove that it satisfies property (a), i.e.~that it exhibits an exponentially long thermalization time scale.
%
Extending to the case of an emergent symmetry then  naturally follows by using the arguments found in Ref.~\cite{Else_1607}.

\emph{Key challenge}---Unfortunately, since the construction found in Ref.~\cite{Abanin_1509} retains no spatial information about $D^*$, one is unable to prove that $U_f^{\mathrm{app}}$ satisfies property (b), i.e.~that the dynamics of local observables are accurately captured.

Crucially, the lack of spatial information about $D^*$ prevents the application of Lieb-Robinson bounds, implying that any bound on the error of local observables diverges with the system size.
To better understand the essential role of the Lieb-Robinson bounds, let us recall that the Floquet unitary is given by the \emph{exact} expression~\cite{Abanin_1509}:
\begin{align}
\label{eq:witherrorterm}
  U_f = \mathcal{T}~ e^{-i \int_0^T dt~D^* + V^*(t)}~,
\end{align}
where $\mathcal{T}$ denotes time ordering and $V^*(t)$ is a time-dependent interaction such that the sum of terms acting on any one site is exponentially small in frequency.
One then builds the approximate unitary evolution, $U_f^{\mathrm{app}}$, by disregarding the role of the exponentially small $V^*(t)$.
%
%

To understand how much error is accrued in this approximation, it is crucial to understand how a local operator $O$ ``spreads'' under the evolution generated by $D^*$.
The bigger the volume of $O$, the larger the number of terms in $V^*(t)$ it can overlap with and whose contribution we are missing when we disregard the role of $V^*(t)$.
%
As such, the rate of error growth is simply bounded by the sum of the local terms of $V^*(t)$ within the support $\Lambda_{O(t)}$ of the operator $O(t)$, while the total error $\delta O(t)$ is the integral: $\delta O(t) \sim e^{-\omega/J_{\mathrm{local}}} \int_0^tdt'~\Lambda_{O(t')}$. 

The role of the interaction range is now apparent. 
If the original Floquet evolution is \emph{short-range}, both the resulting $D^*$ and ${V^*}(t)$ are also short-range and the evolution exhibits a finite Lieb-Robinson velocity $v_{\text{LR}}$.
The volume of the operator $O(t)$ is then bounded by $\propto(v_{\text{LR}}t)^d$, 
and the error $\delta O(t)\sim t^{d+1}e^{-\omega/J_{\text{local}}}$ remains small for an exponentially long time in the frequency.

In contrast, when the original Floquet evolution is \emph{long-range}, the volume of the operator $O$ can grow much faster than $\mathcal{O}(t^d)$.
For example, for interactions decaying with power-laws $\alpha \le 2d$, only an exponential light cone has been proven, $\Lambda_{O(t)} \sim e^{d \eta t}$ \cite{Hastings_0507}. 
In this case, the error $\delta O \sim e^{-\omega /J_{\mathrm{local}} + d \eta t}$ remains small for only a short time \emph{proportional} to the frequency of the drive.
%
%
For $\alpha > 2d$, a power-law light cone has been proven \cite{FossFeig_1410,Matsuta_1604,Tran_1808,Else_1809}, suggesting that if $D^*$ can be shown to exhibit an $\alpha > 2d$ spatial decay, one can immediately apply current Lieb-Robinson bounds.
Of course, we hasten to remind the reader that in order to apply these long-range Lieb-Robinson bounds, one must first extend prior results (in the context of an emergent symmetry \cite{Abanin_1509, Else_1607}) to determine the spatial decay of $D^*$ which, \emph{a priori}, may be quite different from the decay of $H(t)$.



%


%
\emph{Prethermal phases in finite size systems}---Up to now, our discussion has focused on the thermodynamic limit, where Lieb-Robinson bounds are required to prove that local dynamics are captured by $U_f^{\text{app}}$.
However, in finite system sizes, \eqnref{eq:errorUnitaries} can actually be enough to guarantee that the prethermal Hamiltonian properly captures the dynamics.
In particular, by setting the frequency of the drive large enough, i.e.,~$\omega \gg \log \Lambda$, the approximate Floquet unitary is close to the full unitary evolution and the global wavefunction of the system is well approximated, \emph{regardless of the locality of the interactions}.
In this case, \emph{any} observable (local or not) is well captured by the prethermal Hamiltonian until a time scale $\tau_O \sim \Lambda^{-1}e^{\omega/ J_{\text{local}}}$ (which remains smaller than the thermalization time scale $\tau^*$ by a factor of $\Lambda$).
Nevertheless, as long as $\tau_{\text{pre}}$ is smaller than $\tau_O$, the system is guaranteed to approach the Gibbs state of $D^*$ and this intermediate window ($\tau_{\text{pre}} < t < \tau_O$) can host prethermal phases of matter.
%


\subsection{Summary of key analytical results}
\label{sec:summary}
    
    

Our main analytical results are twofold. First, we present a new construction for $D^*$ that explicitly retains information about the spatial locality of the interactions.
Our construction naturally addresses the case where $D^*$ hosts an emergent $\mathbb{Z}_N$ symmetry, extending prior results~\cite{Else_1607} to the case of long-range interactions.
Second, using this novel construction, we are able to apply appropriate long-range Lieb-Robinson bounds to ensure that the prethermal Hamiltonian captures the local dynamics within the prethermal regime [property (b)] and thus, to prove the existence of long-range prethermal phases of matter.

For $\alpha > 2d$, the existence of power-law-light-cone Lieb-Robinson bounds allows us to prove that the local dynamics are accurately captured by $U_f^{\mathrm{app}}$ up to the Floquet heating time scale, $\tau^* \sim e^{\omega /J_{\mathrm{local}}}$ [third row of table in Fig.~\ref{fig1}].
This ensures that within the prethermal regime, the system will approach the equilibrium state of the prethermal Hamiltonian $D^*$; combined with the existence of an emergent symmetry (protected by the time translation symmetry of the drive), this proves the existence of \emph{prethermal phases of matter} [fourth row of table in Fig.~\ref{fig1}(c)].

For $d<\alpha<2d$, we are not be able to directly invoke such power-law-light-cone Lieb-Robinson bounds.
In this case, the equilibration dynamics within the prethermal regime are less clear. 
Nevertheless, one expects that the approximate conservation of energy density means that local observables still relax to the Gibbs state of $D^*$, since this is the state that maximizes the entropy subject to the constraint of conservation of energy.
Under this assumption, we show that the robustness of prethermal phases of matter extends to power-laws $d < \alpha <2d$ as well [fourth row of table in Fig.~\ref{fig1}(c), where the star indicates this additional assumption].
Moreover, in \emph{finite-size} systems, one can prove rigorous statements without making this assumption, as discussed in the previous section.

In summary, our work demonstrates that prethermal phases of matter exist for all extensive power-law interacting systems ($\alpha > d$).

\section{Rigorous statement and proof of prethermalization result in long-range systems}
\label{sec:Proof}

In this section, we describe our novel analytic construction, which extends prior results on prethermal phases~\cite{Abanin_1509,Else_1607} to the long-range interacting case.
At its heart, this construction \emph{exactly} transforms the initial time-dependent Hamiltonian into a new Hamiltonian composed of a static term $D^*$ (with an emergent $\mathbb{Z}_N$ symmetry) in addition to small error terms.
Crucially, this transformation captures two complementary properties:
First, it ensures that the error terms are exponentially small in the frequency of the drive.
Second, it guarantees that $D^*$ and the small error terms inherit the same locality properties as the original Hamiltonian;
if the original Hamiltonian is long-ranged, the transformed Hamiltonian will also be long-ranged.

As discussed in Sec.~\ref{sec:prethermLR}, the first property allows us to prove an exponentially long thermalization time scale, in agreement with previous bounds \cite{Kuwahara_1508,Mori_1509,Abanin_1509,Else_1607}.
Meanwhile, the second property enables us to prove a much stronger statement, namely  that local observables remain well approximated by the long-range prethermal Hamiltonian throughout the prethermal regime (for power-laws $\alpha > 2d$)---a statement which has not been addressed in any prior literature for long-range interacting, prethermal systems with an emergent symmetry.

To guide the reader through this rather technical section, we present a short road map below. 
We begin by providing a careful treatment of previous results on prethermalization (Sec.~\ref{sec:previousResults}).
This introduces the necessary context to discuss the novel ideas required for our construction (Sec.~\ref{sec:Main_Ideas}). 
Next, we will precisely state the key result of our construction in the form of Theorem~\ref{thm:longrangethm} (Sec.~\ref{subsec:thmstatement}).
Finally, we discuss three immediate consequences of our construction (Sec.~\ref{sec:discussion}): (1)
 that local observables are well captured by the approximate Floquet unitary for $\alpha > 2d$ (Theorem~\ref{thm:localdynamics}), (2) how prethermal phases of matter arise even for $\alpha > d$ (Theorem~\ref{thm:localdynamics_short}), and (3) how our ideas can be directly generalizable to static systems with a near integer spectrum.

\subsection{Previous results}
\label{sec:previousResults}

\emph{Analyzing the Magnus expansion}---In Refs.~\cite{Kuwahara_1508,Mori_1509}, the main theoretical tool used to analyze the prethermal regime is the formal Magnus expansion of the single period time evolution operator $U_f$.
This procedure defines the Floquet Hamiltonian $H_F$ as a formal series expansion in the period of the drive $T$:
\begin{equation}
  U_f \equiv e^{iH_FT} \quad \text{ where } \quad H_F = \sum_{m=0}^\infty T^m K_m~,
\end{equation}
with $K_m$ being operators and $m$ the order of the Magnus expansion.
Although such a series will, in general, not converge (otherwise there is a quasi local Hamiltonian $H_F$ which is conserved under the dynamics of the system), understanding its truncation remains very useful.

First, by truncating the Floquet Hamiltonian at the correct order $n_0 = \mathcal{O}(\omega)$, $H_F^{(n_0)} = \sum_{m=0}^{n_0} T^m K_m$, one obtains an \emph{exponentially} good approximation to the full unitary evolution, $U_f \approx e^{-iT H^{(n_0)}_F}$.
This implies that, over a single period of the evolution, the energy density $\langle H^{(n_0)}_F \rangle/\Lambda$ remains exponentially well conserved in the frequency of the drive; this corresponds to property (a) of Sec.~\ref{sec:prethermLR}.
Because this analysis relies only on the few-bodyness of the interaction and the existence of a finite local energy scale, it holds for both short- and long-range interacting systems with $\alpha > d$. 

Second, for power-laws $\alpha > 2d$, one can use Lieb-Robinson bounds with power-law light cones \cite{FossFeig_1410,Matsuta_1604,Tran_1808,Else_1809} to prove that $H_F^{(n_0)}$ is also the approximate generator of the dynamics of local observables for exponentially long times; this corresponds to property (b) of Sec.~\ref{sec:prethermLR}.
Combining these two conclusions, one proves the existence of a long-lived prethermal regime whose dynamics are well captured by the prethermal Hamiltonian for short and long-range interacting systems with power-law $\alpha>2d$ [first and second rows of the table in Fig.~\ref{fig1}(c)].
Again, we emphasize that this construction does not prove the existence of an emergent symmetry in the prethermal regime;
to obtain this result requires (to the best of our knowledge) a different approach.

\emph{Rotating into an appropriate frame}---To this end, a different approach~\cite{Abanin_1509} was pursued which enabled the proof of an emergent symmetry in the prethermal regime \cite{Else_1607}.  The main idea is to find a sequence of frame rotations where each rotation reduces the magnitude of the driven part of the evolution.
Stopping the iteration at the correct step minimizes the driven component and proves the existence of a long-lived prethermal regime.

In more detail, one begins by separating the Hamiltonian $H(t) = H_0(t)$ into two components: a static $D_0$ and a driven $V_0(t)$ term.
Performing a rotation into a new frame, one obtains a new Hamiltonian $H_1(t)$ that \emph{exactly} describes the evolution, but where the norm of the driven term $V_1(t)$ is reduced (while the static component $D_1$ is slightly modified); repeating such a process for $n$ steps reduces the magnitude of the drive $V_n(t)$ exponentially in $n$.
However, much like the Magnus expansion result, this process cannot continue indefinitely or the system would be described by a static local Hamiltonian and thus fail to thermalize to the infinite temperature state. 
The optimal iteration step is given by $n^* \sim \mathcal{O}(\omega/\ln^3\omega)$, leading to the final Hamiltonian $H_{n^*}(t)$:
\begin{align}
  H_{n^*}(t) &= D_{n^*} + V_{n^*}(t) \\
  &\text{where} \quad \|V_{n^*}(t)\| \le \|V_0\| (2/3)^{n^*}~. \notag
\end{align}
Since the local terms of the driven part $V_{n^*}(t)$ are exponentially small, the full evolution is approximately generated by the static component, $U_f \approx e^{-iD_{n^*}T}$.
Analogous to the Magnus expansion approach, one can prove that $D_{n^*}/\Lambda$ remains exponentially well conserved under a single period:
\begin{equation}
  \frac{1}{\Lambda}\|U_f^{-1} D_{n^*} U_f - D_{n^*}\| \le C  T \left(\frac{2}{3}\right)^{n^*} ~,
\end{equation}
for some volume and frequency independent constant $C$;
the thermalization time scale is then exponentially long in the frequency of the drive.

Using this approach, one can also prove that the prethermal Hamiltonian can approximate the dynamics of local operators \emph{provided that the original evolution is governed by a Hamiltonian with short-range interactions}.
The source of this additional restriction is that, unlike the Magnus expansion approach, this construction cannot keep track of the range of interactions due to the way it accounts for the size of the Hamiltonian terms.
More specifically, the proof ensures that any one operator does not grow to act on too many sites, without bounding the distance between the sites it acts on.
In short-range interacting systems, this distinction is unimportant because the two measures of size are proportional; it is then guaranteed that $D_{n^*}$ remains short-ranged and that the appropriate Lieb-Robinson bounds can be used to show it approximately generates the dynamics of local operators.
However, this distinction becomes crucial in long-range interacting systems where these two measures can be very distinct leading to the breakdown of the proof, as explained in more detail in Sec.~\ref{sec:Main_Ideas}.
%

\emph{Generalizing to a prethermal emergent symmetry}---Understanding the limitations of this construction~\cite{Abanin_1509}  is crucial because it provides the only path (to our knowledge) to prove the emergence of symmetries in the prethermal regime \cite{Else_1607}.
The main insight for this generalization is that the previous construction can be slightly modified to preserve the structure of the original Floquet unitary.
Consider a Floquet unitary of the form:
\begin{align} 
  U_f& = \mathcal{T}e^{-i\int_0^Tdt \left[H_0(t) + V(t)\right]} = \\
  &= X ~\mathcal{T} e^{-i\int_0^Tdt\left[D_0 + E_0+ V_0(t)\right]}~,\\
  &\text{ where }\quad \mathcal{T}e^{-i \int_0^Tdt~H_0(t)} = X ~,~ X^N=\mathbb{1}
\end{align}
where $E_0$ corresponds to the static terms of the evolution that \emph{do not} commute with the symmetry $X$.
In this case, $E_0$ and $V_0(t)$ are both the error terms we wish to minimize 
(in this language, the original construction corresponds to the specific case when $N=1$, $X=\mathds{1}$, and $E_0 =0$ \cite{Abanin_1509}).
To adapt their construction, one first rotates the system such that $E_0$ becomes time periodic, while keeping $D_0$ unchanged; the system is now fully characterized by $D_0$ and a new drive $V'_0(t)$.
One can now directly employ the previous construction to reduce the magnitude of the newly defined driven part \cite{Else_1607}. 
The resulting new Hamiltonian contains terms $E_1$ and $V_1(t)$ whose magnitude is reduced and a static $D_1$ whose magnitude slightly increases.
Applying this procedure $n^*$ times reduces the size of $E_{n^*}$ and $V_{n^*}(t)$ optimally, such that the unitary evolution is well approximated by the action of $X$ and an evolution under the final static term $D_{n^*} =D^*$ [\eqnref{eq:approxStatement}].
Let us emphasize that this picture is exact in a slightly rotated frame $\mathcal{U} \approx \mathds{1} + \mathcal{O}(\omega^{-1})$ arising from the small rotation necessary to transform each $E_n$ into a driven term.

Because this analysis follows the results of Ref.~\cite{Abanin_1509},
the results have the same scope with regards to the range of the interactions.
In particular, the heating rate of the system is exponentially slow in frequency for both short and long-range interactions with power-law $\alpha > d$;
however, local observables are only provably well captured by the prethermal Hamiltonian in short-range interacting systems.
Proving this result in full generality is the goal of the next few sections and will open up an entirely new landscape for investigating non-equilibrium phases of matter and their quantum simulation in long-range interacting quantum optical platforms.


\subsection{Main ideas of proof for long-range generalization}
\label{sec:Main_Ideas}

In this section, we outline the novel ideas required to extend prior results~\cite{Abanin_1509,Else_1607} to  long-range interacting systems; our main result is summarized in Theorems~\ref{thm:longrangethm} and \ref{thm:localdynamics}. 
For more details, see Appendix \ref{app:full_proof} for the complete proof.

The main hurdle in generalizing the previous results to long-range interacting systems is to understand how the spatial structure of the interactions changes as one performs the necessary frame rotations.

We highlight, with a simplified example illustrated in Fig.~\ref{fig:OperatorExpansion}, the importance of the range of interactions to the spread of operators.
Although this example uses time evolution, the intuition carries over to the case of a frame rotation generated by some short- or long-range operator.
Consider an operator $O = \sigma^x_i$ and a short-range interacting Hamiltonian $H^{\mathrm{sr}} = \sum_{j} \sigma_j^z\sigma^z_{j+1}$.
At early times, the spread of the operator is given by
\begin{align}\label{eq:ex_sr_Hamil}
  O ~\to~ &e^{itH^{\mathrm{sr}}}Oe^{-itH^{\mathrm{sr}}}  = O +i t[H^{\mathrm{sr}},O]  + \mathcal{O}(t^2) \notag \\ 
  &= \sigma^x_i - 2t\sigma^y_i\left( \sigma^z_{i+1} +\sigma^z_{i-1}\right) + \mathcal{O}(t^2)~.
\end{align}
Crucially, the growth of the operator can happen only where it fails to commute with the Hamiltonian.
Because the Hamiltonian is short-ranged, the range (spatial extent $R$) of the time-evolved operator is proportional to the size of the support of the operator (number of sites $k$ it acts non-trivially on).
This distinction may not seem meaningful for short-range interacting systems, but in long-range systems it becomes crucial. For example, if we consider long-range interactions such as
\begin{equation}
\label{some_H_lr}
 H^{\mathrm{lr}} = \sum_{i \neq j} \frac{\sigma_i^z\sigma^z_j}{|i-j|^\alpha},
\end{equation}
then the spread of the operator is given instead by
\begin{align} \label{eq:ex_lr_Hamil}
  O \to &e^{itH^{\mathrm{lr}}}Oe^{-itH^{\mathrm{lr}}}  = O +i t[H^{\mathrm{lr}},O]  + \mathcal{O}(t^2) \notag \\
  &= \sigma^x_i -2t\sigma^y_i \sum_{j\neq i}\frac{\sigma^z_{j}}{|i-j|^\alpha}  + \mathcal{O}(t^2)~.
\end{align}
In this case, the time-evolved operator immediately becomes a sum of terms that connect two very distant points.
While each term is two-bodied---i.e.~the size of the support remains small with $k=2$---it can connect two points that are arbitrarily far away---i.e.~the range $R$ is arbitrarily large.

We now connect this intuition to a careful analysis of the prethermal Hamiltonian.
Starting from two-body interactions (such as \eqnref{some_H_lr}),
the usual construction  performs a rotation (informed by the driven part of the Hamiltonian) that generates a new Hamiltonian with higher-body and further extended terms~\cite{Abanin_1509,Else_1607}. 
To properly characterize the resulting final prethermal Hamiltonian, it is crucial to account for both the support size $k$ and the spatial extent $R$ of the terms, as these two properties play different roles in our result.

In particular, we need to ensure that terms that have either a large support size or a large range have a small magnitude.
More precisely,
if their magnitude decays exponentially with support size $k$, one can prove that there is a prethermal Hamiltonian exhibiting an exponentially long heating time scale.
If their magnitude \emph{also} decays with $R$ with sufficiently large power-law, one can employ the necessary Lieb-Robinson bounds to prove that the prethermal Hamiltonian is the approximate generator of the dynamics.
In our work, we prove that this condition holds even when there is an emergent symmetry.
%

\figOperatorExpansion

This latter point has eluded previous results \cite{Abanin_1509,Else_1607} because their construction was unable to keep track of the spatial structure of interactions;
in particular, a distinction is not made between an operator that acts on many sites (large $k$) and a few-body interaction that acts on sites far apart (large $R$). 
%

To overcome this issue, our strategy is to imbue the construction with extra structure that enables us to keep track of the range and the size of the operator separately.
To this end, we introduce the definition of an \emph{$R$-ranged set} and use it to build \emph{$R$-ranged operators}.
By representing the Hamiltonian in terms of $R$-ranged operators, we will ultimately be able to keep track of both the range $R$ as well as the size $k$ of the rotated Hamiltonian throughout the construction.

Let us begin by defining an $R$-ranged set.
Schematically, an $R$-ranged set is a union of ``clusters,'' each separated by distance at most $R$.
As a result, any two of its sites are connected via a sequence of ``jumps'' of size at most $R$ through the set, as shown in \rfig{fig:OperatorExpansion}(c).
Formalizing this picture, we define an \emph{$R$-ranged set} as a set $Z_R$ of sites of our system, such that for $x,x'\in Z_R$, there exists a sequence of elements $(x_1,\ldots,x_n)$ with $x_i \in Z_R$ such that $x_1=x$, $x_n =x'$ and dist$(x_i,x_{i+1}) \le R$.

At first sight, this definition appears more involved than simply characterizing a set based on its diameter (i.e.~largest distance between two of its elements).
This is on purpose.
Indeed, our definition of an $R$-ranged set has the following crucial property:
if two $R$-ranged sets have a non-trivial intersection, then their union is itself an $R$-ranged set.
The same is  not true for two sets with diameter at most $R$.

To see the importance of this property, let us first define  an \emph{$R$-ranged operator} as an operator whose (non-trivial) support is an $R$-ranged set.
The previous property of $R$-ranged sets immediately manifests in the following:
if one takes two $R$-ranged operators $A_{R_1},B_{R_2}$, then $e^{A_{R_1}}B_{R_2} e^{-A_{R_1}}$ will be a max$(R_1,R_2)$-ranged operator.
If we consider an operator written as a sum of $R$-ranged terms, then we can easily keep track of the range of each term as we perform a frame rotation (here, corresponding to $e^{A_{R_1}}$).
When applied to the construction of the prethermal Hamiltonian, we can easily keep track of the $R$-rangeness of each term of the original Hamiltonian throughout the different rotations.


The idea now is that we will consider potentials made up of a \emph{hierarchy} of different-ranged interactions, decaying in an appropriate way with range.
Specifically, we introduce a parameter $\sigma > 0$ (the value of which we will choose later), and define a sequences of ranges $R_l = e^{\sigma l}$.
Then we will define a \emph{range-indexed potential} to be a formal sum: 
\begin{equation}
\Phi = \sum_{l=0}^{\infty} \sum_{Z \in \mathcal{Z}_{R_l}} \Phi_{Z,l}~,
\end{equation}
where $\Phi_{Z,l}$ is supported on the $R_l$-ranged set $Z$. Here we have introduced $\mathcal{Z}_{R_l}$, the collection of all possible $R_l$-ranged sets.

Now we introduce a norm whose finiteness ensures our desired  condition, namely, that the strength of the interactions decays exponentially in the size of their support $k$ and as a power-law in the range $R$.
Specifically, we define a norm that depends on two parameters $\kappa,\gamma > 0$ according to
\begin{align}
\label{eq:our_Norm}
\| \Phi \|_{\kappa,\gamma} &= \sup_{x \in \Lambda} \sum_{l=0}^{\infty} R_l^{\gamma} \sum_{Z \in \mathcal{Z}_{R_l}, \, x \in Z} e^{\kappa |Z|} \| \Phi_{Z,l} \|~,
\end{align}
where $\gamma$ characterizes the power-law of the long-range decay. 
This is a generalization of the norm used in Refs.~\cite{Abanin_1509,Else_1607}, 
\begin{align} \label{eq:old_norm}
\| \Phi \|_{\kappa} &= \sup_{x \in \Lambda} \sum_{Z \ni x}  e^{\kappa |Z|} \| \Phi_{Z} \|
\end{align}
which did not keep track of the decay with range.


As an example, we note that for a two-body long-ranged Hamiltonian such as \eqnref{some_H_lr}, our new norm \eqnref{eq:our_Norm} is finite in the thermodynamic limit provided that $\gamma < \alpha - d$.
To see this, note that  we can set
\begin{equation}
\Phi_{Z,l} = \begin{cases} \frac{1}{r^{\alpha}} \sigma_i^z \sigma_j^z & Z = \{ i, j \}, r = \mathrm{dist}(i,j), l = l(r) \\ 0 & \mbox{otherwise} \end{cases},
\end{equation}
where $l(r)$ is the smallest $l$ such that $R_l \geq r$.
Then we have that
\begin{align}
\| \Phi \|_{\kappa,\gamma} &= e^{2\kappa} \sum_{l=0}^{\infty} R_l^{\gamma} \sum_{i,j : R_{l-1} < \mathrm{dist}(i,j)\leq  R_l} \frac{1}{\mathrm{dist}(i,j)^{\alpha}} 
\end{align}
On a $d$-dimensional lattice, we have 
\begin{equation}
\sum_{i,j : r < \mathrm{dist}(i,j)\le r'} \frac{1}{\mathrm{dist}(i,j)^{\alpha}} \leq \frac{C}{r^{\alpha-d}}
\end{equation}
for some constant $C$,
and hence we find
\begin{align}
\| \Phi \|_{\kappa,\gamma} &\leq C e^{2\kappa} \left(R_0^{\gamma} + \sum_{l=1}^{\infty} \frac{R_l^{\gamma}}{R_{l-1}^{\alpha - d}}\right) \\
&= C e^{2\kappa}\left( 1 + \frac{e^{\sigma(\alpha-d)}}{1-e^{\sigma (\gamma - \alpha + d)}} \right) < \infty,
\label{eq:norm_of_two_body}
\end{align} 
provided that $\gamma < \alpha - d$.

However, we emphasize  that our results also hold for Hamiltonians that are not just two-body! The only condition is that they decay fast enough with distance such that the norm in \eqnref{eq:our_Norm} is finite.

\subsection{Statement of the prethermalization theorem for long-range interacting systems}
\label{subsec:thmstatement}
We have now set up all of the requisite tools.
Our key contribution is developing the techniques required to analyze the range of the Hamiltonians produced by the aforementioned iterative construction, which leads to the following two main results (for details, see the appendices).

First, we show that, by revisiting systems with \emph{short-range} interactions, we can obtain stronger bounds by simply replacing the particular sequence of numbers ``$\kappa_n$'' chosen in Ref. \onlinecite{Abanin_1509} with a more optimized version.
Second, by leveraging the properties of $R$-ranged operators and our particular choice of the sequence $R_l$, we encode the information of the two-parameter norm \eqnref{eq:our_Norm}, which captures the long-range nature of the interactions, back into the original one parameter norm \eqnref{eq:old_norm}.
This enables us to make use of the \emph{exact} same analysis as in the short-range
case, while keeping track of the long-range nature of the interactions via this encoding.
Our final result is:


\begin{theorem}
\label{thm:longrangethm}
Suppose we have a time-periodic Hamiltonian $H(t+T) = H(t)$ which induces a Floquet evolution over a period $T$:
\begin{align}
\label{eq:main_doublyinitialForm}
   U_f &= \mathcal{T}\exp\left[-i\int_0^T dt~H(t)\right]\\
   &= X\;\mathcal{T}\exp\left[-i\int_0^T dt\left(D + E + V(t)\right)\right] \label{eq:main_initialForm}
\end{align}
such that $D$ and $E$ are time-independent and
\begin{align}
   X^N &= \mathbb{1}~,\\
   [D,X] &= 0~.
\end{align}

Fix some $\kappa_0, \gamma > 0$, and
define 
\begin{equation}
\lambda = T \max\{ \| D \|_{\kappa_0,\gamma}, \| E \|_{\kappa_0,\gamma}, \| V \|_{\kappa_0,\gamma} \},
\end{equation}
Now fix any $0 < \mathfrak{C} < 1$. Then there exist constants $C_1, \ldots, C_5 > 0$, depending only on $\mathfrak{C}$ and $\kappa_0$, with the following properties.

If $\lambda \leq C_1$ (the high-frequency regime), then there is a unitary transformation $\mathcal{U}$
which transforms the evolution to:
\begin{equation} \label{eq:approxThm_main}
 \mathcal{U}^\dag U_f \,\mathcal{U} = X ~\mathcal{T} \exp\left[-i\int_0^T dt\left(D^* + E^* + V^*(t)\right)\right] 
\end{equation}
where:
\begin{align}
  \| D - D^* \|_{\kappa_*,\gamma_*} T &\leq C_3 \lambda^2, \label{eq:Dstar_main}\\
  \|V^*\|_{\kappa_*,\gamma_*} T &\leq C_2 \lambda^2 \left( \frac{1}{2} \right)^{n_*}, \label{eq:Vstar_main}\\ 
  \| E^* \|_{\kappa_*,\gamma_*} T &\le C_2 \lambda^2 \left(\frac{1}{2}\right)^{n_*}, \label{eq:Estar_main}
\end{align}
and
\begin{equation}
\kappa_* = \mathfrak{C} \kappa_0, \quad \gamma_* = \mathfrak{C} \gamma, \quad n_* = \left\lfloor \frac{C_4}{\lambda} \right\rfloor.
\end{equation}

Moreover, $\mathcal{U}$ is locality-preserving and close to the identity in the following precise sense:
\begin{equation} \label{eq:sizeofU} 
\| \mathcal{U} \Phi \mathcal{U}^{\dagger} - \Phi \|_{\kappa_*,\gamma_*} \leq C_5 \lambda \| \Phi \|_{\kappa_0,\gamma}.
\end{equation}
for any range-indexed potential $\Phi$.
\end{theorem}

We emphasize that, because $\lambda = \mathcal{O}(\omega^{-1})$, we have that $n_* = \mathcal{O}(\omega)$; Eqs.~(\ref{eq:Vstar_main}) and (\ref{eq:Estar_main}) then reflect the exponential suppression (in frequency) of the ``error terms'' $V^*(t)$ and $E^*$.

\subsection{Consequences of Theorem \ref{thm:longrangethm}}
\label{sec:discussion}

\subsubsection{Approximate form of the Floquet unitary}

The end goal of Theorem \ref{thm:longrangethm} is to prove that the discussion in Sec.~\ref{sec:prethermalTC} for realizing prethermal phases of matter (e.g.~the prethermal time crystal) carries over to systems with power-law decaying interactions.

To this end, we build the approximate Floquet unitary evolution, $U_f  \approx \mathcal{U} X e^{-iD^* T}\mathcal{U}^{\dagger}  := U_f^{\mathrm{app}}$, by discarding the exponentially small [in $\lambda^{-1}= \mathcal{O}(\omega)$~] error terms in \eqnref{eq:approxThm_main} [$E^*$ and $V^*(t)$].
As emphasized in Sec.~\ref{sec:prethermLR}, it is important to consider in what sense $U_f \approx U_f^{\mathrm{app}}$ is a good approximation.
In particular, we can consider the difference between the two unitaries
\begin{equation}
  \| U_f^{\mathrm{app}} - U_f \| \leq \Lambda T  \| V^* + E^* \|_{\kappa_*,\gamma_*} = \mathcal{O}(\Lambda 2^{-n_*})~.
\end{equation}
It then follows that property (a) from Sec.~\ref{sec:prethermLR} is satisfied:
the energy density $\langle D^* \rangle/\Lambda$ remains approximately conserved until the heating time $\tau^* \sim 2^{n_*}$.
At this point, this just recovers an already obtainable result (even for long-range interactions) directly from the arguments of Ref.~\cite{Else_1607}, albeit with an improved bound on the heating time since $n_*$ now lacks any logarithmic corrections in $\lambda$.

Crucially, however, our choice of norm \emph{also} guarantees that the interactions in $D^*$ [as well as $E^*$ and $V^*(t)$] remain power-law decaying in space.
This allows us to consider how well $U_f^{\mathrm{app}}$ approximates the dynamics of local observables [property (b) in Sec.~\ref{sec:prethermLR}] which requires the use of Lieb-Robinson bounds.

\subsubsection{Approximation of local observables}
\label{subsub:liebrobinson}

As previously discussed in Sec.~\ref{sec:prethermLR}, proving that local dynamics are well captured by the prethermal Hamiltonian requires the existence of Lieb-Robinson bounds with power-law light-cones.
However, such bounds, in turn, require the prethermal Hamiltonian to exhibit the correct locality properties; its terms must decay, at most, as a power-law of their range.

In our construction, this is guaranteed by the finiteness of our two-parameter norm [captured in Eqs.~(\ref{eq:Dstar_main})-(\ref{eq:Estar_main})], where the power-law decay of each term is characterized by the parameter $\gamma_*$.
Crucially, Theorem \ref{thm:longrangethm} guarantees that $\gamma_*$ can be chosen arbitrarily close to the parameter $\gamma$ that characterizes the power-law decay of the original Hamiltonian of the system.
This ensures that the prethermal Hamiltonian exhibits the same locality properties as the original Hamiltonian.
Let us emphasize, however, that in the case where the original Hamiltonian contains two-body interactions, $\gamma$ does \emph{not} correspond to the exponent $\alpha$ that appears directly in the magnitude of each individual term (as in \eqnref{some_H_lr}); rather, as we found in \eqnref{eq:norm_of_two_body}, $\gamma$ must be smaller than $\alpha - d$.

This language also enables us to immediately use Lieb-Robinson bounds available in the existing literature for multi-body long-range interacting Hamiltonians \cite{Else_1809}.
In particular, as we show in Appendix \ref{app:proof:ApproximationOfLocalObservables}, any long-range interacting Hamiltonian $H$ with bounded norm $\| H \|_{\kappa,\gamma}$ and $\gamma > d$ satisfies the assumptions of Ref.~\cite{Else_1809}, and therefore obeys a power-law-light-cone Lieb-Robinson bound.
We emphasize the requirement of a Lieb-Robinson bound for interactions with \emph{arbitrary} $k$-bodyness since our construction does not guarantee that the $k$-bodyness of the original Hamiltonian is preserved by the prethermal Hamiltonian.

Combining our knowledge of the locality of the prethermal Hamiltonian with the necessary Lieb-Robinson bounds we prove the second main result of our work: all local observables are accurately captured by the approximate unitary $U_f^{\text{app}}$ throughout the entire prethermal regime.
This statement is formalized into the following theorem (see Appendix \ref{app:proof:ApproximationOfLocalObservables} for the proof):
\begin{theorem}\label{thm:localdynamics} 
  {\bf Approximation of local observables}
  Consider the scenario described in Theorem \ref{thm:longrangethm}.
  Define $\widetilde{U}_f = \mathcal{U}^{\dagger} U_f \mathcal{U}$, where $\mathcal{U}$ is the rotation constructed in Theorem \ref{thm:longrangethm}, and define the corresponding approximate unitary $\widetilde{U}_f^{\text{app}} = X e^{-iD^* T}$ by discarding the $E^*$ and $V^*$ terms in \eqnref{eq:approxThm_main}.
  Suppose that $\gamma_* > d$, where $d$ is the spatial dimension.
  Then for any $\eta$ satisfying $\frac{d+1}{\gamma_*+1} < \eta < 1$, and for any local observable $O$ supported on a set $S$, we have
  \begin{multline}
  \| (\widetilde{U}_f^{\mathrm{app}})^{-m} O (\widetilde{U}_f^{\mathrm{app}})^{m} - \widetilde{U}_f^{-m} O \widetilde{U}_f^{m} \|  \\\leq  \mathcal{C}\|O\| m\lambda 2^{-n_*} \left(1 + \tau^{1 + d/(1-\eta)} \right),
\end{multline}
where $\tau = (C_6 \lambda) m$,
where $C_6$ is a constant that depends only on $\kappa_*$ and $\gamma_*$,
and $\mathcal{C}$ is a constant that depends only on the geometry of the system (but not its volume), the spatial dimension $d$, the size of the set $S$, and on $\eta$.
  \end{theorem}

Before concluding this section,
we hasten to emphasize that if novel multi-body Lieb-Robinson bounds can be extended to power-laws $\gamma > 0$, the construction presented in this work will immediately carry over.
Such improvements would be in agreement with previous numerical and experimental results \cite{Richerme_1401,Cevolani_1706,Machado_1708,Luitz_1805}, as well as a recent proof for the particular case of two-body long-range interacting systems in one dimension \cite{Chen_1907}.

%
  

\subsubsection{Prethermal phases for power-laws $d < \alpha < 2d$} 

Unfortunately, we cannot prove a result as strong as Theorem \ref{thm:localdynamics} for $0 < \gamma_* < d$ (corresponding to initial two-body Hamiltonians with $d < \alpha < 2d$).
Nevertheless, we can at least show that the dynamics of local observables are well-approximated by $\widetilde{U}_f^{\mathrm{app}}$ at short times (see Appendix \ref{app:approxobservshort} for the proof).

\begin{theorem}\label{thm:localdynamics_short} 
  {\bf Approximation of local observables (for short times)}.
  Consider the scenario described in Theorem \ref{thm:longrangethm}.
  Define $\widetilde{U}_f = \mathcal{U}^{\dagger} U_f \mathcal{U}$, where $\mathcal{U}$ is the rotation constructed in Theorem \ref{thm:longrangethm}, and define the corresponding approximate unitary $\widetilde{U}_f^{\text{app}} = X e^{-iD^* T}$ by discarding the $E^*$ and $V^*$ terms in \eqnref{eq:approxThm_main}.
  Then for any local observable $O$ supported on a set $S$, we have, for any positive integer $m$ satisfying $m \lambda \leq C_7$,
  \begin{multline}
  \| (\widetilde{U}_f^{\mathrm{app}})^{-m} O (\widetilde{U}_f^{\mathrm{app}})^{m} - \widetilde{U}_f^{-m} O \widetilde{U}_f^{m} \|  \\\leq  
  C_8 \mathcal{C'} \|O\| \lambda^2 2^{-n_*} mT
\end{multline}
where $C_7$ is a constant that depends only on $\kappa_*$, and $C_8$ is a constant that depends only on $\kappa_*$ and the size of the set $S$.
  \end{theorem}
  The assumptions of Theorem \ref{thm:localdynamics_short} differ from Theorem \ref{thm:localdynamics} in that Theorem \ref{thm:localdynamics_short} does not require $\gamma_* > d$, but has an upper bound on the number of periods, $m$, which can be considered. For small enough $\lambda$ (that is, high enough frequency), $m_{\mathrm{max}} = \lfloor C_7/\lambda \rfloor> 1$, so one can at least accurately describe the dynamics of local observables during a \emph{single} driving period.


The consequence of this result is as follows. 
Suppose that at some time $t=nT$, the local observables are approximately described by the Gibbs ensemble of $D^*$, or some spontaneous symmetry broken sector thereof, which we call $\rho$.
As mentioned in Sec.~\ref{sec:summary}, we reemphasize that is a somewhat nontrivial assumption in the absence of a proof that the approximate unitary accurately describes the  dynamics of local observables during the whole approach to thermal equilibrium;
however, it follows if we assume that the system maximizes its entropy subject to the constraint of conserving energy density (which remains true for exponentially long times).
Then, after one more driving period, the local state is approximately described by the rotated Gibbs ensemble $\wUf^{\mathrm{app}} \rho (\wUf^{\mathrm{app}})^{\dagger} = X \rho X^{\dagger}$ (using the fact that $[\rho,D^*] = 0$).
This is all we need to repeat the arguments of Sec.~\ref{sec:prethermalTC} about non-equilibrium prethermal phases of matter.


\subsubsection{Extension to static systems}

The long thermalization time scale of driven systems can also be generalized to static systems whose dynamics are dominated by an operator $P$ with integer spectrum~\cite{Abanin_1509, Else_1607}:
\begin{align} \label{eq:staticCase}
  H = u P + D + V~,
\end{align}
where $[D,P] =0$, while $[V, P] \neq 0$ and $u$ is the largest energy scale.
In this setup, there is a change of frame where $P$ becomes quasi-conversed.
To intuitively understand how this conservation emerges, it is simplest to consider a infinitesimal evolution under $\Delta t = \delta t/u$:
\begin{align} \label{eq:static_intuition}
  U = e^{i\delta t(P + (D+V)/u)} &\approx e^{-i\delta t P} e^{-i\delta t (D+V)/u} \notag \\
  & = X e^{-i\delta t (D+V)/u}
\end{align}
where the integer spectrum of $P$ ensures that $X = e^{-i\delta t P}$ with $N=1/\delta t$.
However, we can make $\delta t$ to be as small as possible, increasing the size of the emergent symmetry.
In the  $\delta t \to 0$ limit, where \eqnref{eq:static_intuition} becomes exact, $N\to \infty$ and we can think of the emergent symmetry as a continuous $U(1)$ symmetry, generated by the ``number'' operator $P$.
Analogously to the driven case, a time-independent change of frame $\mathcal{U}$ ensures that this emergent symmetry is approximately conserved until an exponentially long time in $1/u$.
This was proven in Ref.~\cite{Abanin_1509}, closely following their techniques for driven systems.
%
In a similar fashion, our construction immediately adapts to the proof of the long-lived prethermal regime in static systems, allowing its extension to long-range interactions.
As an application, we note that the existence of a prethermal continuous time crystal in an undriven system \cite{Else_1607}  can now be generalized to systems with long-range interactions.


\section{Long-range prethermal discrete time crystal in one dimension}
\label{sec:Numerics}

We now turn to the example of a non-equilibrium prethermal phase, where long-range interactions are essential to its stability---the disorder free one dimensional prethermal discrete time crystal.
In particular, we study a one dimensional periodically driven spin-$1/2$ chain with long-range interactions decaying with a power-law $d < \alpha < 2d$.
Using massively parallel matrix-free Krylov methods \cite{dynamite,Hernandez__0509, Roman__16, Balay__17}, we compute the late time Floquet dynamics for system sizes up to $L=28$.
This enables us to highlight many of the features of prethermal phases discussed in Sec.~\ref{sec:prethermalTC}.
%
%
First,  by directly comparing short- and long-range interactions, we evince the crucial role of  power-law interactions for stabilizing a 1D PDTC (Fig.~\ref{fig:LongvsShort_evo}). 
Second, by varying the energy density of the initial state, we access the aforementioned transition between the PDTC and the trivial phase (Fig.~\ref{fig:spectrum}).
These two phases can be easily distinguished by the different scaling behavior of the time crystal's lifetime $\tau_{\mathrm{TC}}$:
in the PDTC phase it follows the heating time scale $\tau_{\mathrm{TC}}\sim \tau^* \sim e^{\omega / J_{\mathrm{local}}}$,
while in the trivial phase it is bounded by the prethermalization time scale, $\tau_\mathrm{TC} \lesssim \tau_{\mathrm{pre}}\sim \mathcal{O}(1/J_{\mathrm{local}})$.
We corroborate that our observed finite-size crossover matches the location of the phase transition independently computed via quantum Monte Carlo calculation of the corresponding equilibrium
finite-temperature phase transition.
These results provide insight into the experimental signatures of the PDTC, as well as direct measures of the relevant energy and time scales.





\subsection{Model and Probes}

\figComparison

To generate Floquet dynamics that host a PDTC, the evolution must satisfy two properties: first, it must lead to a prethermal Hamiltonian $D^*$ with a robust emergent $\mathbb{Z}_N$ symmetry, and second, $D^*$ must exhibit a spontaneous symmetry breaking phase.
We engineer a drive, motivated by current generation trapped ion experiments, that exhibits both.

To ensure that the emergent symmetry exists in the prethermal regime, we design a Floquet evolution that matches the form of \eqnref{eq:main_initialForm} in \refthm{thm:longrangethm}. 
In particular, we consider  time evolution under the stroboscopic application of two different Hamiltonians,
$H$ [see Eq.~(\ref{eq:H_ell}) below] and $H_x$, for times $T$ and $T_x$, respectively. 
By choosing  $H_x= \Omega_x \sum_i \sigma^x_i$, with $T_x \Omega_x = \pi/2$ and $\sigma^\nu_i$ the Pauli operator acting on site $i$, the second part of the evolution flips all spins around the $\hat{x}$ direction (in the language of NMR, this part of the evolution corresponds to a global $\pi$-pulse):
\begin{align}
  \exp\left[-iT_x H_x\right] = \prod_{j} i \sigma^x_j = X \quad,\quad X^2 = \mathbb{1}~.
\end{align}
The resulting Floquet evolution then reads:
\begin{equation}\label{eq:Uf}
  U_f = X e^{-iT H}~, 
\end{equation}
matching Theorem~\ref{thm:longrangethm}, with $N=2$ and drive frequency $\omega = 2\pi/T$
\footnote{Although the period of the evolution is given by $T_{\text{tot}} = T + T_x$, the evolution of the system does not depend on our choice of $T_x$. We then choose the limit of $T_x\to 0$ (the global $\pi$-pulse is infinitely fast) where $T_{\text{tot}} \to T$ and the drive-frequency becomes $\omega = 2\pi/T$.}.
We emphasize that $[X,H]\neq 0$; $X$ is \emph{not} a symmetry of the evolution.

Next, to ensure that the associated prethermal Hamiltonian $D^*$ exhibits a spontaneous symmetry breaking phase with respect to $X$, it must include long-range interactions with a power-law $d < \alpha < 2d$.
However, $D^*$ results from the construction in Theorem~\ref{thm:longrangethm} and thus corresponds to a complicated, frequency-dependent object.
Fortunately, as part of Theorem \ref{thm:longrangethm} we saw that $D^*$ remains close (at high frequencies) to $D$, the original static symmetry respecting component of $H$, as defined by \eqnref{eq:main_initialForm}.
Since $H$ is time independent [Eq.~(\ref{eq:H_ell})], $D$ has a very simple form: it precisely contains the terms of $H$ that are even under $X$.
Thus, by including a long-range Ising interaction (which commutes with $X$) directly in $H$, one can guarantee that both $D$ and $D^*$ exhibit a finite-temperature paramagnetic to ferromagnetic symmetry breaking phase transition \cite{Dyson_1969}. 

Combining the long-range Ising interaction with additional generic terms (that help to break integrability) leads to the following long-range Hamiltonian $H$:
\begin{align}
  H &=J \sum\limits_{i<j}^{L-1} \frac{\sigma^z_i\sigma^z_j}{|i-j|^\alpha} + \vec{h}\cdot \sum\limits_{i=0}^{L-1} \vec{\sigma}_i + J_x \sum\limits_{i=0}^{L-2} \sigma^x_i\sigma^x_{i+1}~. \label{eq:H_ell}
\end{align}
When we compare to the ``short-range version'' of this Floquet evolution, we will simply truncate the Ising interaction in $H$ to nearest and next-nearest neighbor; we denote this corresponding short-range Hamiltonian as $H_s$.

For the remainder of this work we consider units where $J=1$ and use the parameters $d < \alpha = 1.13 < 2d$ and  $\{J_x,h_x,h_y,h_z\} = \{0.75,0.21,0.17,0.13\}$ in a spin chain of size $L$ with periodic boundary conditions
\footnote{We slightly modify the long-range profile of the interaction to match the system's periodicity by replacing $|i-j|^{-\alpha}$ with $\left[ (L/\pi)\sin |i-j|\pi/L\right]^{-\alpha}$.};
we have verified that the observed phenomena are not sensitive to this particular choice of parameters.
  We note that, due to our choice of an anti-ferromagnetic coupling $J>0$,
  the ferromagnetic phase occurs at the \emph{top of the spectrum} of $D^*$.

Finally, let us emphasize the role of the field term $h_x \sigma^x_i$ and nearest neighbor interactions $J_x\sigma^x_i \sigma^x_{i+1}$ to the thermalization properties of $D^*$.
While favoring the disordered phase, they also ensure that, to zeroth order in $\omega^{-1}$, $D^*$ is not trivially diagonal and that, at large frequencies, the dynamics under $D^*$ are generic and thermalizing; as a result, both $J_x$ and $h_x$ control the time scale at which the system approaches the prethermal state, $\tau_{\mathrm{pre}}$.

Having described our model, we now introduce the diagnostics used to characterize its Floquet evolution.
First, we consider the energy density of the system.
Naively, one wishes to compute the energy density with respect to the full prethermal Hamiltonian $D^*$;
however, its numerical construction and evaluation is very costly. Therefore, we will instead measure the energy density with respect to $D$, which remains close to $D^*$ at high frequencies.
Second, we consider the half-chain entanglement entropy, $S_{L/2} = -\Tr\left[ \rho_{L/2} \log \rho_{L/2}\right]$ where $\rho_{L/2} = \Tr_{1<i\le L/2}\ket{\psi}\bra{\psi}$, as a probe of the prethermalization and thermalization dynamics of the system.

 To probe time crystalline behavior we wish to consider an observable that can exhibit a subharmonic response to our driving protocol.
  From our discussion in Sec.~\ref{sec:prethermalTC}, a suitable probe should be related to the order parameter of the paramagnetic to ferromagnetic transition in our model's prethermal Hamiltonian; for example,  $\langle \sigma_i^z(t) \rangle$ for some site $i$.
  However, to reduce fluctuations owing to the small support of $\langle \sigma^z_i(t)\rangle$, we find it convenient to average over the different sites of the system; let us then define
\begin{align} \label{eq:Mt}
  M(t) &= \frac{1}{L} \sum_{i=0}^{L-1} \langle \sigma^z_i(0) \rangle \langle \sigma^z_i(t)\rangle~.
\end{align}
It might have seemed more natural to consider instead the average magnetization $\overline{\sigma^z}(t) = L^{-1}\sum_{i=0}^{L-1} \langle \sigma_i^z(t) \rangle$, but $M(t)$, which corresponds to a two-time correlation function, provides a clearer window into the early time decay of the period doubling behavior.
Since we consider initial product states of $\sigma^z$, $M(t=0)$ is guaranteed to be $1$, its maximal value.
After the system prethermalizes to $D^*$ (for $t>\tau_{\mathrm{pre}}$), $M(t)$ approaches a plateau whose sign will change every other period in the PDTC phase.
Crucially, at this point and for translationally invariant systems (like our model), $M(t)$ becomes proportional to the average magnetization $\overline{\sigma^z}(t)$ which itself matches $\sigma^z_i$ (for any $i$).
As a result, $M(t)$ is equally sensitive to the late time decay of the time crystalline behavior (provided that the initial magnetization is nonzero).

  While $M(t)$ is nonzero in the PDTC phase, it can also remains nonzero in the absence of a PDTC, e.g. in the ferromagnetic phase of a static Hamiltonian.
  The true order parameter for the PDTC phase must then measure the \emph{subharmonic (i.e., period doubling) response} of $M(t)$.
  To this end, we introduce the PDTC order parameter:
\begin{equation}
  \Delta M(t) = | M(t+T) - M(t) |~.
\end{equation}
In the PDTC phase, $M(t)$ will remain finite and sign changing every period and thus $\Delta M(t)$ will be nonzero.
By contrast, in the symmetry-unbroken phase, all observables [including $M(t)$] quickly become $T$ periodic and $\Delta M(t)$ approaches zero.


\subsection{Exponentially long-lived PDTC}

Before addressing the long-range PDTC, we begin by exploring the Floquet evolution of its short-range counterpart, $H_s$, where previous results have proven the existence of an exponentially long-lived prethermal regime~\cite{Kuwahara_1508,Mori_1509, Abanin_1509, Abanin_1510, Else_1607}.
As shown in Fig.~\ref{fig:LongvsShort_evo}(a), this is indeed borne out by the numerics: 
the energy density remains approximately constant until a late time $\tau^*_{D^*}$ when $\langle D\rangle/L$ approaches its infinite temperature value of zero.
By increasing the frequency of the drive, one observes an exponential increase in $\tau^*_{D^*}$, in agreement with analytic expectations~\cite{Kuwahara_1508,Mori_1509, Abanin_1509, Abanin_1510, Else_1607} and previous numerical studies \cite{Machado_1708}.
These observations are mirrored in the evolution of the entanglement entropy $S_{L/2}$ [Fig.~\ref{fig:LongvsShort_evo}(d)].
There, the approach to the infinite temperature value, $S_{L/2}^{T=\infty} = [L\log(2) - 1]/2$ \cite{Page_1993}, occurs at $\tau^*_{S_{L/2}}$, which is also exponentially controlled by the frequency of the drive.
The agreement between $\tau^*_{D^*}$ and $\tau^*_{S_{L/2}}$ corroborates the existence of a single thermalization time scale $\tau^*$ that controls the approach to the infinite temperature state.
For the remainder of this work we quantify $\tau^*$ using $\tau^*_{S_{L/2}}$.

Furthermore, $S_{L/2}$ also informs us about the equilibration with respect to the prethermal Hamiltonian $D^*$;
as the system evolves and approaches the prethermal state, the entanglement entropy approaches a plateau that remains constant until the drive begins heating the system at $\tau^*$.
%
The time scale when $S_{L/2}$ approaches this plateau value is frequency independent.
  In fact, the system's prethermalization is well captured by the $\omega\to\infty$ Floquet evolution [black dotted line in Fig.~\ref{fig:LongvsShort_evo}(d)].
  In this limit, we have $U_f \to X e^{-iDT}$; thus, $U_f^2 = e^{-2iDT}$, so the evolution for even periods is \emph{exactly} generated by the static Hamiltonian $D$; for odd periods the wave-function must be rotated by $X$ (which does not affect $S_{L/2}$ or $\langle D\rangle / L$).
  This agreement with the $\omega\to\infty$ limit  highlights that the dynamics within the prethermal regime are indeed well approximated by the prethermal Hamiltonian $D^* \approx D$.


Finally, we turn to $M(t)$, our diagnostic for time crystalline order.
From the discussion in Sec.~\ref{sec:prethermalTC}, the lack of a spontaneous symmetry breaking phase in short-range interacting one-dimensional systems is expected to preclude the existence of the PDTC phase.
In particular, any transient period doubling behavior should quickly decay as the system approaches the prethermal state at $\tau_{\mathrm{pre}}$.
This is precisely what is observed in the dynamics of $M(t)$, as shown in Fig.~\ref{fig:LongvsShort_evo}(g);
while at very early times, even and odd periods exhibit almost opposite $M(t)$, by the time-scale $\tau_{\mathrm{pre}}$, $M(t)$ has decayed to zero and the system no longer exhibits any time crystalline behavior.
Thus, the transient signatures of a time crystal ``melt'' as the system equilibrates to the prethermal Hamiltonian $D^*$, clearly demonstrating the system's lack of a true PDTC phase.

We now contrast this behavior to the long-range case using the same initial state, as evinced in Figs.~\ref{fig:LongvsShort_evo}(b), \ref{fig:LongvsShort_evo}(e) and \ref{fig:LongvsShort_evo}(h).
With respect to the thermalization dynamics---captured by $\langle D\rangle/L$ and $S_{L/2}$ as shown in Figs.~\ref{fig:LongvsShort_evo}(b) and \ref{fig:LongvsShort_evo}(e), respectively---the short-range and long-range models exhibit qualitative agreement; an increase in the frequency of the drive leads to an exponential increase in the thermalization time scale $\tau^*$.
We note, however, an important quantitative difference.
In particular, the value of $J_{\mathrm{local}}$ extracted from the scaling $\tau^* \sim e^{\omega/J_{\mathrm{local}}}$ is larger in the long-range system. This increase is due to the greater number of interaction terms in the Hamiltonian and is in agreement with previous numerical results \cite{Machado_1708}.
In addition, $\tau_{\text{pre}}$ remains frequency independent and the prethermal dynamics are in excellent agreement with the $\omega\to\infty$ time evolution [\rfigP{fig:LongvsShort_evo}{e}].
%

The difference between the short and long-range interacting systems  becomes apparent when considering the PDTC order.
In particular, in the long-range model, the subharmonic response of $M(t)$ survives well beyond $\tau_{\mathrm{pre}}$ and lasts until the heating time scale $\tau^*$.
This behavior is robust.
By increasing the frequency of the drive, the lifetime of the time crystal increases, mirroring the exponential growth of the thermalization time scale; the decay of time crystalline behavior is no longer determined by dynamics within the prethermal window, but rather by  heating toward infinite temperature.


\subsection{Role of the initial state}

Another distinct feature of the PDTC is its sensitivity to the energy density of the initial state.
Unlike the MBL time crystal \cite{Khemani_1508, vonKeyserlingk_1602_b, Else_1603, Yao_1608,vonKeyserlingk_1605,Zhang_1609}, which can exhibit period doubling for all physically meaningful initial states, the stability of the prethermal time crystal relies on the prethermal state's approach to the symmetry broken phase of $D^*$.
As a result, its stability is intimately related to the phase diagram of $D^*$.
Because $\langle D^* \rangle/L$ remains approximately conserved until $\tau^*$, the energy density of the initial state is equal to the energy density of the prethermal state. 
With this in mind, one can then translate the initial energy density into the temperature $\beta^{-1}$ of the prethermal state  via the relation $\langle D^*(t=0)\rangle = \Tr\left[D^* e^{-\beta D^*}\right]/\Tr\left[e^{-\beta D^*}\right]$.
By choosing initial states with different energy densities, one can effectively vary the temperature of the prethermal state across the phase transition; the resulting $M(t)$ dynamics display qualitatively distinct behaviors.

This difference is manifest when we compare the dynamics of a ``cold'' state [near the top of the many-body spectrum \footnote{Owing to our choice of antiferromagnetic coupling, the ferromagnetic phase exists at the top of the spectrum.}, Figs.~\ref{fig:LongvsShort_evo}(b), \ref{fig:LongvsShort_evo}(e) and \ref{fig:LongvsShort_evo}(h)], with the dynamics of a ``hot'' state [near the center of the many-body spectrum), Figs.~\ref{fig:LongvsShort_evo}(c), \ref{fig:LongvsShort_evo}(f) and \ref{fig:LongvsShort_evo}(i)].  
Despite exhibiting the same thermalization behavior to infinite temperature, the period doubling behavior of the ``hot'' state decays significantly faster; indeed, the decay of $M(t)$ [and thus $\Delta M(t)$] is frequency independent and occurs as the system approaches the prethermal state at $t \lesssim \tau_{\mathrm{pre}}$, well before the heating time scale $\tau^*$.
This behavior is directly analogous to that of the short-range model.

\figDecayMain

To directly connect the stability of the prethermal time crystal to the equilibrium phase diagram of $D^*$, we study the decay time scale $\tau_{\text{TC}}$ of the PDTC order parameter $\Delta M(t)$ across the spectrum of $D^*$ (\rfig{fig:spectrum}).
(For details on the extraction of these time scales see Appendix~\ref{app:extraction}.) 


\figSchematic

Crucially, $\tau_{\text{TC}}$ exhibits important differences between the short- and long-range cases [Figs.~\ref{fig:spectrum}(a) and \ref{fig:spectrum}(b), respectively].
In the short-range case, the frequency of the drive has no discernible effect on the lifetime of $\Delta M(t)$ (except for the highest energy state, which we  discuss  below).

In the long-range case, the behavior is significantly richer and modifying the driving frequency has a different effect depending on the energy density [Fig.~\ref{fig:spectrum}(b)].
The most distinct behaviors occur deep in the paramagnetic phase (near the center of the spectrum) and deep in the ferromagnetic phase (near the top of the spectrum).
In the former, we observe the same frequency independent behavior of $\tau_{\text{TC}}$ that characterized the short-range model---the decay time-scale of $\Delta M(t)$ is simply determined by the prethermalization dynamics. 
In the latter, the behavior is dramatically distinct:  $\tau_{\textrm{TC}}$ increases exponentially with the drive frequency, following the thermalization time scale $\tau^*$; in fact, the two time scales approach one another with increasing frequency---this is the key signature of the PDTC phase, namely that the decay of the time crystalline order is limited only by the late time Floquet heating dynamics.

%

Having understood the behavior deep within each phase, we now turn to the transition between the two.
At first glance, it appears that the onset of the exponential frequency scaling (and thus the transition to the PDTC phase) occurs at a lower energy density than what is expected [dark shaded region of Fig.~\ref{fig:spectrum}(b)].
This expectation is based on an independent quantum Monte Carlo calculation for the transition in $D$ (see Appendix~\ref{app:QMC}).
As we explore below, this apparent inconsistency instead corresponds to a small finite frequency effect arising from the slow thermalization dynamics of $D^*$ near the phase transition, as schematically depicted in Fig.~\ref{fig:schematic}.

As a system approaches a phase transition, critical slowing down causes its thermalization time scale to diverge.
As a result, even in the paramagnetic phase, the decay of $\Delta M(t)$ can occur at very late times; we refer to this decay time scale as $\tau_{\text{mag}}$.
In the paramagnetic phase $\tau_{\text{mag}}$ is finite, while  in the ferromagnetic phase, it is infinite.

At low frequencies, if the system is near the phase transition on the paramagnetic side, $\tau_{\text{mag}}$ can be finite but much larger than $\tau^*$. The decay of $\Delta M(t)$ is set by heating rather than the prethermal dynamics of $D^*$ \emph{even though the system is in the trivial phase}.
The situation is resolved upon increasing the frequency of the drive, at which point $\tau^*$ and $\tau_{\mathrm{TC}}$ will both increase exponentially until they reach the magnetization decay time $\tau_{\mathrm{mag}}$;
then, $\tau_{\mathrm{TC}}$  again becomes bounded by $\tau_{\mathrm{mag}}$, losing its frequency dependence, while $\tau^*$ keeps increasing exponentially with  frequency.
Thus, at large enough frequencies, it is always the case that, in the paramagnetic phase, the decay of $\Delta M(t)$ arises from the dynamics of $D^*$.

This behavior is evinced in Fig.~\ref{fig:spectrum}(b) in two distinct ways.
First, by directly simulating the decay of $\Delta M(t)$ in the $\omega \to \infty$ limit (where heating cannot occur), we observe a significant increase of the decay time near the transition.
In particular, in the paramagnetic phase, we observe a decay time scale which diverges around the transition at $\langle D\rangle /L \approx 2.0$---this is direct evidence for the presence of slow prethermalization dynamics near the transition.
%
  Second, near the transition to the ferromagnetic phase,
  the disagreement between $\tau_{\text{mag}}$ (as measured by the decay of the magnetization in the $\omega \to \infty$ evolution) and $\tau_{\mathrm{TC}}$ occurs deeper in the trivial phase  smaller frequencies. 

Interestingly, the above discussion also explains the long thermalization time found in the edgemost state of the short-range model,  Fig.~\ref{fig:spectrum}(a).
  In this case, the initial state is close to the zero temperature ferromagnetically ordered state, leading to a finite, but very large prethermalization time scale.
  This very long prethermal equilibration time might also underlie the recent observations of long lived period-doubling behavior in the prethermal regime of short-range interacting systems \cite{Zeng_1707,Yu_1807,Mizuta_1902}, where no finite-temperature phase transition or stable PDTC should occur. 


\section{Conclusion}
\label{sec:Conclusion}


Using a combination of analytical and numerical results, we demonstrate the existence of prethermal non-equilibrium phases of matter in long-range interacting systems with power-laws $\alpha > d$.
This prethermal approach contrasts with recent MBL-based studies of Floquet phases, since it does not require disorder, nor is it limited by the dimensionality of the system.
We emphasize the generality of our analytic construction, whose limitations arise only from the lack of an appropriate Lieb-Robinson bound for $d < \alpha<2d$.
However, even in this regime, on quite general grounds, we expect the system to approach the Gibbs state with respect to the prethermal Hamiltonian and, thus, for prethermal phases of matter to be well defined.
Finally, we predict the existence of a novel, disorder-free, prethermal discrete time crystal  in one dimension. 
This phase is strictly forbidden in equilibrium, Floquet MBL, and \emph{short-range} interacting prethermal Floquet systems.
%

\emph{Note added.}---Recently, we became aware of a related complementary work on locality and heating in periodically driven, power-law interacting systems \cite{Tran_1908}.

\begin{acknowledgments}
We gratefully acknowledge the insight of and discussions with E. Altman, B. Bauer, S. Choi, G. Pagano, P. Hess,  C. Monroe, B. Ye, and M. Zaletel. This work was supported by the DARPA DRINQS program (award D18AC00033), the David and Lucile Packard Foundation and the W. M. Keck foundation. 
  The numerical work performed in this work used the \texttt{dynamite} package written in Python \cite{dynamite}, which supports a matrix-free implementation of Krylov subspace methods based on the \texttt{PETSc} and \texttt{SLEPc} packages \cite{Hernandez__0509, Roman__16, Balay__17}.
  D. V. E. was supported by the Microsoft Corporation and by the EPiQS Initiative of the Gordon and Betty Moore Foundation, Grant No. GBMF4303.
  D. V. E. also thanks Wen Wei Ho and Philipp Dumitrescu for collaboration on ideas relevant to the present work.
  G. D. K. M. was supported by the Department of Defense (DoD) through the National Defense Science \& Engineering Graduate Fellowship (NDSEG) Program.
\end{acknowledgments}


\appendix

\section{Short-ranged proof}
\label{app:shortranged}
In this appendix, we  prove an improved version of the prethermalization theorem for \emph{short-ranged} Hamiltonians. This improved version will eventually be the key to extending to the case of long-range power-law interactions.

Consider a finite set of sites $\Lambda$ that characterize our system.
Each site is assigned a finite Hilbert space, so the total Hilbert space becomes the tensor product of these local Hilbert spaces.
One can then define any operator, as a sum of terms acting on different parts of the system:
\begin{equation}
    Q = \sum_Z Q_Z
\end{equation}
where $Q_Z$ is an operator that acts on $Z \subseteq \Lambda$. The collection of $Q_Z$ is often referred to as a potential \cite{Abanin_1509}. 
Despite this decomposition not being unique,
our result constructs new potentials from an initial input potential so this ambiguity does not affect our proof.

We begin by introducing a one-parameter norm \cite{Abanin_1509}:
\begin{equation} \label{eq:app_oneParam}
\| Q \|_{\kappa} = \sup_{x \in \Lambda} \sum_{Z \ni x} e^{\kappa |Z|} \| Q_Z \|~. 
\end{equation}
The finiteness of this norm in the limit of infinite volume indicates that the interactions are decaying exponentially with the size of their support.

We can extend this definition to time-periodic potentials $Q(t)$ by considering the time-average of the instantaneous norms:
\begin{align}
  \|Q\|_{\kappa} = \frac{1}{T} \int_0^T dt~ \|Q(t)\|_{\kappa}.
\end{align}

The statement of our theorem is as follows
\begin{theorem}\label{thm:shortrangethm}
Suppose we have a time-periodic Hamiltonian $H(t) = H(t+T)$ which induces a Floquet evolution over a period $T$:
\begin{align}
   U_f &= \mathcal{T}\exp\left[-i\int_0^T dt~H(t)\right]\\
   &= X\;\mathcal{T}\exp\left[-i\int_0^T dt~D + E + V(t)\right] \label{eq:initialForm}
\end{align}
such that $D$ and $E$ are time-independent and
\begin{align}
   X^N &= \mathbb{1}~,\\
   [D,X] &= 0~.
\end{align}

Fix some $\kappa_0 > 0$, and
define 
\begin{equation}
\lambda = T \max\{ \| D \|_{\kappa_0}, \| E \|_{\kappa_0}, \| V \|_{\kappa_0} \},
\end{equation}
Now fix any $0 < \mathfrak{C} < 1$. Then there exist constants $C_1, \ldots, C_5 > 0$, depending only on $\mathfrak{C}$ and $\kappa_0$, with the following properties.

If $\lambda \leq C_1$ (high-frequency regime), then there is a unitary transformation $\mathcal{U}$ which transforms the evolution to:
\begin{equation} \label{eq:approxThm}
 \mathcal{U}^\dag U_f\mathcal{U} = X ~\mathcal{T} \exp\left[-i\int_0^T dt~D^* + E^* + V^*(t)\right] 
\end{equation}
where:
\begin{align}
  \| D - D^* \|_{\kappa_*} T &\leq C_3 \lambda^2,\\
  \|V^*\|_{\kappa_*} T &\leq C_2 \lambda^2 \left( \frac{1}{2} \right)^{n_*}, \\
  \| E^* \|_{\kappa_*} T &\le C_2 \lambda^2 \left(\frac{1}{2}\right)^{n_*}~.
\end{align}
and
\begin{equation}
\kappa_* = \mathfrak{C} \kappa_0, \quad n_* = \left\lfloor \frac{C_4}{\lambda} \right\rfloor.
\end{equation}

Moreover, $\mathcal{U}$ is locality-preserving and close to the identity, in the following precise sense:
\begin{equation}
\| \mathcal{U} \Phi \mathcal{U}^{\dagger} - \Phi \|_{\kappa_*,\gamma_*} \leq C_5 \lambda \| \Phi \|_{\kappa_0,\gamma},
\end{equation}
for any potential $\Phi$.
\end{theorem}

Note that this is very similar to Theorem 1 of Ref.~\cite{Else_1607}. It differs, however, in two important ways.
First, scaling of $n_*$ lacks the logarithm corrections with $\lambda$ (which is proportional to the inverse frequency) found in Ref.~\cite{Else_1607};
as a result the bound on the size of the residual ``error'' terms ($V^*$ and $E^*$) scales more stringently with frequency.
Second, the norm $\| \cdot \|_{\kappa_*}$ with respect to which the final bounds are obtained has a parameter $\kappa_*$ which \emph{does not} depend on $\lambda$.
Roughly, the $\kappa_*$ for which a finite bound can be obtained can be thought of as setting upper an bound on the locality of the Hamiltonians; so the second condition means that $D^*$, $V^*$, and $E^*$ do not become more non-local as the frequency increases (whereas the theorems of Refs.~\cite{Abanin_1509,Else_1607} did not exclude this possibility).


\subsection{The iteration}
\label{app:proof:InnerIteration}
Following Ref.~\cite{Else_1607}, the idea is to construct the necessary rotations iteratively.
At step $n$ of the iteration, there is a slightly rotated frame where the Floquet evolution operator $\Uf$ is in the form
\begin{align}
   \mathcal{U}_n^\dag \Uf \mathcal{U}_n &= \Uf^{(n)} = X~ \mathcal{T} \exp\left(-i\int_0^T dt~ \cH_n(t)\right),\\
   &\text{ with } X^N = \mathds{1}.
\end{align}
We are interested in performing a unitary transformation, such that $\cH_n$ becomes closer to a time independent term which commutes with the symmetry $X$. We begin by writing $\cH_n(t)$ as the sum of two different contributions, $D_n$ and $B_n(t)$. $D_n$ corresponds to the time-independent part of $\cH_n(t)$ which commutes with $X$---the ``good'' part---and it is given by
\begin{equation}
    D_n = \langle \langle \mathcal{H}_n \rangle_T \rangle_X = \frac{1}{N} \sum_{k=0}^{N-1} X^{-k} \left[\frac{1}{T} \int_0^T dt~ \mathcal{H}_n(t) \right] X^{k}
\end{equation}  
where $\langle .\rangle_T$ corresponds to the time averaging across a period:
\begin{equation}
    \langle O \rangle_T = \frac{1}{T} \int_0^T dt~ O(t)~,
\end{equation}
and $\langle . \rangle_X$ corresponds to the symmetrization with respect to $X$, defined as
\begin{equation}
    \langle O \rangle_X = \frac{1}{N} \sum_{k=0}^{N-1} X^{-k} O X^{k}~.
\end{equation}
Together, time averaging and symmetrization  guarantee that $D_n$ is both time independent and commutes with $X$.

$B_n(t)$ is then the remaining ``bad part'' of $\mathcal{H}_n(t)$ and is composed of a time-independent term $E_n$ which does not commute with $X$, and a time-dependent term $V_n(t)$:
\begin{equation}
    B_n(t) = \mathcal{H}_n(t) - D_n = E_n + V_n(t)
\end{equation}
where $V_n(t)$ is chosen such that
\begin{align}
  \langle V_n(t) \rangle_T = 0.
\end{align}
At each step of the iteration we reduce the norm of $B_n(t)$ by performing a transformation informed by $\mathcal{H}_n$. The construction for the iteration is exactly the one described in Ref.~\cite{Else_1607}, and we do not repeat it here. We only differ from Ref.~\cite{Else_1607} in how we analyze the bounds satisfied by the iteration, as we describe in the next section.

\subsection{Analysis of bounds}
Now we prove bounds on the result of the iteration. Our first result is Lemma~\ref{lemma:shortrangeintermediate}, a slightly modified form of Theorem \ref{thm:shortrangethm} (Theorem \ref{thm:shortrangethm} itself will eventually arise as a collorary), in which the constants more explicitly stated.
\begin{lemma}\label{lemma:shortrangeintermediate}

There are order 1 constants $u$ and $v$ (not depending on any other parameters) with the following properties.

Suppose we have a time-periodic Hamiltonian $H(t) = H(t+T)$ which induces a Floquet evolution over a period $T$:
\begin{align}
   U_f &= \mathcal{T}\exp\left[-i\int_0^T dt~H(t)\right]\\
   &= X\;\mathcal{T}\exp\left[-i\int_0^T dt~D + E + V(t)\right]
\end{align}
such that $D$ and $E$ are time-independent and
\begin{align}
   X^N &= \mathbb{1}~,\\
   [D,X] &= 0~.
\end{align}

Fix some $\kappa_0 > 0$, and
define 
\begin{align}
\lambda &= T \| D \|_{\kappa_0}, \\
\mu &= T \max \{ \| V \|_{\kappa_0}, \| E \|_{\kappa_0} \}.
\end{align}
Now fix any $0 < \mathfrak{C} < 1$. Then suppose that
\begin{equation}
b \leq \mathfrak{C}^2 v,
\end{equation}
where
\begin{align}
b &= \frac{1}{\kappa_0^2} 6(N+3) \max\left\{ \frac{12}{u} \left(\lambda + \frac{5}{2} \mu\right),  \mu \kappa_0\right\}.
\end{align}

Then there is a unitary transformation $\mathcal{U}$ which transforms the evolution to:
\begin{equation}
 \mathcal{U}^\dag U_f\mathcal{U} = X ~\mathcal{T} \exp\left[-i\int_0^T dt~D^* + E^* + V^*(t)\right] 
\end{equation}
where:
\begin{align}
  \| D - D^* \|_{\kappa_*} T &\leq \frac{1}{2} \mu, \\
\|V^*\|_{\kappa_*} T &\leq \mu \left( \frac{1}{2} \right)^{-n_*}, \\
\| E^* \|_{\kappa_*} T &\le \mu \left(\frac{1}{2}\right)^{-n_*}, 
\end{align}
and
\begin{equation}
\kappa_* = \mathfrak{C} \kappa_0, \quad n_* = \left \lfloor \frac{ (1-\mathfrak{C}^2)}{b} \right \rfloor.
\end{equation}.

Moreover, $\mathcal{U}$ satisfies
\begin{equation}
\label{eq:Usmallandlocal}
\| \mathcal{U} \Phi \mathcal{U}^{\dagger} - \Phi \|_{\kappa_*} \leq  e^{\mu/2\lambda} \frac{\mu}{2\lambda} \| \Phi \|_{\kappa_0}
\end{equation}
for any potential $\Phi$.
\end{lemma}

\emph{Proof---}To prove Lemma \ref{lemma:shortrangeintermediate}, following Refs.~\cite{Abanin_1509,Else_1607}, we introduce a decreasing sequence of numbers $\kappa_n > 0$. The key difference between our analysis and that of Refs.~\cite{Abanin_1509,Else_1607} is in how we choose this sequence $\kappa_n$. In particular, we choose this sequence in a way that is frequency-dependent, meaning that it depends on the parameters $\lambda$ and $\mu$ that appeared in the statement of the lemma. The higher the frequency (i.e. the smaller $\lambda$ and $\mu$), the slower $\kappa_n$ will decrease, which allows us to run the iteration to a larger order $n_*$.  

First of all, let us define
\begin{multline}
\label{eq:theiterationbounds}
d(n) = \| D_n \|_{\kappa_n}, \quad v(n) = \| V_n \|_{\kappa_n}, \\
 e(n) = \| E_n \|_{\kappa_n}, \quad \delta d(n) = \| D_{n+1} - D_n \|_{\kappa_{n+1}},
 \end{multline}
 We recall the following bounds from Appendix A.4 of Ref.~\cite{Else_1607} (note that these bounds are independent of the choice of $\kappa_n$):
\begin{equation}
2 \delta d(n), v(n+1), e(n+1) \leq \varepsilon_n,
\end{equation}
where
\begin{align}
  \label{eq:epsilon_n}
  \varepsilon_n &= 2T m(n) v'(n) [ d(n) + 2 v'(n) ]~, \\ 
  m(n) &= \frac{18}{(\kappa_{n} - \kappa_{n+1})\kappa_{n+1}}~, \\
    v'(n) &= (N+2) e(n) + v(n)~.
\end{align}
Note that there is an extra factor of 2 in \eqnref{eq:epsilon_n}, which corrects an error \footnote{Specifically,~\cite{Else_1607} neglected to take into account that their modified definition of norm for time-dependent potentials -- the time-average of the instantaneous norm rather than the supremum -- necessitates an additional factor of 2 in the first equation of Section 4.2 in Ref.~\cite{Abanin_1509}}
 in Ref.~\cite{Else_1607}.
These bounds hold provided that
\begin{equation}
\label{eq:condition_on_v}
3 T v'(n) \leq \kappa_{n} - \kappa_{n+1}~.
\end{equation}

These results can be recast in a more intuitive manner as follows. Our eventual goal is to argue by induction. Suppose our induction hypothesis is that, given some $h$ that is independent of the iteration order:
\begin{align}
\label{eq:first_induction_hypothesis}
d(n) + 2 v'(n) &\leq hT^{-1} \\
v(n),e(n) &\leq \left(\frac{1}{2}\right)^n \mu T^{-1}~.
\label{eq:second_induction_hypothesis}
\end{align}

Then we will make sure to  \emph{choose} $\kappa_{n+1}$ in terms of $\kappa_n$ such that the following conditions are satisfied:
\begin{align}
  \frac{1}{2} &\ge 2(N+3) m(n) h, \label{eq:EnsuresConvergence}\\
  \kappa_{n} - \kappa_{n+1}  &\ge 3 (N+3) \mu. \label{eq:sim_to_iteraction_condition}
\end{align}
The point is that \eqnref{eq:sim_to_iteraction_condition}, combined with \eqnref{eq:second_induction_hypothesis}, ensures that \eqnref{eq:condition_on_v} is satisfied, and then \eqnref{eq:EnsuresConvergence} combined with the induction hypothesis ensures that
\begin{equation}
\label{eq:induction_conclusion}
v(n+1),e(n+1),2\delta d(n) \leq \left(\frac{1}{2}\right)^{n+1} \mu T^{-1},
\end{equation}
which, in turn, ensures that \eqnref{eq:second_induction_hypothesis}, one of our induction hypothesis, is satisfied for $n \to n+1$ (we consider the other one later). 

One way to ensure Eqs.~(\ref{eq:EnsuresConvergence}) and (\ref{eq:sim_to_iteraction_condition}) is to define
\begin{equation}
  \kappa_{n+1}= \sqrt{\kappa_n^2 - \epsilon}
\end{equation}
for some $\epsilon > 0$ that we will choose later.
Then,
\begin{align}
  \frac{18}{m(n)} &= \kappa_n \kappa_{n+1} - \kappa_{n+1}^2 \\
  &= \kappa_n^2\left[\sqrt{1 -\frac{\epsilon}{\kappa_n^2}} - \left(1 - \frac{\epsilon}{\kappa_n^2}\right)\right]\\
  &\ge \kappa_n^2 \frac{u\epsilon}{\kappa_n^2}
\end{align}
where $u<1/2$ and $v$ are new constants introduced such that
\begin{equation}
  \sqrt{1-x} - (1-x) \ge ux  \quad \text{ for }\quad 0 \le x \le v \le 1
\end{equation}
Computing explicitly for $v$, one obtains
\begin{align}
  v = \frac{1-2u}{(1-u)^2}.
\end{align}

Equation~\ref{eq:EnsuresConvergence} is then satisfied provided that
\begin{align}
  u\epsilon &\ge 72 (N+3) h,\\
  \epsilon &\le v\kappa_n^2. \label{eq:v_condition}
\end{align}


Meanwhile, for \eqnref{eq:sim_to_iteraction_condition} to be satisfied, we note that
\begin{align}
  \kappa_{n} - \kappa_{n+1} &= \kappa_n \left(1-\sqrt{1-\frac{\epsilon}{\kappa_n^2}}\right)\\
  & \ge \frac{ \epsilon}{2\kappa_n}. \label{eq:new_iteration_condition}
\end{align}
Therefore, \eqnref{eq:sim_to_iteraction_condition} is satisfied provided that
\begin{equation}
\epsilon \geq 6(N+3) \mu \kappa_n~.
\end{equation}


In summary, the conditions on $\epsilon$ are that
\begin{align}
 6(N+3) \max\left\{ \frac{12}{u} h, \kappa_n \mu \right\} \le \epsilon \le v \kappa_n^2 \label{eq:lambdaCondition}.
\end{align}

We choose to only continue the iteration while $\kappa_n \geq \mathfrak{C} \kappa_0$.
Hence, \eqnref{eq:lambdaCondition} is satisfied provided that
\begin{align}
  b \le \epsilon/\kappa_0^2 \le \mathfrak{C}^2 v \label{eq:lambdaCondition2},
\end{align}
where
\begin{equation}
b = \frac{6(N+3)}{\kappa_0^2} \max\left\{ \frac{12}{u} h, \kappa_0 \mu \right\}.
\end{equation}
Accordingly, we will set $\epsilon = b \kappa_0^2$; then \eqnref{eq:lambdaCondition2} requires only that
\begin{equation}
  \label{eq:lambdamax}
b \leq \mathfrak{C}^2 v.
\end{equation}
With this choice, we see that $\kappa_n = \kappa_ 0 \sqrt{1 - b n}$.


Finally, we can complete the argument. The main missing piece is to show that the induction hypothesis \eqnref{eq:first_induction_hypothesis} is satisfied. Indeed, from \eqnref{eq:induction_conclusion} we have that
\begin{align}
d(n) &\leq d(0) + \sum_{n=0}^{\infty}  \left(\frac{1}{2}\right)^{n+2} \mu T^{-1} \\
&\leq \left[ \lambda + \frac{\mu}{2} \right] T^{-1},
\end{align}
and, thus,
\begin{align}
  d(n) + 2 v'(n) &\leq d(n) + 2 v'(0) \\
  &\leq d(n) + 2(N+3) \mu T^{-1}  \\
  &\leq \left[ \lambda + \frac{4(N+3)+1}{2}\mu \right] T^{-1}
\end{align}
Therefore, if we set $h = \lambda + \frac{4(N+3)+1}{2}\mu$, then given the assumptions of Lemma \ref{lemma:shortrangeintermediate}, we can continue the induction up to the maximum iteration order $n_*$.

Finally, we need to prove \eqnref{eq:Usmallandlocal}. From the form of the iteration (see Ref.~\cite{Else_1607}), we have
\begin{equation}
\mathcal{U} = e^{iA_{n_*}} \cdots e^{iA_0},
\end{equation}
where $\| A_n \|_{\kappa_n} \leq N e(n) T$.
Let us define $\Phi_n = e^{iA_n} \Phi_{n-1} e^{-iA_n}$, $\Phi_0 = \Phi$. Then from Lemma 4.1 of Ref.~\cite{Abanin_1509} and Eqs.~(\ref{eq:second_induction_hypothesis}) and (\ref{eq:EnsuresConvergence}), and the fact that $h \geq \lambda$, we obtain
\begin{align}
\| \Phi_{n+1} \|_{\kappa_{n+1}} &\leq \left[ 1 + m(n) N \left(\frac{1}{2}\right)^{n} \mu \right] \| \Phi_n \|_{\kappa_n}\\
&\leq \left[ 1 + \frac{\mu}{4\lambda} \left(\frac{1}{2}\right)^{n} \right]  \| \Phi_{n} \|_{\kappa_{n}} \\
&\leq \exp\left[\frac{\mu}{4\lambda} \left(\frac{1}{2}\right)^{n} \right] \| \Phi_{n} \|_{\kappa_{n}},
\end{align}
and, thus,
\begin{align}
\| \Phi_n \|_{\kappa_n} &\leq \exp\left[\frac{\mu}{4\lambda} \sum_{n=0}^{\infty} \left(\frac{1}{2}\right)^n \right] \| \Phi \|_{\kappa_0} \\
&= e^{\mu/2\lambda} \| \Phi \|_{\kappa_0}
\end{align}
Then, we also have 
\begin{align}
\| \Phi_{n+1} - \Phi_n \|_{\kappa_{n+1}} &\leq \frac{\mu}{4\lambda} \left( \frac{1}{2} \right)^n \| \Phi_n \|_{\kappa_n} \\
&\leq e^{\mu/2\lambda} \frac{\mu}{4\lambda} \left(\frac{1}{2}\right)^n \| \Phi \|_{\kappa_0},
\end{align}
from which we conclude by summation and the triangle inequality that
\begin{equation}
\| \Phi_{n} - \Phi \|_{\kappa_n} \leq e^{\mu/2\lambda} \frac{\mu}{2\lambda} \| \Phi_0 \|_{\kappa_0}.
\end{equation}
This completes the proof of Lemma \ref{lemma:shortrangeintermediate}. \qed

Now let us state how to prove Theorem \ref{thm:shortrangethm}. Lemma \ref{lemma:shortrangeintermediate} (with $\mu \sim \lambda$) already takes us most of the way there, but it does not give the $O(\lambda^2)$ scaling of $\| D - D^* \|_{\kappa_*}$ nor the $O(\lambda)$ scaling of $\| \mathcal{U} \Phi \mathcal{U}^{\dagger} - \Phi \|_{\kappa_*}$. The idea to fix this gap is that one should first do a single iteration of the procedure of Ref.~\cite{Else_1607}, with $\kappa_0 - \kappa_1$ held fixed independently of $\lambda$ (rather than the prescription above, for which $\kappa_1 - \kappa_0 \to 0$ as $\lambda \to 0$).
 In that case, we see from \eqnref{eq:epsilon_n} that $\epsilon_0 = O(\lambda^2)$. Now we apply Lemma \ref{lemma:shortrangeintermediate} to the $D_1, V_1, E_1$ that result from the first iteration. We see that we can set the $\mu$ appearing in the statement of Lemma \ref{lemma:shortrangeintermediate} to be $O(\lambda^2)$. Theorem \ref{thm:shortrangethm} immediately follows.

\section{Proof of Theorem 1}
\label{app:full_proof}

In this appendix, we prove our main theorem, Theorem \ref{thm:longrangethm} from Sec.~\ref{subsec:thmstatement}. One of the principal ingredients is a new version of the prethermalization theorem for \emph{short-range} interactions, which we describe in Appendix \ref{app:shortranged}. Here we extend this proof to range-indexed potentials, as introduced in the main text; recall that these are formal sums,
\begin{equation}
\Phi = \sum_{l=0}^{\infty} \sum_{Z \in \mathcal{Z}_{R_l}} \Phi_{Z,l},
\end{equation}
where we have introduced a sequence $R_l = e^{\sigma l}$, and $\mathcal{Z}_{R_l}$ is the set of all $R_l$-ranged subsets of sites (recall the definition of $R$-ranged set from Sec.~\ref{sec:Main_Ideas}).

We define the formal commutator of two range-indexed potentials according to
\begin{multline}
(\mathrm{ad}_\Phi \Theta)_{Z,l} := [\Phi,\Theta]_{Z,l} \\\:= \sum_{\substack{l_1, l_2 \geq 0 \\ \max\{l_1, l_2\} = l}} \sum_{\substack{Z_1 \in \mathcal{Z}_{R_{l_1}}, Z_2 \in \mathcal{Z}_{R_{l_2}} \\ Z_1 \cap Z_2 \neq \emptyset, Z_1 \cup Z_2 = Z}} [\Phi_{Z_1,l_1},\Theta_{Z_2,l_2}]
\end{multline}
The idea is that we take the commutator of $[\Phi_{Z_1,l_1}, \Theta_{Z_2,l_2}]$ to be supported on $Z_1 \cup Z_2$, and then we observe that in fact, if $Z_1$ and $Z_2$ are non-disjoint $R_{l_1}$- and $R_{l_2}$-ranged sets respectively, then indeed $Z_1 \cup Z_2$ is a $\max \{ R_{l_1}, R_{l_2} \} = R_{\max\{l_1,l_2\}}$-ranged set. This is true because an $R'$-ranged set is also an $R$-ranged set for $R > R'$, and the union of two non-disjoint $R$-ranged sets is also an $R$-ranged set.

 Then, we define the exponential action of one potential on another according to
\begin{equation}
e^\Phi \Theta e^{-\Phi} = \sum_{n=0}^{\infty} \frac{1}{n!} \mathrm{ad}^n_{\Phi} \Theta,
\end{equation}

Recall from the main text that we introduced a two-parameter norm for range-indexed potentials, according to
\begin{equation}
\| \Phi \|_{\kappa,\gamma} = \sum_{l=0}^{\infty} R_l^{\gamma} \sum_{Z \in \mathcal{Z}_{R_l}} e^{\kappa |Z|} \| \Phi_Z \|.
\end{equation}
We will find it convenient to fix some $\kappa_0, \gamma$ and define a one-parameter norm for range-indexed potentials:
\begin{align}\label{eq:app_newoneparamNorm}
\| \Phi \|_{\kappa} &:= \| \Phi \|_{\kappa, \gamma \kappa/\kappa_0} \\ &= \sum_{l=0}^{\infty} \sum_{Z \in \mathcal{Z}_{R_l}}  e^{\kappa(|Z| + \mu l)} \| \Phi_Z \|_{R_l}, \quad \mu = \sigma \gamma/\kappa_0~.
\end{align}
We emphasize that this is \emph{not} the same norm as \eqnref{eq:app_oneParam} for a potential $\Phi$ which does not keep any information regarding the range.

Now we can prove the following key lemma:
\begin{lemma}
\label{Lemma_4_1_extension}
Let $\Phi$,$\Theta$ be range-indexed potentials, and let $0 < \kappa' < \kappa$. Then
\begin{equation}
\| e^{\Phi} \Theta e^{-\Phi} - \Theta \|_{\kappa'} \leq \frac{18}{\kappa'(\kappa - \kappa')} \| \Phi \|_{\kappa} \| \Theta \|_{\kappa}.
\end{equation}
\begin{proof}
  This is analogous to Lemma 4.1 in Ref.~\cite{Abanin_1509}. Indeed, the proof carries through in exactly the same way, line by line, just replacing sums over $Z$ with sums over $(Z,l)$. The key fact for that proof was that for a collection of sets $S_0, \ldots, S_m$ which is connected (i.e. it cannot be separated into non-disjoint subcollections), the size of their union $P = \cup_{k=0}^m S_k$ can be bounded by the sum of the sizes of each $S_i$ as:
  \begin{align}
    |P| \leq -m + \sum_{j=0}^{m} |S_j|~.
  \end{align}
  For us, the analogous fact is as follows. Let $S_0, \ldots, S_m$ be a connected collection of sets, and let $l_0, \ldots, l_m\geq 0$. Then we have that
\begin{equation}
|P| + \mu \max\{l_0, \ldots, l_m\} \leq -m + \sum_{j=0}^m (|S_j| + \mu l_j).
\end{equation}
\end{proof}
\end{lemma}

In fact, Lemma \ref{Lemma_4_1_extension} is already sufficient to allow us to extend Theorem \ref{thm:shortrangethm} to range-indexed potentials!
The reason is that the only two things we needed to prove Theorem \ref{thm:shortrangethm} were the bounds \eqnref{eq:theiterationbounds} and Lemma 4.1 of Ref.~\cite{Abanin_1509}.
However, the only \emph{only} non-trivial property of potentials that was used in deriving \eqnref{eq:theiterationbounds} in Refs.~\cite{Abanin_1509,Else_1607} was Lemma 4.1 of Ref.~\cite{Abanin_1509} itself.

By generalizing Lemma 4.1 of Ref.~\cite{Abanin_1509} to Lemma \ref{Lemma_4_1_extension} (which applied to range-indexed potentials) all of the argumentation in Theorem \ref{thm:longrangethm} from Sec.~\ref{subsec:thmstatement} immediately carries over.

\section{Lieb-Robinson bounds for long-ranged interactions and the approximation of local observables}
\label{app:proof:ApproximationOfLocalObservables}
In this Appendix, we give the proof of Theorem \ref{thm:localdynamics} from Sec.~\ref{subsub:liebrobinson}.

We restrict our attention sets of sites $\Lambda$ that can be embedded in a Cartesian space $\mathbb{R}^d$, such that for any $x \in \Lambda$ there exists $\bm{r}_x \in \mathbb{R}^d$ such that dist$(x,y) = |\bm{r}_x - \bm{r}_y|$.
We also assume that there is a smallest distance $\min_{x,y} \text{dist}(x,y) =a$, which we normalize to be $1$.


The important result that we will use is that there is a Lieb-Robinson bound for time-evolution by range-indexed potentials with bounded norm $\| \cdot \|_{\kappa,\gamma}$, so long as $\gamma > d$.

\begin{lemma}\label{lemma:Lieb} ({\bf Lieb-Robinson bounds for generic graded potentials }
Let $\Phi(t)$ be a (time-dependent) graded potential with $\| \Phi \|_{\kappa,\gamma} < \infty$ for some $\kappa > 0$ and $\gamma > d$. Let $A$ be an operator supported on the set $X \subseteq \Lambda$, and let $B$ be an operator supported on $Y \subseteq \Lambda$. Define the time-evolution $\tau_t(A)$ as the time evolution of $A$ according to $\frac{d}{dt} \tau_t(A) = i[\tau_t(A),\Phi(t)]$. Then for any $\eta$ with $\frac{d+1}{\gamma+1}  < \eta < 1$, there is a Lieb-Robinson bound:
\begin{align}\label{eq:Iterative_Lieb}
    \frac{\|[\tau_t(A),B]\|}{\|A\| \|B\|} \le 2|X|e^{vt - r^{1-\eta}} + K_1 \frac{\tau + \tau^{\beta}}{r^{\eta\gamma}} |X|^{n_* + 2},
\end{align}
where:
\begin{align}
\beta &= 1 + d/(1-\eta), \\
n_* &= \left \lceil \frac{\eta d}{\eta \alpha - d} \right \rceil, \\
\tau &= vt, \\
v &= K_2 \max \left\{e^{-\gamma} \left(\frac{\gamma}{\kappa}\right)^{\gamma}, \kappa^{-1} \right\} \| \Phi \|_{\kappa,\gamma},
\end{align}
and $K_1$ and $K_2$ are constants that depend only on the geometry of the system and on $\eta$, and we have defined
\begin{equation}
\| \Phi \|_{\kappa,\gamma} = \frac{1}{t} \int_0^t ds \| \Phi(s) \|_{\kappa,\gamma}.
\end{equation}





\begin{proof}
  
This is a corollary of Theorem 1 in Ref.~\cite{Else_1809}. To show that the theorem applies,
we need only ensure that the assumptions of Sec.~I of Ref.~\cite{Else_1809} are satisfied.
First observe that there is always a rescaling of time (which might be nonlinear) such that $\| \Phi(t) \|_{\kappa,\gamma}$ becomes independent of $t$ and equal to $ \| \Phi \|_{\kappa,\gamma}$.

Now define $\Phi_Z = \sum_{l=0}^{\infty} \Phi_{Z,l}$ (where we take $\Phi_{Z,l} = 0$ if $Z$ is not an $R_l$-ranged set).
Then we have, for any $x \in \Lambda$, $s \in [0,t]$:
\newcommand{\caa}{c_\infty}
\begin{align}
& \sum_{Z \ni x ; \mathrm{diam}(Z) \geq r} \| \Phi_Z(s) \| \\
&\leq \sum_{l=0}^{\infty} \sum_{x \in Z \in \mathcal{Z}_{R_l} ; \mathrm{diam}(Z) \geq r} \| \Phi_{Z,l}(s) \| \\
&\leq \sum_{l=0}^{\infty} \sum_{x \in Z \in \mathcal{Z}_{R_l} ; \mathrm{diam}(Z) \geq r} e^{\kappa |Z|} e^{-\kappa r/R_l} \| \Phi_{Z,l}(s) \| \\
&\leq \sum_{l=0}^{\infty} \sum_{Z \ni x} e^{\kappa |Z|} e^{-\kappa r/R_l} \| \Phi_{Z,l}(s) \| \\
&= (\kappa r)^{-\gamma} \sum_{l=0}^{\infty} \sum_{Z \ni x} e^{\kappa |Z|} e^{-\kappa r/R_l} (\kappa r/R_l)^{\gamma} R_l^{\gamma} \| \Phi_{Z,l}(s) \| \\
&\leq e^{- \gamma} \gamma^{\gamma} \| \Phi \|_{\kappa,\gamma} (\kappa r)^{-\gamma},
\end{align}
where we used the fact that any $R_l$-ranged set $Z \in \mathcal{Z}_{R_l}$ satisfies $\mathrm{diam}(Z) \leq R_l |Z|$, and the fact that $\max_{x = [0,\infty)} e^{- \kappa x} (\kappa x)^{\gamma} = e^{- \gamma} \gamma^{\gamma}$.

Moreover, for any $x \in \Lambda$:
\begin{align}
  & \sum_{y \in \Lambda} \sum_{Z \ni x,y} \| \Phi_Z(s) \| \\
&\leq  \sum_{Z \ni x} |Z| \| \Phi_{Z}(s) \| \\
&\leq  \sum_{Z \ni x} \frac{1}{\kappa} e^{\kappa|Z|} \| \Phi_{Z}(s) \| \\
&\leq \frac{1}{\kappa} \| \Phi \|_{\kappa,0} \leq \frac{1}{\kappa} \| \Phi \|_{\kappa,\gamma}
\end{align}

Hence, we see that the assumptions of Theorem 1 of Ref.~\cite{Else_1809} are satisfied with
\begin{align}
  J &= e^{-\gamma} (\gamma/\kappa)^{\gamma} \| \Phi \|_{\kappa,\gamma}, \\
  \mathcal{C}_0 &= \frac{1}{\kappa} \| \Phi \|_{\kappa,\gamma}.
\end{align}





Therefore, the Lieb-Robinson bound follows from Ref.~\cite{Else_1809}.
\end{proof}
\end{lemma}

Having proven that Lieb-Robinson bounds apply for range-indexed potentials with bounded norm provided that $\gamma > d$, we can now prove that small (in terms of the norm $\| . \|_{\kappa,\gamma}$) perturbations induce small changes in the dynamics of local observables. This will be encapsulated in Lemma~\ref{lemma:ClosenessOfEvolution}. Combining Lemma \ref{lemma:ClosenessOfEvolution} with Theorem \ref{thm:longrangethm} will then immediately imply Theorem \ref{thm:localdynamics}.

\begin{lemma}\label{lemma:ClosenessOfEvolution}
Let $\Phi_1(t)$ and $\Phi_2(t)$ be two time-dependent range-indexed potentials, such that $\Phi_2$ satisfies Lemma \ref{lemma:Lieb}. Let $U_j(t) = \mathcal{T} \exp\left[-i\int_0^t \Phi_j(t)\right]$ be the corresponding time evolutions, and define $\Delta(t) = \Phi_1(t) - \Phi_2(t)$.

Then, the difference in time evolved local operator $O$ (initially support on the set $X\subseteq \Lambda$)  under $\Phi_1$ and $\Phi_2$ is bounded by:
\begin{align}
  \left\| U_1^\dag (t) O U_1(t) - U_2^\dag(t) O U_2(t)]\right\| \le |X| \|O\|\|\Delta\|_{0,0} t \notag \\
    \times \left\{ K_3(1+ \tau^{d/(1-\eta)}) |X| + K_4 (\tau + \tau^{\beta}) |X|^{n_* + 2} \right \},
\end{align}
where we defined
\begin{equation}
\| \Delta \|_{0,0} = \frac{1}{t}\int_0^t  ds \| \Delta(s) \|_{0,0}.
\end{equation}
Here $K_3$ is another constant that depends only on the geometry of the lattice and on $\eta$ (but \emph{not} the system size), and $K_4$ depends on the geometry of the lattice, on $\eta$, and on $\gamma$, but not the system size. This result holds provided that $\eta$ is as prescribed in Lemma \ref{lemma:Lieb} and also satisfies $\eta \gamma > d$.
\begin{proof}

We write the Lieb-Robinson bound from Lemma \ref{lemma:Lieb} as
\begin{equation}
    \frac{\|[\tau_t(A),B]\|}{\|A\| \|B\|} \leq f(r),
\end{equation}
where $f(r,t) = f_1(r,t) + f_2(r,t) + f_3(r,t)$, with
\begin{align}
f_1(r,t) &= 2 \theta(\xi(t) - r), \\
f_2(r,t) &= 2 e^{vt - r^{1-\eta}} |X| \theta(r - \xi(t)), \\
f_3(r,t) &= K_1 \frac{\tau + \tau^{\beta}}{r^{\eta \gamma}} |X|^{n_* + 2},
\end{align}
Here $\theta$ is the Heaviside step function, $\tau = vt$, $\xi(t) = (vt)^{1/(1-\eta)}$, and we have also invoked the trivial commutator bound $\| [\tau_t(A),B] \| \leq 2 \| A \| \|B ||$.

Now we use the fact that
\begin{align}
    \frac{d}{dt}( U_1 U_2^\dag O U_2 U_1^\dag)
    &= -i U_1 [\Delta, U_2^\dag O U_2] U_1^\dag.
\end{align}
Integrating this result, we obtain:
\begin{align}
&U_1(t) U_2^\dag(t) O U_2(t) U_1^\dag(t) - O = \\
&-i \int_0^t ds~ U_1(s) [ \Delta(s), U_2^\dag(s) O U_2(s) ] U_1^\dag(s),
\end{align}
and, thus,
\begin{align}
    \| U_1^\dag(t) O& U_1(t) - U_2^\dag(t) O U_2(t) \| \\
    &\leq \int_0^t ds \| [ \Delta(s), U_2^\dag(s) O U_2(s) ] \| \\
    &\leq \int_0^t ds \sum_Z \| [ \Delta_Z(s), U_2^\dag (s) O U_2(s) ] \|, 
\end{align}
where we defined $\Delta_Z(s) = \sum_{l=0}^{\infty} \Delta_{Z,l}(s)$.

Now to bound the commutator we consider
\begin{align}
& \int_0^t \sum_Z \| [ \Delta_Z(s), U_2^\dag(s) O U_2(s) ] \| \\
&\leq \int_0^t ds \sum_Z \| \Delta_Z(s) \| \| O \| f(\mathrm{dist}(Z,X),s) \\
&\leq \int_0^t ds \sum_z \sum_{Z \ni z} \| \Delta_Z(s) \| \| O \| f(\mathrm{dist}(z,X),s) \\
&\leq t \| \Delta \|_{0,0} \| O \| \sum_z f(\mathrm{dist}(z,X),t) \\
&\leq t \| \Delta \|_{0,0} \| O \| |X| \sup_x \sum_z f(\mathrm{dist}(z,x),t),
\end{align}
 and used the fact that $f(\cdot,t)$ is monotonic in $t$.

Then we observe that
\begin{equation}
\sum_z f_1(\mathrm{dist}(z,x),t) \leq V\{\xi(t)\},
\end{equation}
where $V\{ \xi(t) \} \leq K_3(1 + \xi(t)^d)$ is the number of points within distance $\xi(t)$ of a given point. Moreover, we also have
\begin{align}
  & \sum_z f_2(\mathrm{dist}(z,x),t) \\
  &= 2|X| \sum_{z, \mathrm{dist}(x,z) \geq \xi(t)} e^{vt - \mathrm{dist}(x,z)^{1-\eta}} \\
&\leq  2|X| \sum_{z,\mathrm{dist}(x,z) \geq \xi(t)} e^{-[\mathrm{dist}(x,z) - \xi(t)]^{1-\eta}} \label{eq:bernoulli} \\
&\leq 2|X| \sum_{y,\mathrm{dist}(x,y) \leq \xi(t)} \sum_z e^{-\mathrm{dist}(y,z)^{1-\eta}} \\
&\leq V \{ \xi(t) \} K_3 |X|,
\end{align}
where in \eqnref{eq:bernoulli} we used Bernoulli's inequality. Finally, we have
\begin{equation}
\sum_z f_3(\mathrm{dist}(z,x),t) \leq K_3 (\tau + \tau^{\beta}) |X|^{n_* + 2}
\end{equation}
where
\begin{equation}
K_4 = K_1 \sup_y \sum_z \frac{1}{\mathrm{dist}(z,y)^{\eta \gamma}},
\end{equation}
which is finite in the thermodynamic limit provided $\eta \gamma > d$.
\end{proof}
\end{lemma}

\section{Approximation of local observables for $\alpha > d$}
\label{app:approxobservshort}
In this appendix, we will deal only with potentials (not range-indexed potentials). Starting from a range-indexed potential we can construct a potential just by defining $\Phi_Z = \sum_{l=0}^{\infty} \Phi_{Z,l}$.

We define the Heisenberg evolution of a (time-independent) 
potential $\Theta$ by a (time-dependent) potential $\Phi(t)$ through the Dyson series for Heisenberg evolution, i.e.
\begin{multline}
\mathcal{E}_{\Phi}(t) \Theta := \sum_{n=0}^{\infty} i^n \int_0^t dt_1 \int_0^{t_1} dt_2 \cdots \int_0^{t_{n-1}} dt_n\\ \times \mathrm{ad}_{\Phi(t_1)} \cdots \mathrm{ad}_{\Phi(t_n)}. \Theta,
\end{multline}
where $\mathrm{ad}_\Phi \Theta = [\Phi,\Theta]$. This satisfies
\begin{equation}
\frac{d}{dt} \mathcal{E}_{\Phi}(t) = i \mathrm{ad}_{\Phi(t)} \mathcal{E}_{\Phi}(t)
\end{equation}

Our key result is as follows.
\begin{lemma}
  \label{lemma:timedependent4_1}
  Consider numbers $0 < \kappa' < \kappa$, and suppose that $3t \| \Phi \|_{\kappa'} \leq \kappa - \kappa'$. Then:
\begin{equation}
\| \mathcal{E}_{\Phi}(t) \Theta - \Theta \|_{\kappa'} \leq \frac{18t }{\kappa'(\kappa - \kappa')} \| \Theta \|_{\kappa} \| \Phi \|_{\kappa},
\end{equation}
Here we defined
\begin{equation}
\| \Phi \|_{\kappa} = \frac{1}{t} \int_0^T \| \Phi(t) \|_{\kappa}.
\end{equation}
\begin{proof}
This is basically a time-dependent version of Lemma 4.1 from Ref.~\cite{Abanin_1509}. The proof proceeds in a nearly identical way. Indeed, we have
\begin{widetext}
\begin{align}
\| (\mathcal{E}_{\Phi}(t) \Theta)_{P} - \Theta_P \| &\leq \sum_{n=1}^{\infty} \int_0^t dt_1 \int_0^{t_1} dt_2 \cdots \int_0^{t_{n-1}} dt_n  \sum_{S_0, \cdots, S_m}^{c,P} \| \Theta_{S_0} \| \prod_{j=1}^n (2\| \Phi_{S_j}(t_j) \| \\
&= \sum_{n=1}^{\infty} \frac{1}{n!} \int_0^t dt_1 \cdots \int_0^t dt_n \sum_{S_0, \cdots, S_m}^{c,P} \| \Theta_{S_0} \| \sum_{j=1}^n (2\| \Phi_{S_j}(t_j) \|) \\
&= \sum_{n=1}^{\infty} \frac{1}{n!} \sum_{S_0, \cdots, S_m}^{c,P} \| \Theta_{S_0} \| \sum_{j=1}^n (2t\| \Phi_{S_j} \|) \\,
\end{align}
\end{widetext}
where we defined $\| \Phi_Z \| = \frac{1}{t} \int_0^t \| \Phi_Z(t) \|$. The rest of the proof proceeds identically to Lemma 4.1 of Ref.~\cite{Abanin_1509}.
\end{proof}
\end{lemma}

A corollary of this (or, in fact, of Lemma 4.1 of Ref.~\cite{Abanin_1509}) is as follows.
\begin{lemma}
  \label{lemma:commbound}
  For any potential $W$, we have
\begin{align}
\| \mathrm{ad}_W \Theta \|_{\kappa'} \leq \frac{18}{\kappa'(\kappa - \kappa')} \| \Theta \|_{\kappa} \| W \|_{\kappa}.
\end{align}
\begin{proof}
Just use the fact that
\begin{equation}
\mathrm{ad}_W = \lim_{t \to 0} \frac{\mathcal{E}_W(t) - \mathbb{I}}{t}.
\end{equation}
\end{proof}
\end{lemma}

Now we can prove a result about approximation of local observables.
\begin{lemma}
  \label{lem:generalapproximation}
  Define $\lambda = \max \{ \| \Phi \|_{\kappa}, \| \Phi' \|_{\kappa}\}$.
  Suppose that $12\lambda t \leq (\kappa - \kappa')$. Then
  \begin{equation}
 \| \mathcal{E}_\Phi(t) \Theta - \mathcal{E}_{\Phi'}(t) \Theta \|_{\kappa'} \leq \mathcal{C}^3 \mathcal{M} t \| \Delta \|_{\kappa},
  \end{equation}
  where we defined $\Delta(t) = \Phi(t) - \Phi'(t)$, $\| \Phi \|_\kappa = \frac{1}{t} \int_0^t \| \Phi(s) \|_{\kappa} ds$ (and similarly
  for $\Phi'$, $\Delta$), and
  \begin{align}
    \mathcal{M} &= \frac{72}{\kappa'(\kappa - \kappa')}, \\
    \mathcal{C} &= 1 + \mathcal{M} \lambda t \leq 1 + \frac{6}{\kappa'}.
  \end{align}

\begin{proof}
We introduce a sequence $\kappa = \kappa_0 > \kappa_1 > \kappa_2 > \kappa_3 > \kappa_4 = \kappa'$, such that $\kappa_j - \kappa_{j+1} = (\kappa - \kappa')/4$.
\begin{equation}
\frac{d}{ds} [\mathcal{E}_{\Phi'}^{-1}(s) \mathcal{E}_{\Phi}(s)] = -i \mathcal{E}_{\Phi'}^{-1}(s) \mathrm{ad}_{\Delta(s)} \mathcal{E}_{\Phi}(s)
\end{equation}
and therefore,
\begin{align}
 & \left\| \frac{d}{ds} \mathcal{E}_{\Phi'}^{-1}(s) \mathcal{E}_{\Phi}(s) \Theta \right\|_{\kappa_3}
 \\&\leq \mathcal{C} \| \mathrm{ad}_{\Delta(s)} \mathcal{E}_{\Phi}(s) \Theta \|_{\kappa_2} \\
&\leq \mathcal{C}\mathcal{M} \| \Delta(s) \|_{\kappa_{1}} \| \mathcal{E}_{\Phi}(s) \Theta \|_{\kappa_1} \\
&\leq \mathcal{C}^2 \mathcal{M} \| \Delta(s) \|_{\kappa} \| \Theta \|_{\kappa},
\end{align}
where we have invoked Lemmas \ref{lemma:timedependent4_1} and \ref{lemma:commbound}. 
This then gives
\begin{align}
  & \| \mathcal{E}_{\Phi'}^{-1}(t) \mathcal{E}_{\Phi}(t) \Theta - \Theta \|_{\kappa_3} \\
  &\leq \mathcal{C}^2 \mathcal{M} \| \Theta \|_{\kappa} \int_0^t ds \| \Delta(s) \|_{\kappa} \\
  &= \mathcal{C}^2 \mathcal{M} \| \Theta \|_{\kappa} \| \Delta \|_{\kappa} t.
\end{align}
 Finally, we obtain
\begin{align}
 & \| \mathcal{E}_{\Phi}(t) \Theta - \mathcal{E}_{\Phi'}(t) \Theta \|_{\kappa'} \\
 &= \| \mathcal{E}_{\Phi'}(t) [ \mathcal{E}^{-1}_{\Phi'}(t) \mathcal{E}_{\Phi}(t) \Theta - \Theta ] \|_{\kappa'} \\
 &\leq \mathcal{C}^3 \mathcal{M} \| \Theta \|_{\kappa} \| \Delta \|_{\kappa} t,
\end{align}
where we invoked Lemma \ref{lemma:timedependent4_1} once more.
\end{proof}
\end{lemma}

An immediate corollary is as follows.
\begin{lemma}
  \label{lem:corollary}
  Define $\lambda = \max \{ \| \Phi \|_{\kappa}, \| \Phi' \|_{\kappa}\}$.
  Suppose that $24 \lambda t \leq \kappa$. Let $O$ be an observable supported on a set $S$. Then
  \begin{equation}
 \| \mathcal{E}_\Phi(t) O - \mathcal{E}_{\Phi'}(t) O \| \leq \mathcal{C}^3 \mathcal{M} e^{\kappa |S|} t \| \Delta \|_{\kappa},
  \end{equation}
  where we defined $\Delta(t) = \Phi(t) - \Phi'(t)$, $\| \Phi \|_\kappa = \frac{1}{t} \int_0^t \| \Phi(s) \|_{\kappa} ds$ (and similarly
  for $\Phi'$, $\Delta$), and
  \begin{align}
    \mathcal{M} &= 288/\kappa^2, \\
    \mathcal{C} &= 1 + \mathcal{M} \lambda t \leq 1 + \frac{12}{\kappa}.
  \end{align}

\begin{proof}
  We define $\kappa' = \kappa/2$ and treat $O$ as a potential with a single term $O_S = O$. Then $\| O \|_{\kappa} = e^{\kappa |S|} \| O \|$. Moreover, we observe that $\delta := \mathcal{E}_{\Phi} O - \mathcal{E}_{\Phi'} O$, considered a potential, only takes nonzero values on sets $Z$ that contain $S$. Therefore, given some $s \in S$, we have
  \begin{align}
    \| \delta \| \leq \sum_Z \| \delta_Z \| = \sum_{Z \ni s} \| \delta_Z \| \leq \| \delta \|_0 \leq \| \delta \|_{\kappa'},
  \end{align}
  and then the result follows from Lemma \ref{lem:generalapproximation}.
\end{proof}
\end{lemma}

Lemma \ref{lem:corollary} then immediately implies Theorem \ref{thm:localdynamics_short} in the main text.

\section{Further Data on the prethermalization to $D^*$}
\label{app:prethermalization}

\begin{figure*}
  \centering
  \includegraphics[width = 7.2in]{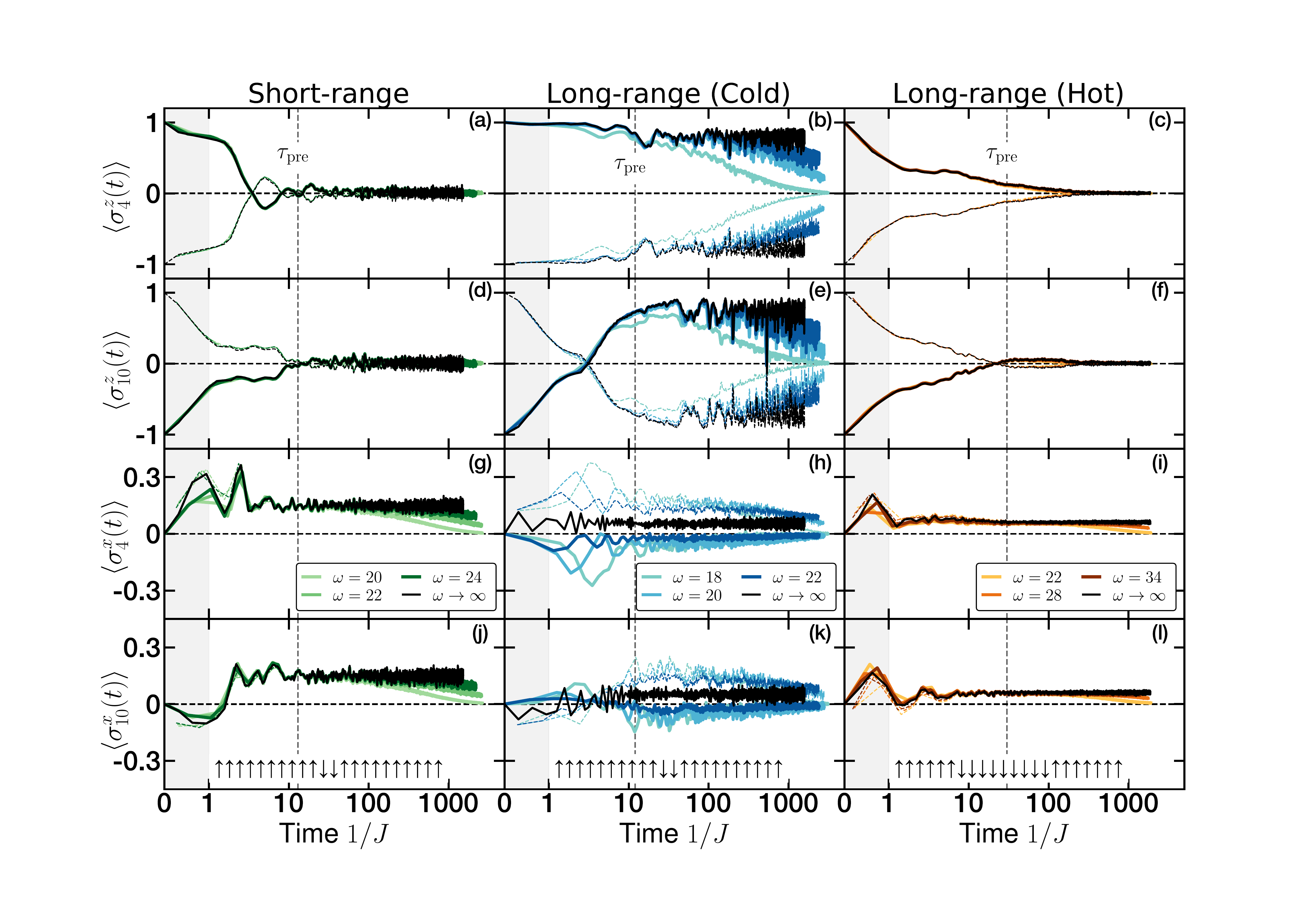}
  \caption{
    Analysis of the evolution of different single spin operators---$\sigma^z_4$, $\sigma^z_{10}$,$\sigma^x_4$ and $\sigma^x_{10}$---for the different conditions considered in \rfig{fig:LongvsShort_evo}: the short-range model (a), (d), (g) and (j), a ``cold'' initial state in the long-range model (b), (e), (h) and (k), and a ``hot'' initial state in (e), (f), (i) and (l).
    On the different single spin observables, we observe the approach to a position independent constant within the prethermal regime, consistent with the plateau observed in the $\omega\to\infty$ limit Floquet evolution, further suggesting that the system has approached a thermal state of the prethermal Hamiltonian $D^*$.
    By increasing the frequency of the driven system, we observe this agreement extending to later time, highlighting that the disagreement occurs due to the late time heating which becomes meaningful at $\tau^* \sim e^{\omega/J_{\mathrm{local}}}$.
    We also note that this simple picture is more complex in the case of $\sigma^x$.
    In this case, one needs to account for the small frame rotation $\mathcal{U}$ which can induce a finite overlap between  $\mathcal{U}\sigma^x\mathcal{U}^\dag$ and an observable that fails to commute with $X$.
  }
  \label{fig_ap:SingleSpin}
\end{figure*}

\begin{figure*}
  \centering
    \includegraphics[width=7.2in]{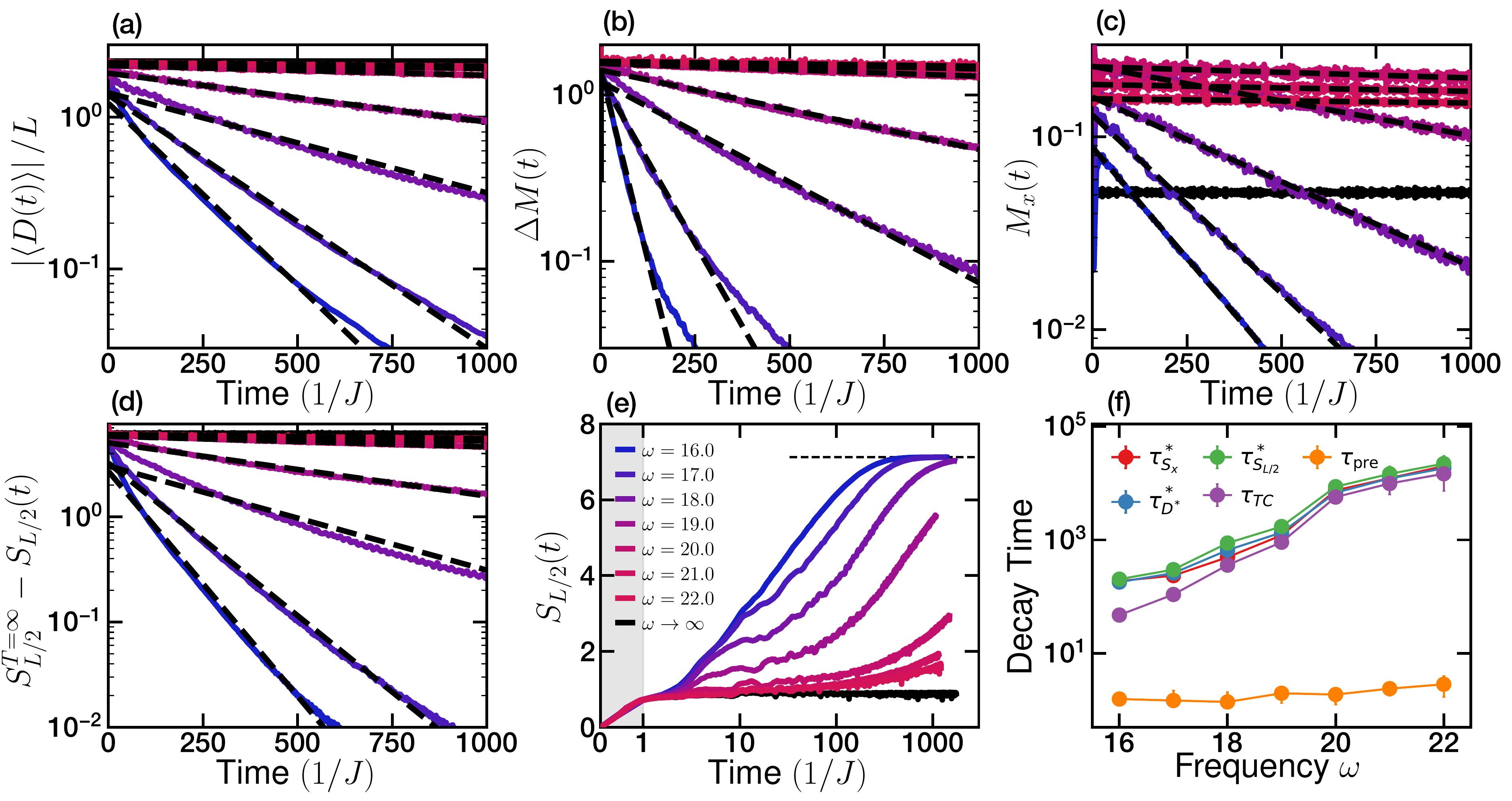}
    \caption{Example of the fitting procedure for extracting the decay times for a particular initial state evolved with the long-range Floquet evolution.
      We apply the same procedure to all initial states in both the short- and the long-range model.
      We observe that a simple exponential decay captures the approach of different observables to their thermal values:
      {\bf (a)} energy density $\langle D(t)\rangle/L$,
      {\bf (b)} time crystalline order parameter $\Delta M(t)$,
      {\bf (c)} $\hat{x}$ magnetization $M_x(t)$ (here plotted with a moving average over five points for clarity), and
      {\bf (d)} and half-chain entanglement entropy $S_{L/2}(t)$.
      {\bf (e)} The entanglement entropy provides an extra time scale $\tau_{\mathrm{pre}}$ which captures the approach to the prethermal state. The x-axis in the shaded region is linear with time to emphasize the early time entanglement entropy behavior.
      {\bf (f)} Comparison of the different decay times.
      The decay time of the energy density $\tau^*_{D^*}$, entropy $\tau^*_{S_{L/2}}$ and $\hat{x}$ magnetization $\tau^*_{S_x}$  provide different estimates of the true thermalization time scale of the system $\tau^*$.
      Because this particular initial state is a ``cold'' state of the long-range model, it hosts a prethermal time crystal; the decay of the time crystalline order parameter also occurs at $\tau^*$.
      The agreement of all these time scales further corroborates the existence of a prethermal time crystal and the existence of a single thermalization time scale.
      Finally, we observe that $\tau_{\mathrm{pre}}$  occurs at a much earlier, frequency independent time scale.}
    \label{fig:Extraction}
\end{figure*}

In Fig.~\ref{fig:LongvsShort_evo} of the main text, we studied the late time Floquet dynamics of different initial states.
The main feature, that underlies much of our results is the existence of a long-lived prethermal plateau, where the system approaches an equilibrium state with respect to the prethermal Hamiltonian $D^*$.
In the main text, we studied the system's equilibration via the dynamics of energy density, entanglement entropy, and global magnetization (where the latter two exhibit long-lived plateaus consistent with the evolution under $D$, the zeroth term of $D^*$)
In this appendix, we supplement this analysis with the dynamics of \emph{local} observables where we observe the approach of the dynamics to that of the prethermal Hamiltonian.
Curiously, by studying the dynamics of the $\sigma^x$ operator, we observe evidence of the small, but finite, rotation of frame $\mathcal{U}$ that appeared in the statement of our theorem.

Our results are summarized in Fig.~\ref{fig_ap:SingleSpin}, where we consider the dynamics of $\sigma^z_4$, $\sigma^z_{10}$, $ \sigma^x_4$, $\sigma^x_{10}$ for the initial states considered in the main text, Fig.~\ref{fig:LongvsShort_evo}.
We focus on the dynamics of even (full lines) and odd periods (thin dashed lines) independently in order to highlight any time crystalline behavior the local observables might possess (indeed this behavior is clear in the dynamics of $\sigma^z$).
We also consider the time evolution in the $\omega\to\infty$ limit where $U_f = Xe^{-iDT}$ (thin dashed line).
This evolution enables us to see how well the full Floquet dynamics is captured by $D^*$ within the prethermal regime.

In particular, we wish to emphasize three different features in the dynamics of local observables.
First, for the initial states that fail to approach the symmetry broken prethermal phase, first and third column of Fig.~\ref{fig_ap:SingleSpin}, we observe that the dynamics of local observables under the Floquet evolution closely follows the dynamics of local observables under $D$ until a late time approach to their infinite temperature value.
By increasing the frequency of the drive, we observe this agreement extending to longer and longer times, emphasizing that $D^*$ is indeed the generator of the local dynamics of the system in the prethermal regime and that deviations occur due to the heating at a time scale $\tau^* \sim e^{\omega/J_{\mathrm{local}}}$.

Second, this picture is not so clear when considering the initial state which approaches symmetry broken state in the prethermal regime, second column of Fig.~\ref{fig_ap:SingleSpin}.
While the dynamics of $\sigma^z$ in this case are also very well described by $D$, the same is not true when considering $\sigma^x$. We can attribute this to the effect of the small change of frame $\mathcal{U}$; in the original lab frame, the system is really evolving under $\mathcal{U} D^* \mathcal{U}^{\dagger}$ rather than $D^*$. Hence, measuring $\sigma^x$ in the lab frame is equivalent to measuring $\mathcal{U} \sigma^x \mathcal{U}^{\dagger}$ in the rotated frame (where the evolution is governed by $D^*$). The latter has some overlap with $\sigma^z$, which has large expectation value in the spontaneous symmetry broken phase of $D^*$ (but zero expectation value in the symmetry-unbroken phase). Hence, since $\mathcal{U}$ is $O(1/\omega)$ close to the identity, one finds that there is an $O(1/\omega)$ contribution to the expectation of $\sigma^x$ in the lab frame, which disappears as $\omega \to \infty$ as can be observed in the numerics.
[Note that other observables in principle could display the same effect, both inside and outside of the prethermal time crystal phase, but one can check by explicitly computing the perturbative expansion for $\mathcal{U}$ that the $O(1/\omega)$ corrections happen to be much smaller in those cases.]
These $O(1/\omega)$ corrections also differ between odd and even periods (i.e., they exhibit time-crystalline behavior), which is consistent with the picture that they arise from the overlap of $\mathcal{U} \sigma^x \mathcal{U}^{\dagger}$ with $\sigma^z$.

\begin{figure*}
  \centering
  \includegraphics[width=7.2in]{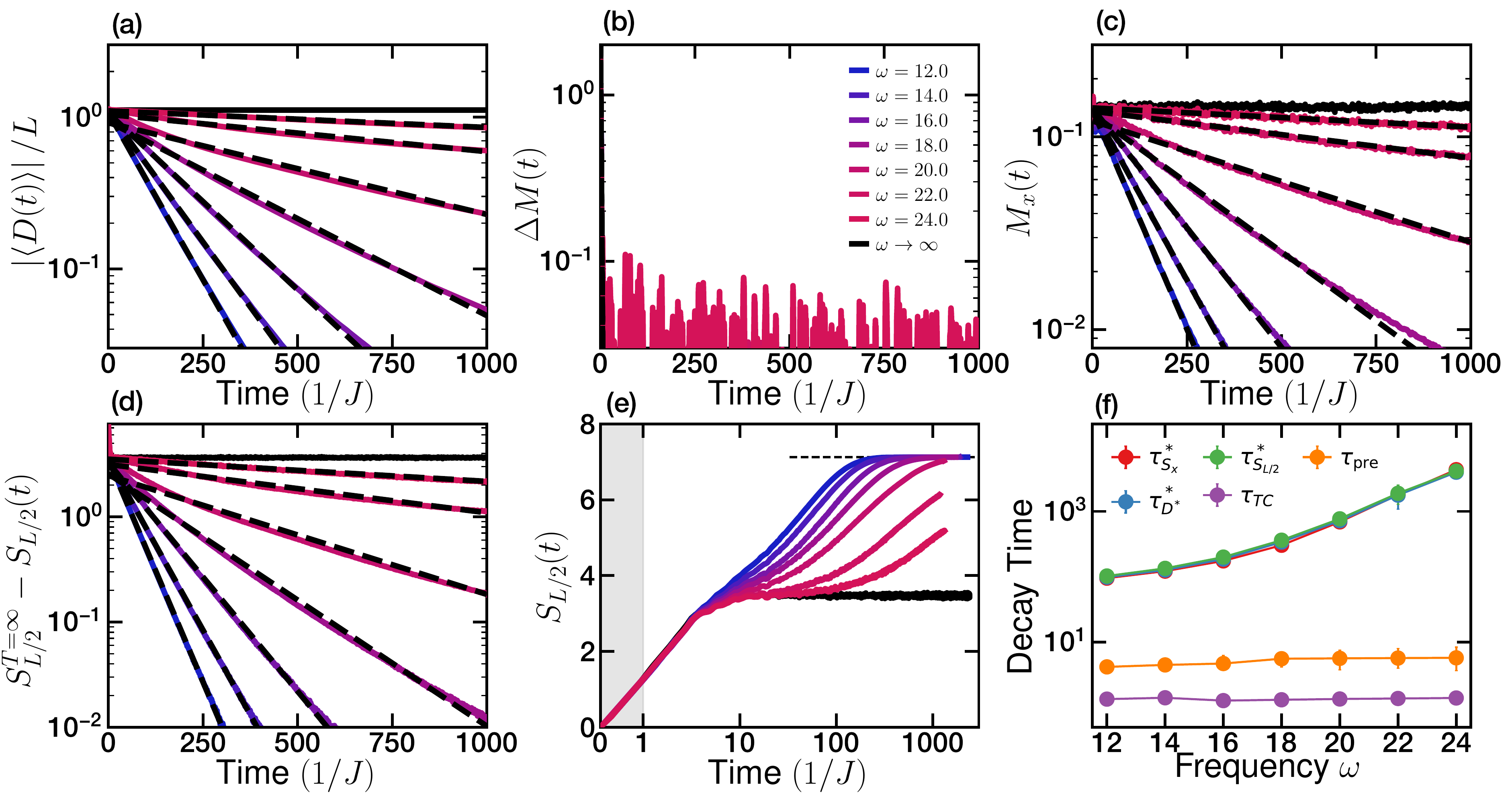}
  \caption{
    Analogous to Fig.~\ref{fig:Extraction}, but considering an initial state time evolved with the short-range Floquet evolution.
    As in \rfig{fig:Extraction} we observe that a simple exponential decay captures the broad features of the approach of the different quantities to their thermal values.
    Moreover, we also observe a good agreement between the $\tau^*_{D^*}$, $\tau^*_{S_{L/2}}$, and $\tau^*_{S_x}$ as measures of the thermalization time $\tau^*$.
    However, unlike the long-range case, the time crystalline order parameter [{\bf (b)}] decays at a much faster, frequency independent, time scale.
    This time is on the same order of $\tau_{\mathrm{pre}}$ further corroborating that, in this case, the decay of the time crystalline order arises from the dynamics of the prethermal Hamiltonian.}
    \label{fig:Extraction_2}
\end{figure*}

Finally, by comparing the dynamics of $\sigma_4^z$ and $\sigma^z_{10}$, we can directly observe the local prethermalization of the system.
In our choice of states, these two observables take opposite initial values, yet the translation invariance of our system implies that they must thermalize to the same value. In particular, in the symmetry-broken phase, the thermal value of $\sigma^z$ is large, and so the sign of one of the local observables must change.
Since the chain is mostly pointing up, $\sigma^z_{10}$, which started with a negative value, must prethermalize to a finite positive value, matching the magnetization of the remaining spins (including $\sigma^z_4$).
This is indeed what we observe, supporting the claim that the system approaches the prethermal state and that we are indeed observing the prethermal time crystalline phase.

\section{Extraction of the thermalization time scales}
\label{app:extraction}

In order to better understand the thermalization dynamics of our Floquet evolution, we quantify the time scale at which  different quantities approach their late time thermal values.
In particular, we focus on the following quantities: the energy density of the system $\langle D (t)\rangle / L$, entanglement entropy $S_{L/2}(t)$, time crystalline order parameter $\Delta M(t)$ and the average magnetization in the $\hat{x}$ direction $M_x(t)$, where the latter is defined as:
\begin{align}
  M_x(t) &= \frac{1}{L} \sum_{i=0}^{L-1} \langle \sigma_i^x(t)\rangle
\end{align}
We define the associated decay times as $\tau_{D^*}^*$, $\tau^*_{S_{L/2}}$, $\tau^*_{\mathrm{TC}}$ and $\tau^*_{S_x}$, respectively.

Although the complete dynamics of each quantity $O(t)$ is non-trivial, at late times the system is in a local thermal state with respect to $D^*$ and their dynamics become much simpler.
In particular, we observe that they exhibit an exponential approach to their infinite temperature value $O^{T=\infty}$:
\begin{align}
  |O(t) - O^{T=\infty}| \approx O_c e^{-t/\tau}~.
\end{align}
Although this prescription is not exact and small deviations are observed, it provides a simple and robust way of extracting the thermalization time scale associated with each quantity.

This functional form motivates the following fitting procedure
\begin{itemize}
\item We consider the evolution information at every other period, so as to avoid any systematic effects of the period doubling behavior on the fits.
  The only observable where this effect is significant is the $\hat{x}$ magnetization $M_x(t)$ (as discussed in Appendix~\ref{app:prethermalization}).
  Nevertheless, we observe that the extracted time scales are consistent regardless of the parity of the period considered.
\item We restrict the data for the fit to the regime where $|O^{T=\infty} - O(t)| > \epsilon$ for some small $\epsilon$ ($\epsilon = 0.05$ for energy density, $\epsilon=0.1$ for time crystalline order parameter and entanglement entropy and $\epsilon = 0.015$ for $\hat{x}$ magnetization).
  We found this cutoff necessary to ensure that the fitted curves captured the correct approach and were not dominated by the very small late time fluctuations close to the thermal value.
\item We fit the linear relation $y=x/a+b$ to $\log|O^{T=\infty} - O(t)|$ as a function of $t$.
  The decay time scale is immediately given by the extracted value of $a$.
\item Finally we estimate the error of the procedure by partitioning the data in five regions and performing the same fitting procedure. The error is given by the weighted standard deviation of these results with respect to the global fit.  
\end{itemize}

Before moving on, let us note a small detail regarding the entropy time scale.
Near infinite temperature $\beta^{-1}$, the entanglement entropy scales as $\beta^2$ as opposed to $\beta$ like the other observables.
As a result, to ensure that $\tau^*_{S_{L/2}}$ is capturing the same heating time scale $\tau^*$, the extracted value must be multiplied by a factor of 2 (for more details, see the appendix of Ref.~\cite{Ye_1902}).

Finally, the time evolution of the entanglement entropy also provides one more time scale: the time at which the system has approached the prethermal state $\tau_{\mathrm{pre}}$.
Unfortunately, the entropy dynamics are much more complex, so the above detailed fitting procedure does not apply.
As a result, we follow a different procedure.
Using the evolution of the initial state under the static Hamiltonian $D$, we obtain an approximation to the prethermal entanglement entropy value $S_{L/2}^{\mathrm{pre}}$, averaging the entanglement entropy value at late times.
The time at which the driven system reaches $0.9\times S_{L/2}^{\mathrm{pre}}$ provides an estimate for $\tau_{\mathrm{pre}}$.
The error of this procedure is estimated by measuring the times at which the evolution reaches $(0.9 \pm 0.05) S_{L/2}^{\mathrm{pre}}$.

We summarize the both fitting procedures in Figs.~\ref{fig:Extraction} and Fig.~\ref{fig:Extraction_2}, where we consider an initial state evolved under the long- and the short-range model, respectively.
The resulting decay times are plotted in the bottom right-hand panel, where we see agreement between all measures of the heating time scale $\tau^*$, as well as the existence of a much earlier, frequency independent, decay time associated with the approach to the prethermal regime $\tau_{\mathrm{pre}}$.

\section{Further Evidence of Critical Slowing Down}
\label{app:Critical_Slowing}

\begin{figure}
  \centering
\includegraphics[width = 3.2in]{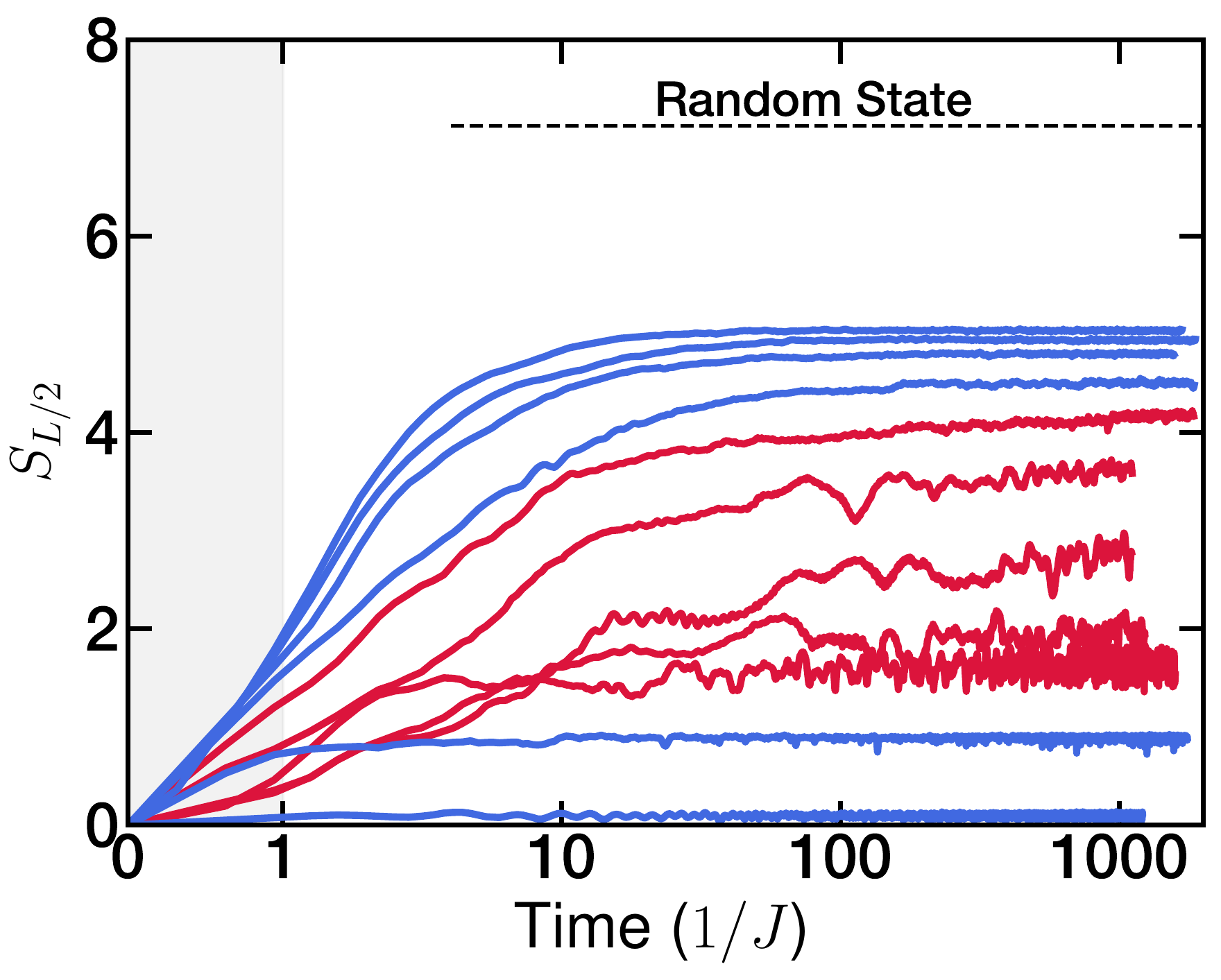}
\caption{
  Evolution of the half-chain entanglement entropy $S_{L/2}(t)$ for different initial states evolved under the long range $D$.
  We observe that, for states away from the phase transition (blue lines), the evolution is characterized by a fast approach to a well defined constant plateau.
  However, for initial states near the phase transition (red lines), the approach takes a very long time, displaying a slowly growing entropy for very long times and displaying large fluctuations.
  The initial states marked in red, correspond to the state lying in the transition region in Fig.~\ref{fig:spectrum} of the main text.
}
  \label{fig:Entropy_SlowingDown}
\end{figure}

As we approach the phase transition of $D^*$ from the paramagnetic side, we begin to observe the extension of the life time of the time crystalline order parameter, despite the system being in the trivial phase.
This does not correspond to the breakdown of the prethermal phase, but rather extra physics in the equilibration dynamics under the prethermal Hamiltonian $D^*$.
In particular, this corresponds to the known phenomena of critical slowing down.
When one is close to the phase transition, small fluctuations in energy alter significantly the system's tendency to order or not; the system is unable to efficiently ``choose'' which side of the transition it actually is and equilibration takes a long time.
This results in significant fluctuations in the dynamics and an enhancement in the timescale at which the system approaches the prehtermal state $\tau_{\mathrm{pre}}$.

We can corroborate this hypothesis by investigating the dynamics of different initial product states evolving under the static Hamiltonian $D$.
We focus on the entanglement entropy as its behavior has the simplest expectation; starting from zero, we expect the entanglement entropy to monotonically increase and approach a well-defined plateau corresponding to the equilibrium state.
This is exactly what we observe, for initial states far away from the phase transition, blue curves in Fig.~\ref{fig:Entropy_SlowingDown}.
For initial states near the phase transition (on either side), red curves in Fig.~\ref{fig:Entropy_SlowingDown}, we observe a slower rate of entropy growth, plagued by much larger fluctuations.
Moreover, these states also exhibit a very late approach to a well-defined plateau; some curves have yet to approach such a plateau although we are considering very late time dynamics, $t\gtrsim 1000/J$.

\section{Quantum Monte Carlo calculation}
\label{app:QMC}

One of the requirements for a prethermal time crystal is a spontaneously symmetry broken phase of the prethermal Hamiltonian; as long as the system thermalizes to a spontaneous symmetry broken phase of $D^*$, it will exhibit long-lived time crystalline behavior.
As such, whether the system is in the prethermal time crystal phase is dependent on the temperature $\beta^{-1}$ of the system as it prethermalizes to $D^*$.
In particular, as the system crosses the critical temperature $T_c$, the system transitions from the prethermal time crystal phase to the prethermal trivial phase.

In order to estimate $T_c$ and by extension the critical energy density of the initial state $\epsilon_c$, we perform a quantum Monte Carlo simulation to understand the transition temperature of $D^*$.
Unfortunately, the full $D^*$ depends on the frequency of the drive.
Fortunately, since we are working in the large frequency regime, we expect the transition to be dominated by the zeroth order term of $D^*$, given by $D$:
\begin{align}
  D = J \sum_{i<j}  \frac{\sigma^z_i\sigma^z_j}{|i-j|^\alpha} + J_x\sum_{i=0}^{L} \sigma^x_i\sigma^x_{i+1} + h_x \sum_{i=0}^L \sigma^x_i
\end{align}
For ease of the numerical methods, for this analysis we invert our Hamiltonian by taking $J$ to be negative, inverting the spectrum of the system.
In this case, the bottom of the spectrum corresponds to the ferromagnetic ordered regime we observe at the top of the spectrum in the numerical calculations of \refsec{sec:Numerics} of the main text.
We note that $h_x$ and $J_x$ are kept positive to ensure that the Hamiltonian is sign problem free.
Since we expect the nature of the transition to be classical, we believe the difference of sign in these couplings does not significantly change the position or properties of the transition.
In fact, when comparing our quantum Monte Carlo results to the classical model with $J_x=h_x=0$, the location of the transition does not change; we believe flipping the sign of these couplings will not alter the stability and location of the phase.

To accommodate the periodic boundary condition of our problem, we modify the simple power law behavior to the closest periodic function that describes a long-range decay,
\begin{equation}
  \frac{1}{|i-j|^\alpha} \to \left(\frac{\pi/L}{\sin |i-j|\pi/L}\right)^\alpha ~,
\end{equation}
as it avoids any discontinuity in the derivative of the interaction.

\begin{figure}[t]
  \centering
  \includegraphics[width=3.2in]{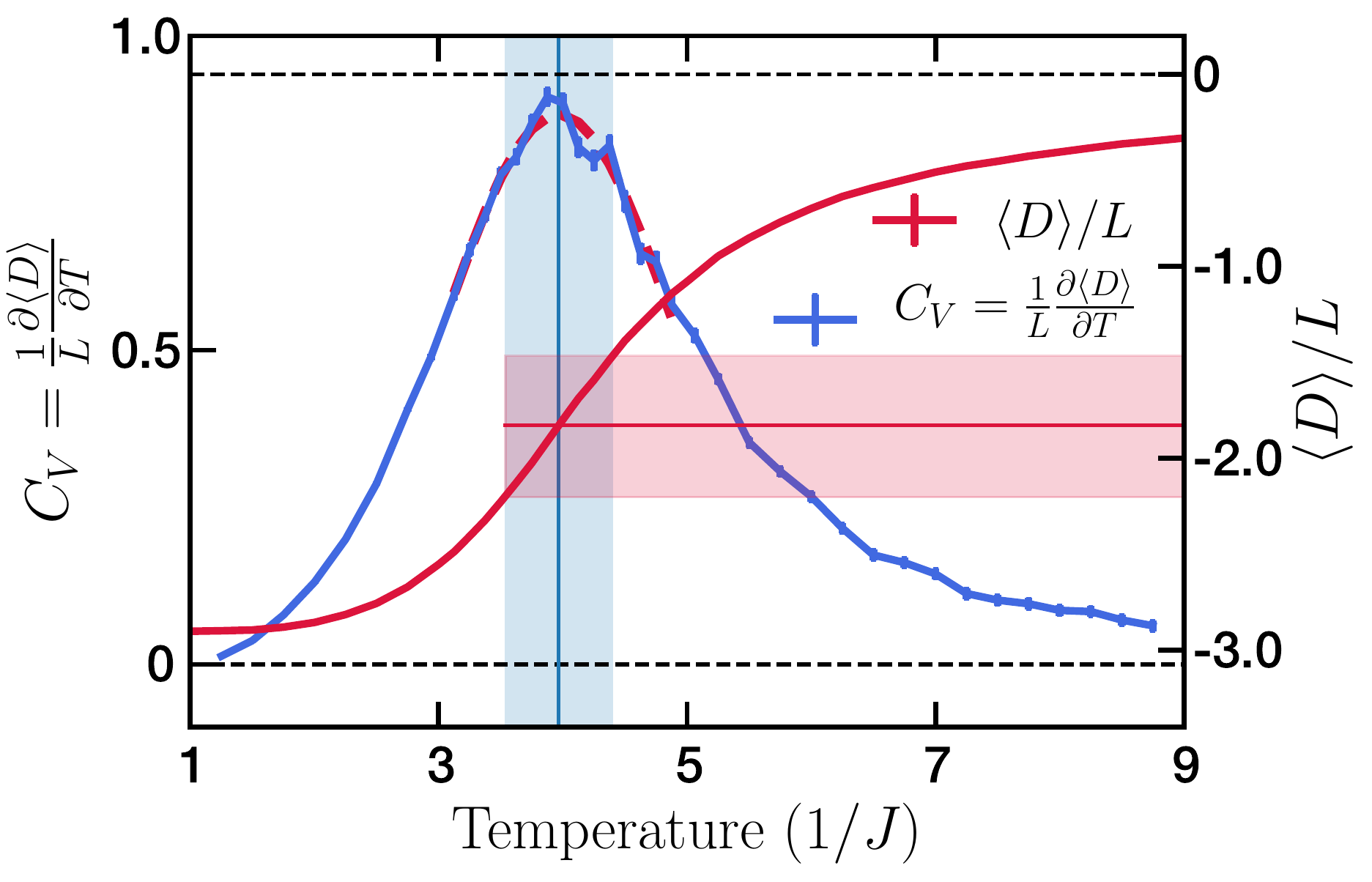}
  \caption{
    Quantum Monte Carlo calculation exhibiting a peak in the heat capacity.
    We identify the position of the peak as the position of the finite size transition or crossover between paramagnetic and ferromagnetic phases.
    This occurs at $T_c^{L=22}/J = 3.97 \pm 0.44$.
    Using the measured value of the energy density $\langle D\rangle / L$ as a function of temperature, we can directly obtain the critical energy density $\epsilon_C^{L=22} = -(1.83 \pm 0.37)$.
  }
  \label{fig:QMC_Cv}
\end{figure}

For the case of this numerical investigation, we are interested in the finite size crossover regime between the ferromagnetic and paramagnetic phases.
This is of particular importance to correctly estimate the critical temperature, as long-range interacting systems often exhibit significant finite size effects.

To diagnose the crossover, we make use of the heat capacity of the system which should present a divergence in the thermodynamic limit.
In the finite system case, no true divergence occurs, but the presence of a peak in $C_V$ corresponds to a finite size transition or crossover.
The position of such a peak can then be used for estimating the critical temperature of the finite system $T_C^{L=22}$.

Using the information about the energy density of the system, as illustrated in \rfig{fig:QMC_Cv}, we numerically differentiate the data with respect to temperature to obtain the heat capacity of the system.
The position of the transition is then obtained by fitting the top of the peak in heat capacity to a Gaussian distribution.
We estimate the uncertainty region associated with $T_c^{L=22}$ as the region where the Gaussian distribution remains above $90\%$ of its peak value (blue shaded region), leading to the estimation:
\begin{align}
  T_c^{L=22}/J = 3.97 \pm 0.44~.
\end{align}

Finally, we can use the energy density curve to translate between critical temperature $T_C^{L=22}$ and the critical energy density $\epsilon_C^{L=22}$ (red shaded region):
\begin{align}
  \epsilon_C^{L=22}/J = -(1.83 \pm 0.37)~.
\end{align}


\bibliography{ref-autobib,ref-manual}

\begin{thebibliography}{103}
\bibitem [{fta()}]{fta}%
  \BibitemOpen
  \href@noop {} {}\bibinfo {note} {Counter-examples of this broad phenomenology
  are integrable models or those displaying a many-body localized
  phase.}\BibitemShut {Stop}%
\bibitem [{ftb()}]{ftb}%
  \BibitemOpen
  \href@noop {} {}\bibinfo {note} {Previous proofs that the dynamics of local
  observables are correctly captured by the prethermal Hamiltonian (within the
  prethermal regime) can be divided into two categories: those applicable in
  prethermal regimes with no emergent symmetry, and those applicable in
  prethermal regimes with an emergent symmetry. In the former, proofs exist for
  short \cite{Kuwahara_1508,Abanin_1509,Else_1607} and long-range interacting
  systems with power-law $\alpha > 2d$ \cite{Kuwahara_1508}. In the latter,
  proofs only exist for short-range interacting systems \cite{Else_1607}. Our
  work considers the latter category, extending current results to long-range
  interacting systems with power-law $\alpha > 2d$.}\BibitemShut {Stop}%
\end{thebibliography}

\end{document}